\documentclass[10pt]{article}
\usepackage{graphicx}
\usepackage{epsfig,color}

\newcommand{\be}{\begin{eqnarray}}
\newcommand{\ee}{\end{eqnarray}}

\def\nn{\nonumber}
\def\ltap{\ \raisebox{-.4ex}{\rlap{$\sim$}} \raisebox{.4ex}{$<$}\ }

\begin{document}

\title{Neutrinoless Double Beta Decay and Heavy Sterile Neutrinos
}
\author{
Manimala Mitra$^a$ \thanks{email: \tt manimala.mitra@lngs.infn.it},~
Goran Senjanovi\'c$^b$ \thanks{email: \tt goran@ictp.it},~
Francesco Vissani$^a$ \thanks{email: \tt francesco.vissani@lngs.infn.it}
\\\\
{\normalsize \it$^{a}$ INFN, Laboratori Nazionali del Gran Sasso, Assergi (AQ),  Italy }\\
{\normalsize \it$^b$ ICTP,  Trieste, Italy }\\ \\ 
}
\date{\today}

\maketitle
\begin{abstract}

The experimental rate of neutrinoless double beta decay
can be saturated by the exchange of virtual sterile neutrinos,  
that mix with the ordinary neutrinos and are 
heavier than 200 MeV.
Interestingly, this hypothesis is subject
only to marginal experimental constraints, because of the new 
nuclear matrix elements.
This possibility is analyzed in the context of
the Type I seesaw model, performing also 
exploratory investigations of the 
implications for heavy neutrino mass spectra, rare decays of mesons as well as neutrino-decay search, LHC,
and lepton flavor violation.
The heavy sterile neutrinos can saturate the rate
only when their masses are below some 10 TeV, but
in this case, the suppression of the light-neutrino
masses has to be more than the ratio of the electroweak scale and the
heavy-neutrino scale; i.e., more suppressed than the naive seesaw expectation.
We classify the cases when this condition holds true in the minimal
version of the seesaw model, showing its compatibility (1)~with
neutrinoless double beta  rate being dominated by heavy neutrinos and (2)~with
any light neutrino mass spectra. 
The absence of
excessive fine-tunings and the radiative stability of light neutrino mass matrices, together with 
a saturating sterile neutrino contribution, imply
an upper bound  on the heavy neutrino masses of about 10 GeV. 
We extend our analysis to the Extended seesaw scenario, 
where the light and the heavy sterile neutrino contributions are
 completely decoupled, allowing the sterile neutrinos to saturate the present experimental bound on neutrinoless double beta decay. 
In the models analyzed, 
the rate of this process is not strictly connected with 
the values of the light neutrino masses,  and a fast transition rate is 
compatible with neutrinos lighter than 100 meV.

\end{abstract}

\newpage

{\small
\centerline{\bf\large Contents}
\begin{enumerate}
\item[\bf 1.] {\bf Introduction \hfill 2}
\item[\bf 2.] {\bf Light and heavy neutrino exchange \hfill 4}\\[-4ex]
\begin{enumerate}
\item[2.1] Parameters of the $0\nu 2\beta $ amplitude \hfill 4
\item[2.2] Role of the nuclear matrix elements \hfill 5
\end{enumerate}
\item[\bf 3.] {\bf Type I seesaw and the nature of the $0\nu 2\beta $ transition \hfill 7}\\[-4ex]
\begin{enumerate}
\item[3.1] Notation \hfill 8
\item[3.2] Naive expectations for $0\nu 2\beta $ in Type I seesaw model \hfill 9
\item[3.3] Departing from the naive expectations for neutrino masses \hfill 12
\item[3.4] A dominant role of heavy neutrino exchange in $0\nu 2\beta$\hfill 15
\end{enumerate}
\item[\bf 4.] {\bf Going beyond Type I seesaw \hfill 24}\\[-4ex]
\begin{enumerate}
\item[4.1]  Extended seesaw \hfill 25
\item[4.2]  Extended seesaw and $0\nu 2\beta$ transition \hfill 28
\item[4.3]  Constraining $m-m_s$ parameter plane \hfill 30
\end{enumerate}
\item[\bf 5.] {\bf Summary and discussion \hfill 32}
\item[] {\bf Appendices \hfill 34}
\item[] {\bf References \hfill 39} 
\end{enumerate}
}

\section{Introduction} 

The study of neutrinoless double beta decay ($0\nu 2\beta$)  transition  $ (A,Z) \to (A,Z+2)+2e^{-}$ 
 has a special relevance for testing the physics beyond the standard model. 
On the experimental side, there is a very lively situation \cite{epj, cuoricino,igex, nemo,Gerda,Cuore} 
and even an experimental claim, that the $0\nu 2\beta$ transition has been already 
measured at present \cite{klapdor}; but even postponing a judgment on these findings, 
it is remarkable that there are realistic prospects for order-of-magnitude 
improvements in the search for the $0\nu2\beta$ 
lifetime 
\cite{Gerda, Cuore, exo,supernemo,Majorana0,lucifer,snop,klz,cobra,next}. 
On the theoretical side, accepting that neutrinos have masses, see  
\cite{solar,kl, atm,macro, soudan, k2k,t2k,minos,chooz} and 
\cite{book1, book2, book3, book4, bookm, bookk,  rev-ponte, review-osc-petcov, 
mik-shap, rev-koshiba, gelmini-rev, kail, nir-garcia-rev, zuber,  rev-moha-smir, visreview, Senjanovic:2011zz,  Fogli2011osc, limits}, 
the investigation of $0\nu 2\beta$ becomes the most natural option; moreover, the observation of lepton-number violating processes would be a cogent manifestation of incompleteness of the standard model, and could be even considered as a step toward the understanding of the origin of 
the matter.\footnote{In fact, the $0\nu 2\beta$ process  can correctly be described as a nuclear transition in which some basic constituent of the ordinary matter--i.e., two electrons--are created. Furthermore, once recognized the existence of transitions among different leptonic flavors (again, neutrino oscillations) one should conclude that any global symmetry of the standard model excepting at most $B-L$ is broken; thus,  the observation of a lepton number violating process as $0\nu 2\beta$ would imply that also the baryon number is at some level violated.}

However, once we enter the theoretical discussion, one has to stress immediately an evident but essential point, that the meaning of $0\nu 2\beta$ depends on the model. To naive eyes, the fact that no neutrino is emitted in this transition leads one to wonder what is the link of $0\nu2\beta$ with neutrinos. The most popular theoretical answer is that, the hypothesis that neutrinos have Majorana mass \cite{Majorana}  suggests that the exchange of virtual, light neutrinos is a plausible mechanism for the occurrence of $0\nu 2\beta$ \cite{0nu2beta-old}. An additional (and more recent) theoretical argument is that, listing the effective operators that obey the standard model gauge symmetry, Majorana neutrino masses arise already as dimension-five operators \cite{dim5}; thus any further contribution to  $0\nu2\beta$ (or, say, to proton decay)  will be due to the  higher-dimensional operators, and, as such,  are expected to be suppressed. However, there is an implicit  assumption  underlying this approach: namely, that the new physics is at very high scale. This assumption may or may not hold. In fact, the possibility that $0\nu2\beta$ is mostly due to mechanisms different from the conventional one (light neutrino exchange) has been proposed since long   \cite{feinberg} and it is actively discussed, see e.g., \cite{ms0nu2beta, mwex, pion-ex, hirsch,  allanach, vogel, choi,  tello, Ibarra, blennow}.

Among the simplest renormalizable extensions of the standard model, that are able to account for neutrino masses,  the addition of heavy sterile  neutrinos--i.e., so called 
Type I seesaw \cite{seesawM, seesaw-Goran, seesaw-Yan,seesaw-Ramond}--is largely considered as the minimal option. These new particles, if lighter than a some 10 of TeV (see below) can act 
as new sources of lepton number violation, or in other terms, as 
potential additional contributors to $0\nu 2\beta$. The  main goal of the present paper is a systematic study of this possibility, considered occasionally in the past \cite{Ibarra}.
Note incidentally that the possibility that the heavy neutrinos are not ultra-heavy, and thus can be potentially tested experimentally, is the first one that has been considered in \cite{seesawM, seesaw-Goran}.\footnote{A theoretical objection to the hypothesis of Type I seesaw, defined introducing the heavy sterile neutrinos as pure gauge singlets  as in \cite{seesaw-Yan}, is that the mass of right-handed neutrinos is not related to any gauge symmetry, differently from the masses of all other known particles.  But, as pointed out originally \cite{moha} and  stressed  recently \cite{tello},  
even if such a gauge symmetry is introduced (most plausibly through a SU(2)$_R$ group) similar considerations hold: the mechanism of $0\nu 2\beta$  is not necessarily light neutrino exchange. The phenomenology of this type of models is however different and to some extent richer than the one we will describe.}

\bigskip
 
   The outline of the paper is the following: In Sect.~\ref{gen}, we review  the basics of light as well as heavy neutrino exchange in  $0\nu 2\beta$ process. The most important result of this section is given  in Sect.~\ref{nme}; using the updated nuclear matrix element of reference \cite{f2010}, the bound on the active-sterile mixing  coming from $0\nu 2 \beta$ transition is re-examined.  The improvement in the uncertainty of the nuclear matrix element  leads the bound to be one order of magnitude tighter than the  existing one \cite{f2005}. On the face of this analysis, the bounds coming from 
   other potentially relevant experiments, see \cite{atre} for a review, 
   have become relatively less significant. Also see \cite{mohabnd} for a specific realization, where the bound on sterile neutrino mass and mixing has been obtained from $0\nu2\beta$, astrophysical and cosmological informations. 

Following this, we provide  the detailed  analysis  on  the nature of $0\nu 2 \beta$ transition  for the simplest extensions of the standard model with heavy sterile neutrinos.  In Sect.~\ref{typeI-intro}, we first concentrate on the usual   Type I seesaw,  and  later in Sect.~\ref{ex},  we extend the  discussion  to the other seesaw scenarios as well, namely  Extended seesaw  \cite{Kang-Kim, Parida}. 

In Type I seesaw, the generic naive expectation (as precisely defined in Sect.~\ref{m-M}) leads us to believe that the heavy sterile neutrino contribution in $0\nu 2\beta$ process is much smaller  than the light neutrino contribution. For one generation of light and and heavy sterile neutrino state,  this naive expectation is established  by the very basic seesaw structure (see Sect.~\ref{on}). Going beyond one generation, it is however possible to reach the opposite extreme; i.e., one can obtain a dominant and even saturating \cite{epj} sterile neutrino contribution, which is not inherently linked with the light neutrino contribution. The systematic study of this possibility requires, that the light neutrino contribution should be smaller than the naive expectation suggested by seesaw. This consideration is  carried out in  Sect.~\ref{see0}, where we analyze  the vanishing seesaw condition and its perturbation,  leading  to small neutrino masses. We classify the different cases, where the light neutrino spectra is not necessarily degenerate, and even possibly hierarchical.  All these cases can provide a dominant sterile neutrino contribution in $0\nu 2\beta$ process, as discussed   in Sect.~\ref{multi-contact}. We derive an useful parameterization in Appendix \ref{ma} to study these cases analytically. For completeness we  also provide explicit numerical example in  Sect.~\ref{num-ex}. 

Assuming a saturating contribution \cite{epj}  from heavy sterile neutrino exchange, in Sect.~\ref{nv}, we provide
a naive estimation on  the prospect of heavy Majorana neutrino search at 
LHC \cite{aguila-LHC-d,aguila-LHC-m,  smirnov}, as well as in lepton flavor violating process \cite{mutoegamma}; these prospects  turn out to be  weak due to the stringent constraints coming from $0\nu 2\beta$ process. The possible issue, like radiative stability of the light neutrino mass matrices,  below the naive seesaw expectation,  is  discussed in Sect.~\ref{qft}.  This, along with the request of  a dominant heavy sterile neutrino contribution in $0\nu 2 \beta$ process gives an upper bound of about $10$ GeV on the heavy sterile neutrino mass scale (Sect~\ref{upbndem}) and  also an upper bound on the necessary perturbation  of the vanishing seesaw condition.

Following the analysis of Type I seesaw, in Sect.~\ref{ex}  we then consider a natural seesaw extension, 
namely Extended seesaw scenario \cite{Kang-Kim, Parida}. We describe  the basics of Extended seesaw  
in Sect.~\ref{ex-see} and after that quantify  the different sterile neutrino contributions  in 
$0\nu 2\beta$ process (Sect.~\ref{ex-0}). As for the Type I seesaw, the sterile neutrino states 
in this case can also give significant contributions in $0\nu 2\beta$ process. In this particular 
seesaw scenario, the light  neutrino contribution depends on a small lepton number violating parameter, 
while to the leading order, the active-sterile neutrino mixing is independent of that parameter.  Due to 
this particular feature,  the sterile neutrino contribution to $0\nu 2 \beta$ process is totally  independent 
of the light neutrino contribution.   In the next section i.e., Sect.~\ref{cons-ex}, we discuss the possibility 
of obtaining a saturating contribution \cite{epj} from the sterile neutrino states, the scope of finding sterile neutrinos  at LHC \cite{aguila-LHC-d,aguila-LHC-m,smirnov}, and as well as the possibility of  obtaining  a rapid lepton flavor violation\cite{mutoegamma}.  Possible
issues, such as, the higher-dimensional correction to the active-sterile
mixing angle and  $0\nu 2\beta$ transition amplitude, that  will arise  due to the small lepton number violating scale,
 has also been discussed. The details of higher-dimensional correction to the mass and
mixing matrix has been evaluated in the Appendix \ref{diagonalization-extended}. Finally, in Sect.~\ref{sum}, 
we present the summary of our work.

The analysis presented in this paper clearly  shows that the heavy sterile neutrino states can 
certainly dominate the $0\nu 2\beta$ transition; this  possibility has the potential to overcome any  
conflict \cite{fogli-05} between the cosmological bound and the experimental hint on $0\nu 2\beta$ obtained by Klapdor and  collaborators \cite{klapdor},
or more in general it permits to reconcile a fast (and potentially observable) 
rate of $0\nu 2\beta$ and small neutrino masses. 
In addition, for Type I seesaw, 
  the demand of radiatively  stable light neutrino mass matrices  lowers the mass scale of the heavy sterile 
neutrino states below 10 GeV. Note that, based on the updated nuclear matrix elements \cite{f2010}, 
the bound  from $0\nu 2\beta$ is now much more stringent than the previous consideration \cite{f2005, atre}.  
Further improvement in meson as well as neutrino-decay experiments  have 
a certain potential to  provide us with more information on (and possibly a measurement of) 
the active-sterile mixings, which is similar to the conclusion obtained in the
$\nu$MSM model by Shaposhnikov and collaborators \cite{mesus}.

\section{Light and heavy neutrino exchange\label{gen}}
The phenomenological possibility that some heavy  neutrino state contributes 
to $0\nu 2\beta$ transition amplitude  has been remarked since long:
see \cite{MYRG} for a model that however contradicts the current understanding of neutrino masses 
and interactions,  \cite{moha} for the first modern discussion within gauge theories, 
\cite{0nu2beta-pet} for a further earlier contribution. 

The relevant discussion is summarized in this section, emphasizing: 
1) the large number of free parameters, Sec.~\ref{par};
2) the role of (and the remarkably large uncertainties in) 
nuclear matrix elements for heavy neutrino exchange, 
Sec.~\ref{nme}. 
Subsequently, we will apply  the results of this  discussion 
to the specific model of interest for heavy neutrinos,  Type I seesaw, 
where the number of free parameters is smaller.

\subsection{Parameters of the $0\nu 2\beta$ amplitude\label{par}}
As we recall, several experiments testify that  the usual left 
neutrinos $\nu_\ell$ ($\ell=e,\mu,\tau$) are subjected  to flavor 
transformations, as expected if they have mass. Considering only the  minimal case 
these neutrinos have Majorana mass; this minimal ansatze  amounts to postulate that the particle content  of the 
SU(2)$_L\times$U(1)$_Y$ standard model theory remains  the same, while  the Lagrangian  is 
endowed with a non-renormalizable 
term, which  after spontaneous symmetry breaking reads
$\frac{1}{2} \nu_{\ell} (M_\nu)_{\ell\ell'} \nu_{\ell'}+$h.c.,  with
\be\label{pn}
M_\nu=U^* \mathrm{diag}{(m_i)} U^\dagger.
\ee
In the above, the unitary matrix $U$ is the leptonic mixing matrix, $\nu_\ell= U_{\ell i} \nu_i$, $\nu_l$ and $\nu_i$ are 
respectively the flavor and mass basis; the physical masses 
$m_i$ of the neutrinos 
  are real and non-negative;  the possible Dirac and Majorana  phases are included into $U_{\ell i}$. 

This hypothesis not only accounts for oscillations, but also 
has some predictive power for the lepton number violating neutrinoless double beta decay process. Indeed,  
the ee-element of the mass matrix: 
\be
|(M_\nu)_{ee}|=|   \; \sum_i U_{ei}^2\; m_i\; |,
\ee
contributes to the neutrinoless double beta decay rate (see \cite{oli-rev}, 
\cite{vogel-rev}, \cite{bilenky-rev}, 
\cite{werner-rev} for recent reviews). Evidently this quantity cannot exceed $\sum_{i} |U_{ei}^2| m_i$,
and since the differences of neutrino mass squared are measured and 
the relevant elements of the mixing matrix are sufficiently well-known, 
there is a upper bound on $|(M_\nu)_{ee}|$ as a function of the lightest 
neutrino mass  $m_{\mbox{\tiny min}}$ \cite{visreview, vuso, vis-fer}, or equivalently
of the other mass scales,  such as the mass probed in direct search for   neutrino masses 
$m_\beta^2=\sum_i |U_{ei}^2| m_i^2$,
 or    the sum of the neutrino masses  $m_{\mbox{\tiny cosm}} =\sum_{i} m_i$ 
probed in cosmology  (see \cite{Wmap7} and e.g., \cite{stru,  cosmobound}).  
  For the lightest neutrino mass scale   $m_{\mbox{\tiny min}}>0.1$ eV, relevant to the case of present 
experimental sensitivities, one can approximate the bound as,
\be\label{foo1}
m_\beta\approx m_{\mbox{\tiny cosm}}/3\approx m_{\mbox{\tiny min}} >|(M_\nu)_{ee}|,
\ee 
with an accuracy better than 10\%, that moreover improves for normal mass hierarchy.
It is interesting that the experimental hint on $0\nu 2\beta$ obtained by Klapdor and 
collaborators \cite{klapdor}, according to \cite{fogli-05}, challenges  the bound obtained in cosmology, though 
this conclusion depends on {\em which} cosmological bound is considered and {\em which} nuclear matrix
element is used, see below for a more quantitative statement.

Of course, in less minimal models new contributions to $0\nu 2\beta$ are expected and 
 the   conflict between the cosmological bound and the experimental hint 
on $0\nu 2\beta$ \cite{klapdor} can be overcome. 
This can happen when
the usual left-handed neutrinos contain heavy neutrino components, too, 
\be \label{conva}
\nu_\ell=\sum_{i=1}^3 U_{\ell i} \nu_i + \sum_{i=1}^{n_h} V_{\ell i} N_i,
\ee
where we  have considered  $n_h$ heavy neutrinos $N_i$ with masses $M_i$ 
and small mixing $|V_{\ell i}|\ll 1$ (this condition is discussed later). 
In fact, the amplitude of $0\nu 2\beta$ is proportional to
\begin{equation}
\mathcal{A}=  \frac{U_{ei}^2\ m_i}{p^2} -\frac{V_{ei}^2}{M_i}, 
\label{amp}
\end{equation} 
which includes the contribution due to heavy neutrino exchange
(also called direct contribution, or contact term).
Here, $p^2$ is the virtuality of the exchanged neutrino. 
We have $p^2 <0$, for the  time component $p_0$ is of the order
of the $Q$ value of the reaction, few MeV, wheres its space  component is much larger and essentially 
determined by the separation between neutrons, $|\!|\; \vec{p}\; |\!|\sim \hbar c/\mbox{fm}\approx 
200$ MeV. In the above expression for the amplitude, we considered the case of interest, 
\begin{equation}\label{wih}
m_i\ll 200\ \mathrm{ MeV}\ll M_i.
\end{equation}
Indeed, if we have a virtual Majorana neutrino with mass $\mu$ and with momentum $p$,  
its propagator implies that the amplitude is proportional to:
\begin{equation}
\mu/(p^2-\mu^2),
\label{prop}
\end{equation} 
which in the limits of $\mu\ll  200\ \mathrm{ MeV}$ or $\mu\gg  200\ \mathrm{ MeV}$ reduces
to $\mu/p^2$ and $-1/\mu$ respectively.

\subsection{Role of the nuclear matrix elements\label{nme}}

A traditional expression for the $0\nu 2\beta$ half-life is:
\begin{equation}
\frac{1}{T_{1/2}}=G_{0\nu} \left| \mathcal{M}_\nu \eta_\nu +\mathcal{M}_N \eta_N \right|^2,
\label{lif}
\end{equation} 
where the parameters of light and heavy neutrinos, as defined in  Eq.~\ref{amp},
are presented through  the complex dimensionless  parameters $\eta_\nu=U_{ei}^2 m_i/m_e$ 
and $\eta_N=V_{ei}^2 m_p/M_i$, and, conventionally, the reference mass scales are 
chosen to be the electron and the proton masses $m_e$ and $m_p$. In the above, $\mathcal{M}_{\nu}$ and 
$\mathcal{M}_N$ are the nuclear matrix elements corresponding to the light and heavy neutrino exchange in 
$0\nu 2\beta$ process.

Whenever necessary, we will assume for the nuclear matrix elements 
the values given in  \cite{f2010}, where we read that, in the case of $^{76}$Ge 
$0\nu 2\beta$ transition,
the `phase space' factor  is  
$G_{0\nu}=7.93\times 10^{-15}$ yr$^{-1}$ and the matrix elements are 
$\mathcal{M}_\nu=5.24\pm 0.52$ and $\mathcal{M}_\nu=363\pm 44$. 
These values imply, for instance, that the result of Klapdor, 
$T_{1/2}(^{76}\mathrm{Ge})=2.23^{+0.44} _{-0.31}\times 10^{25}$ yr \cite{klapdor} would be consistent with
$\eta_\nu=4.5\times 10^{-7}$ and $\eta_N=0$;
with $\eta_\nu= 0$ and $\eta_N=6.5\times 10^{-9}$;
or with linear combinations of these limiting cases (possibly allowing for an overall sign).  
The first possibility,   where only the neutrino mass is present, 
would imply
\be
|(M_\nu)_{ee}|=0.23\pm 0.02\pm 0.02\mbox{ eV}, 
\ee
namely, a degenerate neutrino spectrum, partly testable 
with KATRIN experiment and 
 of great interest for cosmological investigations, since it implies
 $m_{\mbox{\tiny cosm}}>0.69$~eV (see Eq.~\ref{foo1}).
In the following, we will be especially interested to explore the 
opposite limit, when the $0\nu 2\beta$ is dominated by the second term, and the neutrino mass spectrum
is not necessarily degenerate, and  even possibly  hierarchical.

The traditional expression in Eq.~\ref{lif}
can be recast into the following equivalent form:
\begin{equation}\label{greb}
\frac{1}{T_{1/2}}=K_{0\nu} \left| \frac{U_{ei}^2 m_i}{\langle p^2 \rangle} - \frac{V_{ei}^2}{M_i}  \right|^2,
\label{kk}
\end{equation}
where we set $K_{0\nu}=G_{0\nu} (\mathcal{M}_N m_p)^2$ and following \cite{tello} 
we defined,
\begin{equation}\label{mumo}
\langle p^2 \rangle\equiv -m_e m_p \frac{\mathcal{M}_N}{\mathcal{M}_\nu}.
\label{psq}
\end{equation}
 Note the resemblance  of Eq.~\ref{kk} with  the expression of the amplitude given in Eq.~\ref{amp}.
With the values of \cite{f2010}, we get $\langle p^2 \rangle=-(182\mbox{ MeV})^2$, in remarkable 
accordance with the rough expectations for $p^2$ described in Sec.~\ref{par}.

An alternative  presentation of the lifetime, valid for generic values of the neutrino masses, is 
\begin{equation}\label{gnr}
\frac{1}{T_{1/2}}=G_{0\nu} \left| \Theta_{ei}^2\ \mathcal{M}(\mu_i)\ \mu_i/m_e \right|^2.
\end{equation} 
This agrees with Eq.~\ref{lif} if
one identifies the following sets of parameters, 
\be
(\mu_i,\ \Theta_{ei})=
\left\{
\begin{array}{ll}
(m_i,\ U_{ei}) & \mbox{ when }\mu_i\to 0,\\[1ex]
(M_i,\ V_{ei}) & \mbox{ when }\mu_i\to \infty
\end{array}
\right.
\ee
and, at the same time, the following limits hold: 
\begin{equation}
 \lim_{\mu\to0} \mathcal{M}(\mu)=\mathcal{M}_\nu \ \mbox{and}\ 
 \lim_{\mu\to\infty} \mu^2  \mathcal{M}(\mu)=m_e m_p\ \mathcal{M}_N 
 \equiv -  \langle p^2\rangle \mathcal{M}_\nu,
\end{equation}
the scale of comparison being $-\langle p^2\rangle \sim $ (200 MeV)$^2$.
A simple and useful analytical approximation of the 
general expression of Eq.~\ref{gnr}
has been proposed in \cite{kovalenko}: 
\begin{equation}\label{sa}
\frac{1}{T_{1/2}}=K_{0\nu} \left| \Theta_{ei}^2 \frac{ \mu_i}{\langle p^2 \rangle-\mu_i^2}  \right|^2.
\end{equation}
The advantage  of this formula is  that it 
emphasizes the role of the neutrino propagator given in Eq.~\ref{prop} and 
allows one to switch easily from  the regimes of light and heavy  neutrino exchange
(compare with Eq.~\ref{greb} in the limit of light and heavy neutrinos),
even being slightly inaccurate in the region where $-\langle p^2 \rangle\sim \mu_i^2$ 
\cite{blennow, f2005}.

\begin{figure}[t]
\begin{center}
\includegraphics[width=0.5\textwidth, angle=270]{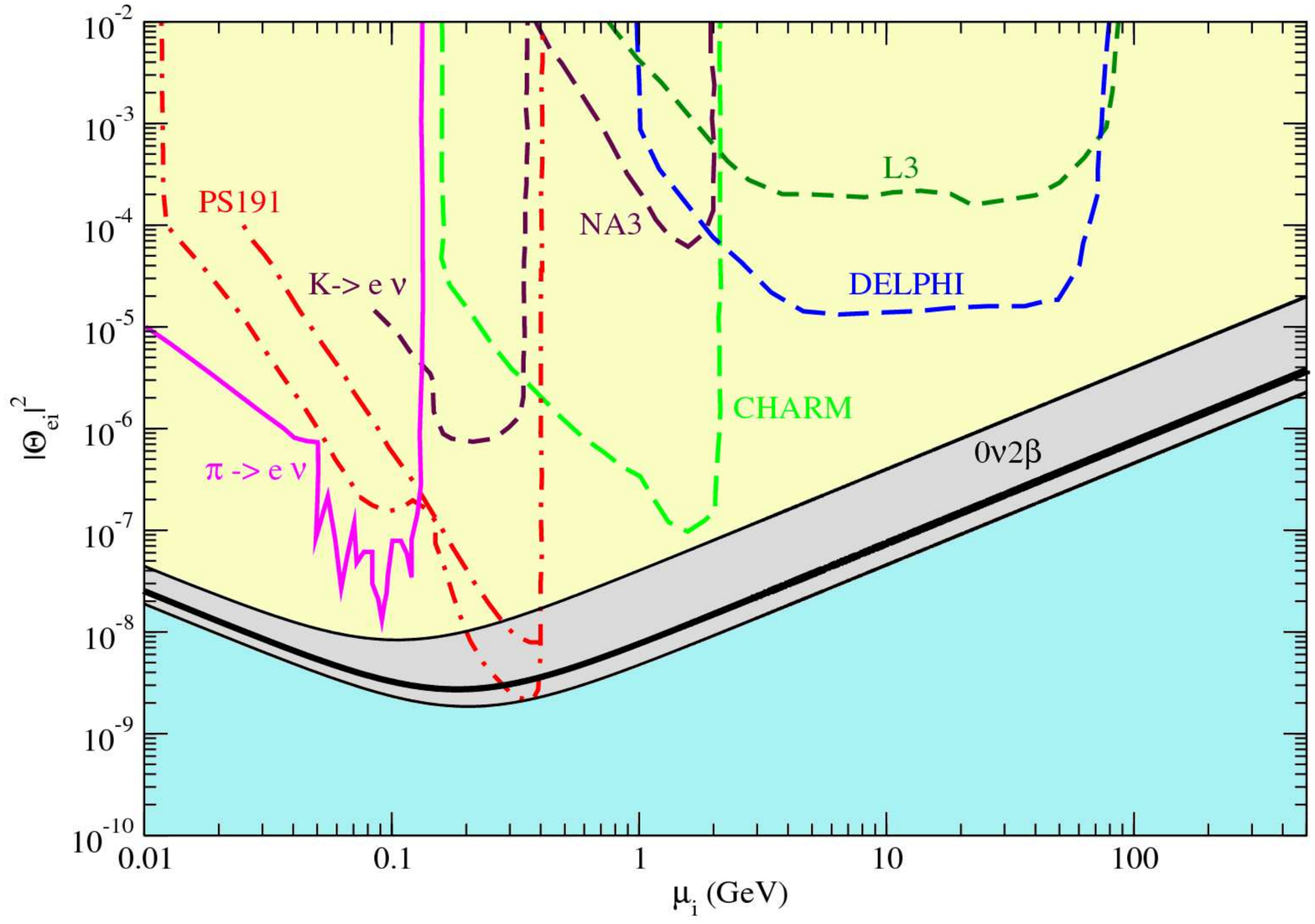}
\end{center}
\caption{\label{plut}   Bounds on the mixing between the electron neutrino and a (single) 
heavy neutrino as obtained from Eq.~\ref{sa}. The upper thin black line corresponds to the result of 
ref.~\cite{f2005}, the thick black  one to ref.~\cite{f2010}, while the lower thin black line  is an 
attempt to convey a conservative assessment on the residual uncertainty. For comparison, 
we also show other experimental constraints
as compiled in \cite{atre}. See text for details.}
\end{figure}

Using the last formula and the present experimental bound on $0\nu2\beta$
lifetime
$T_{1/2}>1.9\times 10^{25}$ yr \cite{epj}, we obtain the upper  bound on
the mixing $|\Theta_{ei}|^2$, which is shown  in Fig.~\ref{plut}. When
this is compared with the other experimental constraints on the model,
compiled by \cite{atre}, it is quite evident that $0\nu 2\beta$ play the
most important role. The  details of the figure are as follows,

\begin{itemize}

\item

The upper yellow region is disallowed from neutrinoless double beta decay consideration. Part of this 
region is as well constrained from different meson decays, neutrino decay-searches  as well as  other experiments, shown explicitly in the figure. 
The lower blue region is the allowed one from $0 \nu 2\beta$ as well as the various above mentioned  experiments.
\item
The middle grey band which has been obtained considering the exchange of a single heavy neutrino
 with mass $\mu_i$ in $0\nu 2\beta$ process, corresponds to the uncertainty of the nuclear matrix elements 
$\mathcal{M}_{\nu}$ and $\mathcal{M}_{N}$.  For the thick black line in this grey band,  we have adopted the 
parameters of \cite{f2010}, $\mathcal{M}_{\nu}=5.24$ and $\mathcal{M}_{N}=363$, 
i.e., $\langle p^2 \rangle=-(182\mbox{ MeV})^2$; for the upper thin black line $\mathcal{M}_\nu=3$
and $\mathcal{M}_N=69$, namely, $\langle p^2 \rangle=-(105\mbox{ MeV})^2$; while for the lower
thin line $\mathcal{M}_\nu=7$ and $\mathcal{M}_N=600$, namely, $\langle p^2 \rangle=-(203\mbox{ MeV})^2$. 
The upper line agrees numerically with the results of \cite{f2005} (see also 
\cite{kovalenko}); the large difference with \cite{f2010} should be attributed to the 
new short range correlations and improved  nucleon form factors.\footnote{We thank
F.~\v{S}imkovic for clarifying discussions on this issue.}
The lower line, instead, is meant to convey a conservative idea of the uncertainties;  see 
\cite{s2011,s2011bis}  for  the most stringent upper bound on $0\nu 2 \beta$ we could have at present, 
depending on the size of the nuclear matrix elements.  For each of these black lines, the region above the 
line is disallowed from $0\nu 2\beta$ transition.

\item
The span of values of $\mathcal{M}_\nu$ used in Fig.~\ref{plut} is much more conservative
than the one of \cite{f2010}, quoted above. It corresponds to the 
range given in the compilation \cite{gomez}, see their Fig.\ 1. By comparing with a similar compilation
of about 10 years ago  \cite{vis-fer}, see their Fig.\ 2, one understand that the new nuclear physics
calculations obtained a 
reduction of the uncertainty of a factor of two for the regime of light neutrino exchange.
This improvement is 
of enormous importance: the lifetime scales only as the square of the nuclear matrix elements,
while in presence of background, the improvement of the bound on the lifetime scales as the square 
root of the exposure.

\item

The limits from $0\nu2\beta$ which has been derived using the result of  \cite{f2010} and presented in 
Fig.~\ref{plut},  are significantly 
tighter than the previous limits  on mass and mixing  given in \cite{f2005} (the result of \cite{f2005} 
has also been adopted in recent global analysis \cite{atre}).
Conversely, the impact of other constraints, in particular those from meson 
decays, neutrino-decay searches  and other experiments, becomes relatively less important: See again \cite{atre} 
(and in particular their Fig.\ 2)  where full 
reference to the original literature is provided.

\end{itemize}

\section{Type I seesaw and the nature of the $0\nu 2\beta$ transition \label{typeI-intro}}

Type I seesaw is in many regards the simplest
extension of the standard model: only 
 heavy sterile neutrino states are added to 
the spectrum of the SU(3)$_C\times$SU(2)$_L\times$U(1)$_Y$ theory 
\cite{seesawM,seesaw-Goran, seesaw-Yan,seesaw-Ramond}, with a primary purpose to 
account for light neutrino masses in a 
renormalizable gauge model.  However, these heavy states might lead to measurable effects, 
in particular, for the neutrinoless double beta decay.

In this section we discuss the nature of $0\nu 2\beta$ transition within the Type I seesaw \cite{Ibarra, blennow},
emphasizing the possibility discussed occasionally in the literature  that the 
heavy neutrino exchange contribution plays the main role for  $0\nu 2\beta$. 
In the present study, we 
analyze in greater detail the parameter space of Type I seesaw.

Let us describe in detail the outline and scope of this section.
First, we recall the basic notations for the model (Sect.~\ref{bn}).
In Sect.~\ref{m-M} we provide a precise formulation of  a 
naive and widespread expectation: within Type I seesaw, 
the contribution of the heavy neutrino states to the $0\nu 2\beta$ decay is smaller than the one due 
to light neutrino states.
Actually, 
for one generation 
 this naive estimation works perfectly well (see Sect.~\ref{on})
but for more than one generation, it is 
possible to obtain  a large and dominant contribution to $0\nu 2\beta$
from the heavy neutrino states, which is {\it not necessarily inherently} 
linked with the light neutrino contribution. 
This will be discussed in detail, after the mathematical premise of Sect.~\ref{see0}, aimed 
at outlining the cases when the light neutrino masses are 
much smaller than suggested by the naive expectations from seesaw. 
Finally, we discuss in Sect.~\ref{multi-contact}  
the possible cases when heavy neutrino exchange provide us with a large effect in $0\nu 2\beta$. 
We exhibit explicit examples  when this happens.  We prove 
that this possibility can be implemented within Type I seesaw, 
without occurring into limitations on the structure of the light neutrino mass matrix.
As an extreme possibility, we show that it is possible to arrange a 
large contribution from the heavy Majorana neutrino exchange, even when 
the light neutrino contribution to $0\nu2\beta$ is negligible. 
We discuss the possible issues, like radiative stability in Sect.~\ref{qft}, 
the possibility of relatively less  fine-tuning in Sect.~\ref{s},  and  derive bounds on heavy neutrino  
mass scale  and fine-tuning parameter  in Sect.~\ref{upbndem}. Finally, in Sect.~\ref{num-ex} we present 
an explicit numerical example, where  the heavy neutrino contribution  is the dominant one  and the light 
neutrino contribution is negligibly  smaller  than the heavy neutrino contribution. 

\subsection{Notation\label{bn}}
In our subsequent discussion of $0\nu 2\beta$ process  and its relation with  Type I seesaw, we denote  the 
standard model flavor neutrino states by $\nu_L$ and the heavy  Majorana neutrinos by $N_L$. The  Lagrangian 
describing the mass terms is the following,
\be
{\it L}=-\frac{1}{2} \pmatrix{\nu_L & N_L } \pmatrix {0 & M_D^T \cr M_D & M_R} \pmatrix{\nu_L \cr N_L} +\rm{h.c}.
\label{eq:seesaw1}
\ee
For three  generation of standard model neutrinos $\nu_L$ and $n_h$ generation of sterile neutrino state $N_L$, $M^T_D$ and $M_R$ will  be of 
$3 \times n_h$ and $n_h \times n_h$ dimension. From the  above Lagrangian one obtains this following  neutral lepton mass matrix,
\be
M_n=\pmatrix{0 & M^T_D \cr M_D & M_R}.
\label{eq:seesawfull}
\ee
The  neutrino flavor state $\pmatrix{\nu_L & N_L}^T$ is related to the neutrino mass state $\pmatrix{\nu_m & N_m}^T$ by the unitary mixing matrix $\mathcal{U}$ where, 
\be
\pmatrix {\nu_L \cr N_L}=\mathcal{U} \pmatrix{\nu_m \cr N_m}.
\ee
The mixing matrix $\mathcal{U}$  which diagonalizes the above mentioned neutral lepton mass matrix  satisfies the following relation 
$\mathcal{U}^T M_n\; \mathcal{U}=M^d_n$, where $M^d_n$ is the diagonal neutrino mass matrix. We denote $M^d_n$ as follows,  
\be
M^d_n=\pmatrix{\mbox{diag}(m_i) & 0 \cr 0 & \mbox{diag}(M_i) }, 
\ee
where $m_i$ and $M_i$ represent the light and heavy neutrino masses respectively.
It is convenient to introduce a couple of auxiliary matrices $\mathcal{U}_{1,2}$ as $\mathcal{U}=\mathcal{U}_1\mathcal{U}_2$.
 The first matrix $\mathcal{U}_1$  block-diagonalizes $M_n$, namely it satisfies
$\mathcal{U}_1^TM_n\: \mathcal{U}_1=M_{bd}$. 
Subsequently, $M_{bd}$ is further diagonalized by the matrix $\mathcal{U}_2$, that
satisfies the relation $\mathcal{U}_2^TM_{bd}\ \mathcal{U}_2=M^d_n$. Let us denote the block-diagonalized matrix as,
\be
M_{bd} = \pmatrix{M_{\nu} & 0 \cr 0 & M_N}.
\label{masud}
\ee
It is possible to operate a systematic expansion of 
$M_{bd}$ and $\mathcal{U}_1$ is powers of $M_R$  \cite{Grimus-Lavoura}, thus enforcing the 
seesaw approximation, $M_R \gg M_D$. 
Up to leading order in powers of $M_R$, 
we have simply $M_N=M_R$ for the heavy neutrino mass matrix, 
while the  light neutrino mass matrix reads,
\be
M_{\nu}=-M_D^T M_R^{-1} M_D.
\label{mabel}
\ee
Keeping terms up to 2nd order in $M_R^{-1}$, the mixing matrix $\mathcal{U}_1$ is,
\be\label{cr2}
\mathcal{U}_1=\pmatrix{ 1-\frac{1}{2} M^{\dagger}_D{M^{-1}_R}^*M^{-1}_RM_D & M^{\dagger}_D {M^{-1}_R}^* \cr -M^{-1}_RM_D & 1-\frac{1}{2} M^{-1}_R M_D {M^{\dagger}}_D{M^{-1}_R}^*}.
\label{mexel}
\ee
Next, we denote the mixing matrix $\mathcal{U}_2$ as follows, 
\be
\mathcal{U}_2=\pmatrix{U & 0 \cr 0 & W},
\label{mexelb}
\ee
where the mixing matrices  $U$ and $W$ diagonalize  the light and heavy neutrino mass matrices $M_{\nu}$ and $M_R$ respectively: $U^TM_{\nu}\: U=\mbox{diag}(m_i)$ and $W^TM_R\: W=\mbox{diag}(M_i)$.  From Eq.~\ref{mexel} and Eq.~\ref{mexelb}, one immediately obtains, 
\be
\mathcal{U}=\pmatrix { (1-\frac{1}{2} M^{\dagger}_D{M^{-1}_R}^*M^{-1}_RM_D)U & M^{\dagger}_D {M^{-1}_R}^* W \cr -M^{-1}_RM_D U & (1-\frac{1}{2} M^{-1}_R M_D {M^{\dagger}}_D{M^{-1}_R}^*)W}.
\label{itsu}
\ee
To the leading order, the mixing matrix $\mathcal{U}$  is simply, 
\be\label{cr1}
\mathcal{U} = \pmatrix{U & M^{\dagger}_D {M^{-1}_R}^* W \cr -M^{-1}_RM_D U & W}.
\ee
Finally and quite importantly, we note that the mixing between light and heavy neutrino states is $M^{\dagger}_D{M^{-1}_R}^*W$.  According to  convention of Eq.~\ref{conva}, this mixing matrix  
is denoted as $V$, namely
\be
V=M^{\dagger}_D{M^{-1}_R}^*W.\label{bunul}
\ee
In the basis where the heavy Majorana neutrino mass matrix is diagonal, 
$M_R=W^*\mbox{diag}(M_i)W^\dagger$, 
we rewrite the mixing matrix $V$ as follows, 
\be
V=\hat{M}^{\dagger}_DM_i^{-1},\label{bunal}
\ee 
 and the Dirac mass matrix in this basis is simply,
\be\label{dd}
\hat{M}_D=W^TM_D.
\ee
We note in passing, that Eq.~\ref{conva} is actually valid only when $|V_{\ell i}|\ll 1$; the 
deviations that should be expected (due to the unitarity constraints) 
are formally evident from Eq.~\ref{itsu}, and are usually small.

\subsection{Naive expectations for $0\nu2\beta$ in Type I seesaw model\label{m-M}}
In this section, we  aim to define precisely  which are the 
{\em naive expectations} from the Type I seesaw model for various 
interesting measurable quantities. 
Subsequently, we 
argue for the interest in exploring alternative possibilities.

\subsubsection{Heavy neutrino exchange in the single flavor case\label{on}}
Let us begin the analysis by showing that,  
 a single flavor 
Type I seesaw
implies that the light neutrino exchange in $0\nu 2\beta$ process 
is never sub-leading. In other words, it is not possible to attribute the $0\nu 2\beta$ transition to the heavy neutrino exchange  with one light and one heavy neutrinos only.

Assume  one light and one heavy Majorana neutrino $\nu_L$  and $N_L$ with masses
$m_1$ and $M_1$ respectively. The flavor state $\nu_e$ is mixed with the 
mass states as follows,
\be
\nu_e=U_{e1} \nu_1 + V_{e1} N_1.
\ee
and thus the $0\nu 2\beta $ transition amplitude receives a contribution proportional to,
\be
U_{e1}^2 \frac{m_1}{p^2} + 
V_{e1}^2 \frac{M_1}{p^2-M_1^2}.
\label{eq:amp}
\ee
The expression is valid  whatever value the mass $M_1$ has. 
Now, since by hypothesis we are considering Type I seesaw, the left-left 
 element of the mass matrix given in Eq.~\ref{eq:seesawfull} is zero. Hence we have,
$U_{e1}^2 m_1+ V_{e1}^2 M_1=0$. Thus we conclude that the amplitude of Eq.~\ref{eq:amp}  can be rewritten as:
\be
U_{e1}^2 \frac{m_1}{p^2} \times    \frac{M_1^2}{M_1^2-p^2}.
\label{eq:1f}
\ee
Since  $p^2<0$, we see that the 
effect of the heavier state can only {\em reduce} the strength of the $0\nu 2\beta$ transition; this becomes negligible, leaving only the contribution due the light neutrino exchange, in the limit when $M^2_1\gg -p^2$. 

It is also useful to note that,  upon expanding Eq.~\ref{eq:1f} in powers of ${p^2}/{M_1^2}$ we get the following,
\be
\frac{U^2_{e1}m_1}{p^2} \times \left(1+\frac{p^2}{M_1^2}\right).
\ee
As  expected from Eq.~\ref{amp}, the new contribution $U^2_{e1}m_1/M^2_1$ has the form of 
a contact term--i.e., it is a constant. It is clearly evident from above that for the limit $M_1^2 \gg -p^2$, the second term within the bracket is  much smaller than unity. 
 
With this, we conclude that for one generation case, 
the contribution  of the {\it heavy} sterile Majorana neutrino state to $0\nu 2\beta$ process is always much smaller than the one of the light Majorana state. This implies that, in order to have a large contribution to $0\nu 2\beta$
within Type I seesaw, we  need to consider the multi-flavor case. However, such a property is not generic of a multi-flavor case, as shown later.

\subsubsection{Naive expectations: $0\nu 2\beta$, colliders and lepton flavor violation\label{nv}}
Next, we consider the  naive expectations from Type I seesaw. Since this discussion is 
quite important for the following discussion, we begin by providing a precise definition of 
what  is meant by `naive expectations'.
Most of these expectations correspond to simple scaling laws, obtained 
replacing the Dirac mass matrix $M_D$ in Eq.~\ref{eq:seesawfull} 
with a single mass scale $m$,  and likewise the Majorana mass matrix $M_R$ with
a single mass scale $M$. In other words, here we assume that  
all heavy neutrino masses are of the order of $M$, all light neutrino masses of order of 
$m^2/M$ (see Eq.~\ref{mabel})
all mixing angles with heavy neutrinos $V_{\ell i}$ 
(with $\ell=e,\mu,\tau$ and $i=1,2,3$) are 
of the order of $m/M$ (see Eqs.~\ref{bunul} and \ref{bunal});
and, when we speak of ``seesaw'', we simply mean that 
we restrict to the case $M\gg m$.

Fig.~\ref{xxx} shows the relevant portion of the $(m-M)$-plane. 
The three gray bands correspond to the following boundaries: 
(1) $M>200$ MeV, i.e., 
heavy sterile  neutrinos are assumed to act as point-like 
interactions in the nucleus, as discussed in Sect.~\ref{par}, see in particular Eq.~\ref{prop};
(2) $m<174$ GeV, in order to ensure perturbativity of the Yukawa couplings;
(3) $M>m$, namely, to the seesaw in a conventional sense.
The $(m-M)$-plane is divided in various regions (half-planes) by the 
three oblique lines, that corresponds to
power laws in log-log plot, and are defined as follows:
\begin{itemize}
\item The leftmost oblique line separating white and blue region corresponds to the condition that the 
naive formula for neutrino mass gives a small enough result: $M_\nu\sim m^2/M=0.1$ eV.
The inequality  $m^2/M>0.1$ eV, excluded experimentally, 
corresponds to the region that lies below the line.
(We will discuss below the cases when this bound  can be evaded).

\begin{figure}[t]
\begin{center}
\includegraphics[width=0.5\textwidth, angle=0]{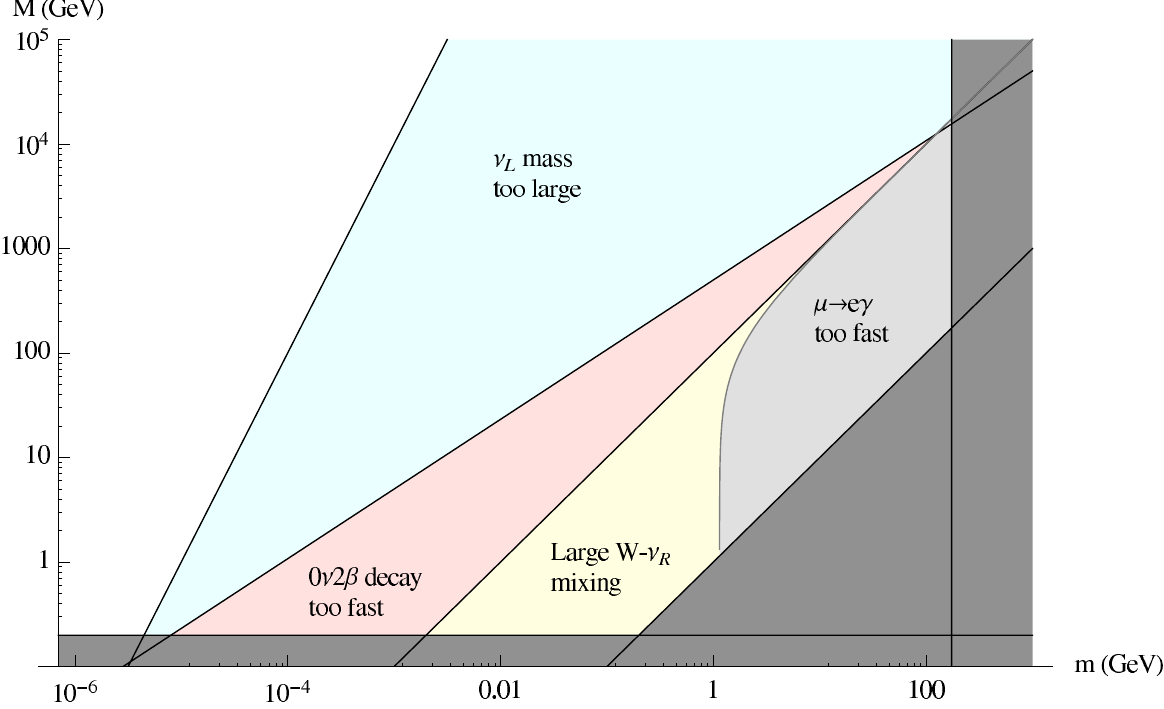}
\end{center}
\caption{\label{fig:$m-M$}\label{fun}
Naive expectations on Type I seesaw model are  displayed  on 
the $(m-M)$-plane. The constraints from $0\nu 2 \beta$ transition 
heavy Majorana neutrino searches in colliders 
and lepton flavor violating decays are shown. See the text for detailed 
explanation.}
\label{xxx}
\end{figure}

\item The condition that the naive contribution from  the 
heavy neutrino exchange to $0\nu 2\beta$ amplitude: $V_{e i}^2/M_i\sim m^2/M^3$
saturates the present experimental bound \cite{epj}
gives the line close to the diagonal and which separates the pink and blue region of the parameter space. The half-plane below this line defines the region where 
this contribution is larger than allowed experimentally.

\item The area below the  oblique line separating the pink and yellow region is the region in $m-M$ plane, where the production of the heavy Majorana  neutrinos in colliders is not  suppressed by a small coupling. The oblique line corresponds to the  heavy and light neutrino mixing angle  $V_{\mu i}\sim {m}/{M} \sim 10^{-2}$, and the region below this line corresponds to $V_{\mu i} >10^{-2}$: the muon flavor refers to the possibility to have a same sign di-muon signal \cite{Keung:1983uu}.
\end{itemize}
Note that the last region is divided in two parts by the $\mu\to e\gamma$ bound  that applies to 
 $V_{e i}^* V_{\mu i} \phi(M_i/M_W)$, with $\phi(x)=x/2 (1-6 x+3 x^2+2 x^3-6 x^2 \ln x)/(1-x)^4$, and that translates into $m^2/M^2 \phi(M/M_W)<10^{-4}$ \cite{tommasini}. This excludes the rightmost region; the almost vertical line corresponds to the fact that when $M<M_W$, the limit is relaxed. For one generation of standard model neutrino $\nu_L$ and one heavy sterile state $N_L$, of course the bound would be absent.

If taken rather literally, the naive expectations  on Type I seesaw (as defined here) would suggest that the existing bounds on neutrino masses imply  that there is no room for large contributions to $0\nu 2\beta$ from heavy neutrinos, or to produce heavy neutrinos at colliders, or to expect a rapid $\mu\to e\gamma$ transition.

\subsubsection{Alternative possibilities in the multi-flavor case}
However, as it is known in the literature and as we discuss in 
great details later, the naive expectation on $M_\nu$ discussed in the previous section, can be strongly relaxed 
in certain multi-flavor cases. 

This offers interesting possibilities for the phenomenology of Type I seesaw and in particular for the investigations of the nature of $0\nu 2\beta$ transition within this model. 
Indeed, the impression that one receives from Fig.~\ref{xxx} is that, even after evading the constraint from neutrino mass, the constraints from $0\nu2\beta$ are likely to be relevant in a large portion of the parameter space, which agrees with the conclusion of \cite{aguila-LHC-m}.
We will show in the following that this impression is confirmed 
by a detailed analysis in a large portion of the parameter space. 
When we depart from the condition that the mixing angles are all the same, $V_{e i}\sim V_{\mu i}\sim m/M$ we can decouple the size of the various amplitudes: in fact, the amplitude of $0\nu 2\beta$ depends on $V_{e i}^2$, the one of same-charge di-muon signal depends instead on $V_{\mu i}^2$, and finally $\mu \to e\gamma$ depends on $V_{e i}^* V_{\mu i}$. This fact, already, provides a large freedom to phenomenological investigations.  However, in this work we prefer to proceed systematically, and will be mostly concerned to classify which cases (which matrices)  evade the constraint from neutrino mass, showing that at the same time, the heavy neutrino exchange contribution can play a relevant role for the $0\nu 2\beta$ decay process. This can be  achieved  if light neutrino mass is strongly suppressed  than the naive expectation from seesaw.

With these phenomenological motivations in mind, we proceed to investigate in detail the different cases  when 
the neutrino masses are much smaller than suggested by the naive expectations.

However, we would like to sketch out in passing an important
theoretical consideration, that will be developed in the following.
Consider the case when the tree-level neutrino
mass-matrix is very suppressed or zero.
We expect that the radiative corrections will provide us with non-zero
mass-matrix; let us say, for definiteness, of the order of ${g^2}/{(4\pi)^2}
\times {m^2}/{M}$, where $g$ is some order-one gauge coupling. Thus, the
tighter constraint  depicted in Fig.~\ref{xxx}--the one denoted `$\nu_L$
mass too large'--can be relaxed, but only by a couple of orders of
magnitude. This implies that the sterile (i.e., heavy) neutrino exchange
contribution to $0\nu 2\beta$ can have a dominant role only in a limited
region of the
parameter space, when the sterile neutrinos are not too heavy. Stated
differently, the lighter the sterile neutrinos, the less problematic is to
reconcile a dominant role of the sterile neutrinos for the $0\nu 2\beta$
transition with the values of the masses of the ordinary neutrinos.

\subsection{Departing from the naive expectations for neutrino masses \label{see0}}

We are interested to the cases when the naive expectation 
for the light neutrino mass, $M_D^T M_R^{-1} M_D \sim m^2/M$, 
typical of Type I seesaw, does not hold.  
Thus,  we first proceed in Sect.~\ref{m} by a  mathematical analysis of 
the vanishing seesaw condition  $M_D^T M_R^{-1} M_D=0$ and then we classify in Sect.~\ref{cases} which are 
the perturbations of this condition that permit us to have neutrino masses much smaller than suggested 
by this naive expectation.

\subsubsection{Solving the condition $M_D^T M_R^{-1} M_D=0$: Light-neutrino masses=0 \label{m}}
We describe here a direct procedure to find the 
non-trivial solutions to the matricial condition $M_D^T M_R^{-1} M_D=0$. 
Consider the three flavor scenario with two $3 \times 3$ matrices $M_D$ and $M_R$, where  $M_R$ is 
assumed to be invertible. Using a 
suitable bi-unitary transformation, we go to the basis where the Dirac mass matrix is diagonal, i.e., $M_D={\rm diag}(q,n,m)$.
This basis is very convenient but only to solve the condition
 $M_D^T M_R^{-1} M_D=0$; eventually, we have to return to the original flavor basis, where 
 the weak interactions and the charged lepton mass matrix are diagonal.
The condition $M_D^T M_R^{-1} M_D=0$ is compatible  with an invertible matrix $M_R$ if $q=n=m=0$, that is 
the trivial solution, but also if one diagonal element in $M_D$ is non-zero;
all other cases are excluded.\footnote{If $q$, $n$ and $m$ are all non-zero,
we immediately find $M_R^{-1}=0$. If, e.g., only $q$ and $n$ are non-zero, we have 
$(M_R^{-1})_{11}= (M_R^{-1})_{12}= (M_R^{-1})_{22}=0$, which implies
${\rm det}(M_R^{-1})=0$, that is again incompatible with the existence of the inverse of $M_R$.
In mathematical terms, recalling that the characteristic of a matrix is basis invariant, 
we conclude that $M_D$ has characteristic 0 or 1.}
Let the non-zero element be the third one, i.e., $q=n=0$ and 
$m\neq 0$: We have to satisfy $(M_R^{-1})_{33}=0$. This means that  the $2\times 2$ block including $(M_{R})_{11}$,
$(M_{R})_{12}$ and $(M_{R})_{22}$  has zero determinant, i.e., 
 \be
(M_{R})_{11}{(M_R)}_{22}-{(M_R)}^2_{12}=0.
\ee 
Rotating  this $2 \times 2$ block of $M_R$ into  a diagonal form, it has one diagonal element equal to zero. 
We conclude that, in a given basis, we are dealing with the following Lagrangian including the Dirac and the Majorana masses:
\be
{ \mathcal L}=\frac{1}{2} 
\pmatrix{\nu_{L_1},\nu_{L_2},\nu_{L_3},N_{L_1},N_{L_2},N_{L_3}}
\left(
\begin{array}{ccc|ccc}
0 & 0 & 0 & 0 & 0 & 0 \\
0 & 0 & 0 & 0 & 0 & 0 \\
0 & 0 & 0 & 0 & 0 & m \\ \hline
0 & 0 & 0 & 0 & 0 & M_1 \\
0 & 0 & 0 & 0 & M_2 & M_3 \\
0 & 0 & m & M_1 & M_3 & M_4 \\
\end{array} \label{ciclone}
\right)
\pmatrix{\nu_{L_1} \cr \nu_{L_2} \cr \nu_{L_3} \cr N_{L_1} \cr N_{L_2} \cr N_{L_3}}
\ee
where $M_1$ and $M_2$ are non-zero since ${\rm det}(M_R)=-M_1^2 M_2$, while
$M_3$ and $M_4$ are free parameters.

Evidently, the characteristic of this $6\times 6$ matrix is 3: 
there are three null eigenvalues. More in details, we see that
operating a rotation in the $\nu_{L_3}$ and the $N_{L_1}$ fields, we can transform it   
into  a mass matrix where the new `Dirac' block is just zero.
The physical meaning of these mathematical results 
is that the condition 
$M_D^T M_R^{-1} M_D=0$ always implies that the mass matrix of the light neutrinos is 
zero to all order, and that this condition can be realized in a non-trivial manner only arranging for a 
`large' mixing (i.e., order $m/M $) between the left and the right neutrinos.
(This conclusion has been 
derived previously using a systematic expansion of the neutrino mass matrix \cite{Grimus-Lavoura, Pilaftsis} see also \cite{ARC, Rodejohann, 0-others}.) 
 From the above proof, it is easy to understand that, 
up to change of basis, the previous non-trivial solution of the condition $M_D^T M^{-1}_R M_D=0$ is the most 
general one.

\subsubsection{Perturbing the condition $M^T_D M_R^{-1} M_D=0$: Light-neutrino masses$\neq 0$\label{cases}}

Here, we perturb both $M_D$ and $M_R^{-1}$, maintaining the first one diagonal.
From the previous discussion, it is pretty evident that   to satisfy the 
 vanishing seesaw condition $M^T_DM^{-1}_R M_D=0$ with an invertible $M_R$, one should have at most one non-zero 
diagonal element in $M_D$. Hence, we want to keep only   one large diagonal element in the Dirac  mass matrix; 
while the other two diagonal elements appear due to  perturbation: 
In formulae, we write
\be
M_D=m\ {\rm diag}(\epsilon_1,\epsilon_2,1),
\ee
with $\epsilon_1$ and $\epsilon_2 \ll 1$.
In the following, we denote by $\epsilon$ a small parameter, which  we will use to 
explicitly tune the smallness of the light neutrino masses. We will  
show how it is possible to organize the elements of 
$M_R^{-1}$ in powers of  $\epsilon$ in order to maintain a special  
suppression of the light neutrino mass matrix.
 To simplify the notation, 
we refrain from writing  explicitly   
the coefficients of  ${\mathcal{O}}(1)$ 
of the mass matrices
$M_D$, $M_R^{-1}$ and $M_\nu=-M_D^T M_R^{-1} M_D$, 
but we will use the symbol $\stackrel{\tiny\mathcal{O}(1)}{=}$ to keep 
track of this simplification; i.e., to say, in each of the matrix elements of $M_D$, $M^{-1}_R$ and $M_{\nu}$, 
we show only the leading order. In short, in the following formulae 
we emphasize the necessary suppressions of 
the matrix elements in powers of the small parameter $\epsilon$.
We identified 3 main cases:

\paragraph{Case A:} Consider the following Dirac and Majorana mass matrices:
\be
M_D
\stackrel{\tiny\mathcal{O}(1)}{=}
m\ 
\mbox{diag}(0,\epsilon,1)\ \ \
M_R^{-1}
\stackrel{\tiny\mathcal{O}(1)}{=}
M^{-1}
\left(
\begin{array}{ccc}
1 & 1 & 1 \\
1 & 1 & 1 \\
1 & 1 & \epsilon
\end{array}
\right)
\Rightarrow
M_\nu
\stackrel{\tiny\mathcal{O}(1)}{=}
\frac{m^2}{M}
\left(
\begin{array}{ccc}
0 & 0 & 0 \\
0 & \epsilon^2 & \epsilon \\
0 & \epsilon & \epsilon
\end{array}
\right)
\label{casea}.
\ee
(The elements (1-2) and (1-3) of $M_R^{-1}$ could be much smaller without affecting the argument.) 
The analysis of this 
case essentially reduces to analysis of the $2\times 2$ matrices.
This case yields one massless and two massive light neutrinos
and will be discussed in details later, being a prototypical case.

\paragraph{Cases B:} A similar situation is realized  for the following mass matrices:
\be
M_D
\stackrel{\tiny\mathcal{O}(1)}{=} m\ 
 \mbox{diag}(\epsilon,\epsilon,1);\ \ \
M_R^{-1}
\stackrel{\tiny\mathcal{O}(1)}{=}
M^{-1} 
\left(
\begin{array}{ccc}
1 & 1 & 1 \\
1 & 1 & 1 \\
1 & 1 & \epsilon
\end{array}
\right)\ , \ 
M^{-1} \left(
\begin{array}{ccc}
1 & 1 & 1 \\
1 & 1 & \epsilon \\
1 & \epsilon & \epsilon
\end{array}
\right)
\ , \
M^{-1} \left(
\begin{array}{ccc}
1 & 1 & \epsilon \\
1 & 1 & 1 \\
\epsilon & 1 & \epsilon
\end{array}
\right),
\label{caseb}
\ee
that correspond to the following light neutrino mass matrices:
 \be
M_\nu
\stackrel{\tiny\mathcal{O}(1)}{=}
\frac{m^2}{M} 
\left(
\begin{array}{ccc}
\epsilon^2 & \epsilon^2 & \epsilon \\
\epsilon^2 & \epsilon^2 & \epsilon \\
\epsilon & \epsilon & \epsilon
\end{array}
\right)\ , \ 
\frac{m^2}{M} 
\left(
\begin{array}{ccc}
\epsilon^2 & \epsilon^2 & \epsilon \\
\epsilon^2 & \epsilon^2 & \epsilon^2 \\
\epsilon & \epsilon^2 & \epsilon
\end{array}
\right)
\ , \
\frac{m^2}{M} 
\left(
\begin{array}{ccc}
\epsilon^2 & \epsilon^2 & \epsilon^2 \\
\epsilon^2 & \epsilon^2 & \epsilon \\
\epsilon^2 & \epsilon & \epsilon
\end{array}
\right)
\label{casebb}.
\ee
The analysis of these cases is pretty similar to the analysis of the previous one.
It is easy to see that, for all of them:
\begin{enumerate}
\item The elements of the neutrino mass matrix are {\it at most} of the 
order of $\epsilon$. Thus  all neutrino masses are more 
suppressed than the what naive seesaw formula would suggest.
\item However, the determinant of the light neutrino mass matrix  is ${\mathcal{O}}(\epsilon^4)$. This essentially 
implies  that the lightest neutrino mass is order $\epsilon^2$, i.e., very small.
\end{enumerate}
Both features of Case B are in common with those of Case A, where the lightest neutrino mass is just zero.

\paragraph{Case C:}
Finally we consider an interesting case, which 
is in favor of  non-suppressed lightest neutrino mass, i.e.,
\be\label{petreus}
M_D
\stackrel{\tiny\mathcal{O}(1)}{=} m\ 
 \mbox{diag}(\epsilon^2,\epsilon,1)\ \ \
M_R^{-1}
\stackrel{\tiny\mathcal{O}(1)}{=} M^{-1}\ 
\left(
\begin{array}{ccc}
1 & 1 & 1 \\
1 & 1 & \epsilon \\
1 & \epsilon & \epsilon^2
\end{array}
\right)
\Rightarrow
M_\nu
\stackrel{\tiny\mathcal{O}(1)}{=}
\frac{m^2}{M}
\left(
\begin{array}{ccc}
\epsilon^4 & \epsilon^3 & \epsilon^2 \\
\epsilon^3 & \epsilon^2 & \epsilon^2 \\
\epsilon^2 & \epsilon^2 & \epsilon^2
\end{array}
\right)
\label{casec}.
\ee
Now the elements of the light neutrino mass matrix are at most of the order of 
${\mathcal{O}}(\epsilon^2)$ (that is the same as before up to 
the redefinition $\epsilon^2 \to \epsilon$)  but the determinant of the mass matrix is ${\mathcal{O}}(\epsilon^6)$. 
Hence, depending on $\epsilon$, it is possible to have a  lightest neutrino mass, which  is not small {\em a priori}.

\paragraph{Remark}
Before passing to the discussion of the previous 
mass matrices, we note that the presence of a very light 
(or just massless) neutrino, common to Cases 
 A and B--but not necessarily to Case C--could be tested experimentally by future 
 cosmological measurements \cite{stru, cosmobound}: In fact, 
 in the case of normal (resp., inverted) hierarchy, we would 
 expect that the sum of neutrino mass is (resp., twice) the atmospheric mass scale $\sqrt{\Delta m^2_{atm}}\approx 50$ meV.

\subsubsection{Quantification of the fine-tuning\label{qft} and lower  bound on $\epsilon$}
The common feature of the neutrino mass matrices classified in the above  section is that they 
are smaller than suggested by the naive seesaw formula; 
 using the symbols as in Sects.~\ref{nv} and \ref{cases},
\be
M_\nu\sim \epsilon\ \frac{m^2}{M},
\label{ppo1}
\ee
where $\epsilon$ is a small parameter.\footnote{ For case C, this requires  a redefinition of $\epsilon^2 \to \epsilon$.}  
The question arises, 
whether such a structure is stable under radiative corrections.

\begin{figure}[t]
\begin{center}
\epsfig{file=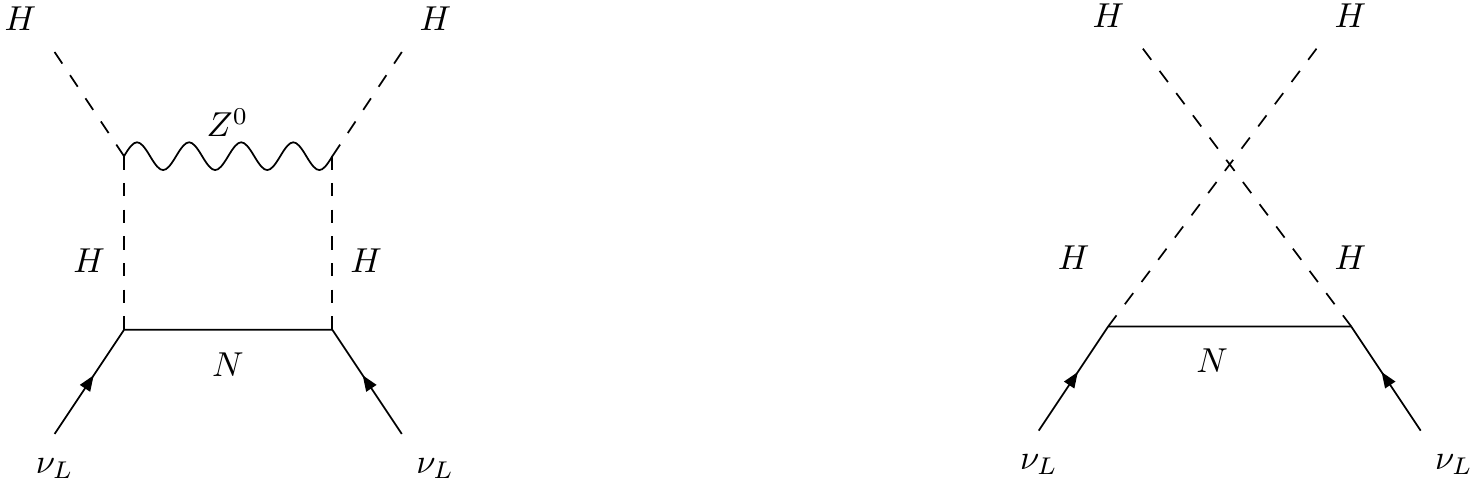, width=0.7\textwidth}
\caption{ \label{bb3} One-loop correction to the $\nu_L$ mass}
\end{center}
\end{figure}

It has been remarked in \cite{smirnov} that, for non-supersymmetric Type I seesaw, the decoupling of 
two heavy neutrinos with different masses $M_{1}$ and $M_2$ 
produces a correction to the neutrino mass matrix of the order of 
\be\label{rge}
\delta M_\nu\sim \frac{g^2}{(4\pi)^2} \frac{m^2}{M} \log(M_1/M_2),
\ee
where $g$ is a gauge or Higgs coupling, since the renormalization group evolution of the effective operator 
differs from the evolution of the Yukawa couplings (or Dirac mass), as shown in \cite{rg}. 
This can be seen as a minimum natural size of the  coefficient in Eq.~\ref{ppo1},
\be\label{cza}
\epsilon > g^2/(4\pi)^2\sim 10^{-2},
\ee
unless we want to accept very fine-tuned mass matrices,
an unattractive possibility that we could however consider, 
if the data should force us to do so: 
see Sect.~\ref{hphp} for a discussion.

\vspace*{0.2cm}
Besides the possibility to enforce $M_1=M_2$, by 
an (approximate) global symmetry \cite{smirnov,ingelman} to put to zero the 
radiative correction of Eq.~\ref{rge}, there are also other cases when the resulting 
condition on $\epsilon$ (Eq.~\ref{cza}) can be, 
if not avoided, at least relaxed. First, it is known \cite{vs} that in supersymmetry, the 
effective operator and the neutrino mass matrix receive the same radiative correction, so that the 
conclusion of Eq.~\ref{rge} does not hold. Second,  also in non-supersymmetric model, heavy neutrinos with 
masses  below the electroweak scale, $M<M_{ew}$,  will not give 
logarithmic corrections, but smaller polynomial  corrections only:
\be\label{nrge}
\delta M_\nu\sim \frac{g^2}{(4\pi)^2} \frac{m^2}{M}  \frac{M^2}{M_{ew}^2}
\ee
simply because the electroweak physics 
should decouple.\footnote{When $M\to 0$ we enforce 
lepton number conservation in the model; this agrees with 
the fact that Eq.~\ref{nrge} vanishes in this limit.} 
These corrections can be 
attributed to finite diagrams, where the usual tree level operator for neutrino mass 
is dressed by $Z$ (or by higgs boson) exchanges (see Fig.~\ref{bb3}). 
The condition of radiative stability of the tree level
neutrino mass matrix, $M_\nu>\delta M_\nu$ 
bounds $\epsilon$ from below:
\be
\epsilon > ( M/1  \mbox{ TeV})^2.
\label{lowbnd-ep1}
\ee
The  discussions in this section  can be summarized as   follows, 
\be
\epsilon >
\left\{
\begin{array}{ll}
\displaystyle \left({M}/{\mbox{1 TeV}}\right)^2 & \mbox{if }M<M_{ew} \\[1ex]
10^{-2} & \mbox{if }M>M_{ew}
\end{array}
\right.
\label{lowbnd-ep}
\ee
where $M_{ew}\approx 100$ GeV; 
as we see in the following, it is the regime where the corrections are smaller, 
$M<M_{ew}$,  the one in which we will be mostly interested.

\subsection{A dominant role of heavy neutrino exchange in $0\nu 2\beta$\label{multi-contact}}
In this section we discuss 
how it is possible that   heavy neutrino exchange provides us with a dominant contribution to $0\nu2\beta$
within Type I seesaw models.  First, we state precisely in Sect.~\ref{n} what is 
the role of the heavy neutrino exchange for the amplitude of 
$0\nu2\beta$, and we consider in Sect.~\ref{s} the case  
when this contribution is large building on the mass matrices discussed in Sect.~\ref{see0}.
Then, we pursue a detailed investigation of this case for two flavors   (Sect.~\ref{c2})
and as well as for three flavor mass matrices (Sect.~\ref{c3}).
We discuss the compatibility with fine-tuning issues in Sect.~\ref{upbndem} and conclude in 
Sect.~\ref{num-ex} with numerical example.

\subsubsection{The amplitude of $0\nu2\beta$ and the role of heavy neutrino exchange\label{n}}
\begin{figure}[t]
\begin{center}
\epsfig{file=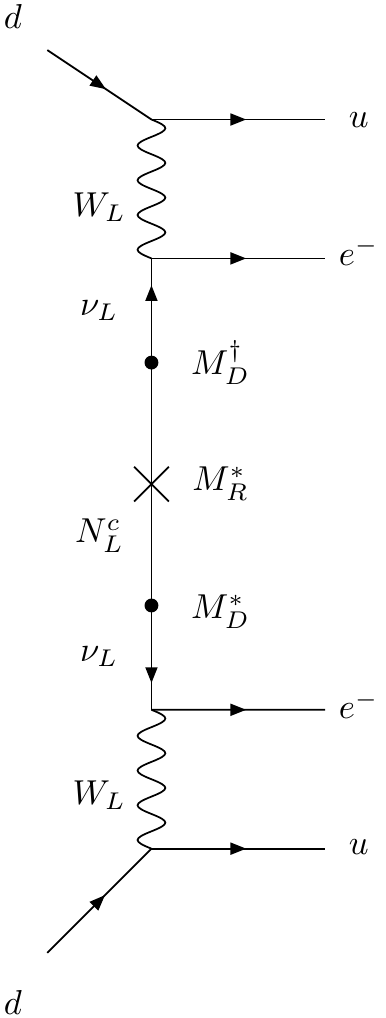, width=0.2\textwidth}
\caption{ \label{bb11} Feynman diagram of the $0\nu 2 \beta$ process for Type I seesaw
at leading order in $M_D$.}
\end{center}
\end{figure}

The mixing between the heavy  sterile  neutrino state  $N_L$ and the  
 standard model neutrino state  $\nu_L$ is caused by the Dirac mass matrix  $M_D$ (see Eq.~\ref{eq:seesaw1}). As 
clear from Fig.~\ref{bb11}, the 
 amplitude for the $0\nu 2 \beta$ process is proportional to the following factor 
 stemming from the vertices and the propagator, 
\be
 \left[ \frac{1}{\not{p}} \hat{M}_D^{\dagger} 
\ \mbox{diag}\!\left(
\frac{1}{\not{p}-M_i} \right) \hat{M}^*_D \frac{1}{\not{p}}\right]_{ee}
\ee
where we consider  the expressions in leading (second) order in $M_D$ and
$\hat{M}_D$ is the Dirac mass matrix in the basis where the 
heavy neutrinos are diagonal, see Eq.~\ref{dd}.
Reminding that this expression in sandwiched between chiral projectors,  its 
contribution to the $0\nu 2\beta$ amplitude is just,
\be
\mathcal{A}= \left[ M_D^{\dagger}W^* \mbox{diag}\!\left(
\frac{1}{M_i} \left( \frac{1}{p^2-M_i^2}  - \frac{1}{p^2} \right)\right) W^{\dagger}M^*_D\right]_{ee}
\label{mrl}
\ee
Focussing again on the case
$M^2_i \gg |p^2| \sim (200)^2\mbox{ MeV}^2$, 
we can  expand Eq.~\ref{mrl} as follows:
\be
{\cal A}^* =
 \left[ \frac{M_\nu}{p^2} 
-M_D^T M_R^{-1}{M_R^{-1}}^*M_R^{-1} M_D  
+ \cal{O}(\rm{\it {M_R^{-\rm{5}}}})\right]_{ee}
\label{eq:ampl}
\ee
where we have used the diagonalizing relation $M^{-1}_R=W\mbox{diag}\!\left( M_i^{-1}\right)W^T$. 
The first term in brackets,  evidently due to the exchange of light Majorana neutrinos,
is the usual one; the second one is the effect of heavy neutrino exchange in which we are interested. For 
real $M_R$, it can be written simply (up to the sign) as, 
\be
(M_D^T M_R^{-3} M_D)_{ee}
\label{zdt}
\ee
a quantity with dimension of an inverse mass.
In the following, we will refer often to this as a `contact term', having in mind the nature of  the operator 
that induces the $0\nu 2\beta$ transition.  
Using Eqs.~\ref{mabel} and  \ref{bunul}, it is easy to verify that 
Eq.~\ref{eq:ampl} coincides with Eq.~\ref{amp} fully; but 
the new expression is more convenient  
for the subsequent theoretical analysis of 
Type I seesaw model.

In the following, we will speak of a `saturating contribution' \cite{epj} when the 
heavy neutrino exchange dominates the transition. In formulas and using the
notations of Sect.~\ref{nme}, this implies that
the (absolute value) of the contact term in Eq.~\ref{zdt} is equal to  
${1}/{\sqrt{ K_{0 \nu} T_{1/2}}}$, or, in numerical terms
  \be
| (M^T_D M^{-3}_R M_D)_{ee} | 
= 7.6 \times 10^{-9}\  \mathrm{ GeV}^{-1}  \times  
\left(
\frac{363}{\mathcal{M}_N}
\right) 
\times   
\left(
\frac{ 1.9 \times 10^{25}\ \rm{yr}}{T_{1/2}} 
\right)^{\!\! 1/2}
\ee

\subsubsection{The case for a large contact term\label{s}}
Being ready to consider the special class of neutrino mass matrices classified in Sect.~\ref{see0}, 
it is possible to understand the cases when the naive expectations
of Type I seesaw discussed in Sect.~\ref{typeI-intro} (and in particular Sect.~\ref{nv}) do not work.
These special cases are of great phenomenological interest, especially for neutrinoless double beta decay, but 
have also some theoretical interest, in view of the fact that they are based on the simplest renormalizable 
extension of the standard model, that includes massive neutrinos.

In formal terms, and using the symbols as in Sects.~\ref{nv} and \ref{cases}, we are considering the possibility 
that the neutrino mass matrix is smaller than suggested by the seesaw formula 
and at the same time the contribution of heavy Majorana neutrino states in $0\nu 2\beta$ process, 
\be\label{ppo2}
(M_D^T M_R^{-3} M_D)_{ee}=\kappa\ \frac{m^2}{M^3}\label{kapa}
\ee
where  $\kappa$ is a coefficient  which depends 
 on the specific particle physics model, and  will be 
determined later in this section for the cases of interest. 
We are particularly interested in  
heavy neutrino masses that  saturate  the $0\nu2\beta$ experimental bound \cite{epj},
\be\label{bomba}
M=16\mbox{ TeV}\times \left( \frac{T_{1/2}}{1.9\times 10^{25}\mbox{yr}}\right)^{1/6} \left( \frac{\mathcal{M}_N\times \kappa}{363\times 1}\right)^{1/3} \left( \frac{m}{174\mbox{ GeV} }\right)^{2/3}
\label{16tev}
\ee
where the nuclear matrix element and the half-life apply both to ${}^{76}$Ge
(see Sect.~\ref{gen}). It is important to note that by reducing the mass scales, the need of 
fine-tuning (i.e., too small $\epsilon$) diminishes; indeed Eq.~\ref{ppo1} and \ref{ppo2} are left unchanged by the scaling 
\be\left\{
\begin{array}{lcl}
M & \to & \alpha\times M \\[1ex]
m & \to  & \alpha^{3/2}\times m \\[1ex]
\epsilon & \to  & \alpha^{-1}\times \epsilon
\end{array}
\right. 
\ee
where $\alpha<1$ implies that  $m$ will be  
smaller than $174$ GeV, maintaining perturbativity of the   Yukawa couplings. (Actually, the 
fact that decreasing $m$ and $M$ we need less fine-tuning 
can be understood also from an inspection of Fig.~\ref{fun}.)
Also note the very mild dependency of the heavy neutrino mass scale in Eq.~\ref{bomba}
on the true value of the half-life; if the central value found  by Klapdor and collaborator is  confirmed, this 
would change by only 3\%,  but even if 
the true lifetime should  turns out to be 100 times larger, i.e., $T_{1/2}=1.9\times 10^{27}$ yr,
the mass $M$ in Eq.~\ref{bomba} would just be  doubled.
The dependence on the matrix elements is also quite mild. Finally, even
stretching the perturbativity condition on $Y_D=m/(174$ GeV), from $Y_D<1$ to
$Y^2_D/(4 \pi)<1$, the mass $M$ would increase only by a factor of 2.3.

\subsubsection{Dominating heavy-neutrino contribution with two flavors\label{c2}}
Here we analyze a prototypical two flavor case. We start from rather specific mass matrices,
given in the basis where the Dirac mass matrix is diagonal, and then switch to the 
flavor basis (where the charged current interactions and the 
charged lepton masses are diagonal). We show that, in this way, we 
can obtain information on the structure of the contact term.

\paragraph{Structure of the contact term in the basis where $M_D$ is diagonal}

Suppose that we have the $2\times 2$ matrices
\begin{equation}
M_R
\stackrel{\tiny\mathcal{O}(1)}{=}
M \left(
\begin{array}{cc}
\epsilon & 1 \\
1 & 1
\end{array}
\right)
\mbox{ and }
M_D
\stackrel{\tiny\mathcal{O}(1)}{=}
m \left(
\begin{array}{cc}
\epsilon & 0 \\
0 & 1
\end{array}
\right)
\label{e1}
\end{equation}
(compare with Sect.~\ref{cases}, Case~A,  Eq.~\ref{casea}). 
We derive immediately the neutrino mass matrix and the contact term:
\begin{equation}
M_D^T M_R^{-1} M_D
\stackrel{\tiny\mathcal{O}(1)}{=}
\frac{\epsilon m^2}{M} \left(
\begin{array}{cc}
\epsilon & 1 \\
1 & 1
\end{array}
\right)
\mbox{ and }
M_D^T M_R^{-3} M_D
\stackrel{\tiny\mathcal{O}(1)}{=}
\frac{ m^2}{M^3} \left(
\begin{array}{cc}
\epsilon^2 & \epsilon \\
\epsilon & 1
\end{array}
\right)
\label{cond}
\end{equation}
Using the results of  \cite{Grimus-Lavoura}, 
the next higher order contribution to the neutrino mass is found to be
\be
\delta M_{\nu}
\stackrel{\tiny\mathcal{O}(1)}{=}
\frac{m^4 \epsilon}{M^3}\pmatrix{0 & 1 \cr 1 & 1}+ {\mathcal O}(\epsilon^2)
\ee
 This contribution is suppressed as compared to the leading contribution by a factor 
${m^2}/{M^2} \ll 1$; for $m=174 $ GeV and $M \sim 16 $ TeV, the suppression is in fact  $10^{-4}$. Hence,  we do 
not take into account  the sub-leading contribution any further.\footnote{This bound can be obtained as follows:
denoting the heavy sterile  neutrino contribution to $0\nu2 \beta$ as 
${m^2}/{M^3}=k$, we find  
${m^2}/{M^2}=k M$; thus, we maximize ${m^2}/{M^2}$  
when (1)~$k$ is as large as possible,
(i.e., when we saturate the experimental upper bound) and when 
(2)~$M$ is as large as possible, that happens  when $m=174 $ GeV and $M=16$ TeV,
as seen in the previous section.}

\paragraph{The limit $\epsilon\to 0$}
Keeping only the leading order entries in $\epsilon$, the light neutrino mass matrix becomes:
\be
M_{\nu}
\stackrel{\tiny\mathcal{O}(1)}{=}
\frac{\epsilon m^2}{M} \left(
\begin{array}{cc}
0 & 1 \\
1 & 1
\end{array}
\right) + {\mathcal O}(\epsilon^2)
\mbox{ and }
M_D M_R^{-3} M_D
\stackrel{\tiny\mathcal{O}(1)}{=}
\frac{m^2}{M^3} \left(
\begin{array}{cc}
0 & 0 \\
0 & 1
\end{array}
\right)+ {\mathcal O}(\epsilon) 
\label{pm}
\ee
The interpretation is the following: 
\begin{enumerate}
\item The parameter $\epsilon$ can be used to diminish 
the value of neutrino masses to the desired value. As an example, for $m=174$ GeV and $M=16$ TeV, 
eV-neutrino masses can be generated for $\epsilon \sim 10^{-9}$; for smaller 
values of $m$ and $M$, we need less fine tuning, i.e.,  the parameter $\epsilon$ increases.
\item Generically, we expect two neutrino masses of similar order and with large mixing (note that the first 
statement is basis independent, while the second is not; we are still in the basis where the Dirac mass matrix is diagonal).
\item The size of the contact term has been made independent 
from the size of neutrino masses. In other words, the contact term can take whatever value.

\end{enumerate}
By a suitable choice of phases and restoring the coefficients of the order of 1 from here on, 
we see that the first non-trivial terms in $\epsilon$ of the mass matrix in Eq.~\ref{pm}
can be rewritten as:
\be
M_\nu \approx \left(
\begin{array}{cc}
0 & \sqrt{m_1 m_2} \\
\sqrt{m_1 m_2} & m_2-m_1
\end{array}
\right) 
\ee
where $0\le m_1 \le m_2$ are the mass eigenstates; think e.g., to the solar doublet.

\paragraph{Structure of the contact term in the flavor basis}
Let us then go to the flavor basis for the neutrino mass and see what happens to the contact term
(note that both neutrino mass matrix and the contact term  rotate in the same manner when we change basis). First, we apply 
a rotation by an angle
$$
\tan\theta=\sqrt{\frac{m_1}{m_2}}
$$
and, in this way, go to  the basis where the neutrino mass matrix reduces to the 
diagonal form ${\rm diag}(-m_1,m_2)$.
Then we perform a second rotation with a solar mixing angle $\theta_\odot$
to go to the flavor basis, and include a Majorana phase $\phi$ 
to consider the most general mass matrix. 
In this way, we  recover the well-known expression for 
ee-element of the neutrino mass matrix, relevant to the $0\nu 2\beta$: 
\be
(M_\nu)_{ee}=\sin^2\theta_\odot\ m_2-\cos^2\theta_\odot\ m_1\ e^{i 2 \phi}
\label{contone}
\ee
and in the same basis (flavor basis), the contact term is
\be
(M_D^T M_R^{-3} M_D)_{ee} ^{\mathrm{(Fl.)}}= \xi \frac{m^2}{M^3}
\frac{(\sin \theta_\odot\ \sqrt{m_2}+\cos\theta_\odot\ \sqrt{m_1}\ e^{i  \phi})^2}{m_1+m_2}
\label{contwo}
\ee
where $\xi$ is a factor of the order of 1 (further discussed below) and   in the above  we used the freedom to redefine the phase $\phi$ as $\phi \to \phi+ \pi$. Thus, in general the latter is non
zero and it is correctly estimated dimensionally as $m^2/M^3$. 
This concludes the proof of the principle in the case of two flavors. 

\medskip
Before passing to the three flavor case, let us consider a couple of 
particular noticeable two flavor cases: 

\paragraph{Additional suppression of light neutrino contribution to $0\nu 2\beta$ \label{sym}}
An interesting special case is when we 
arrange $(M_\nu)_{ee}=-(M_D^T M_R^{-1} M_D)_{ee}=0$, a condition that can be realized in  
normal mass hierarchy but not in inverted mass hierarchy,
as noted in \cite{vuso, vis-fer}. From Eq.~\ref{contone} 
we see that for $(M_{\nu})_{ee}=0$, we need to have $\phi=0$ or $\pi$ and 
\be
\left\{
\begin{array}{l}
m_2=\sqrt{\Delta m^2_{\odot}} \frac{\cos^2\theta_\odot}{\sqrt{\cos 2\theta_\odot}}\\[2ex]
m_1=\sqrt{\Delta m^2_{\odot}} \frac{\sin^2\theta_\odot}{\sqrt{\cos 2\theta_\odot}}
\end{array}
\right.
\ee
i.e., $m_1\approx 4.5$ meV as
calculated with $\Delta m^2_\odot=7.6\times 10^{-5}$ eV$^2$ and 
$\theta_\odot=34^\circ$.
If $\phi=\pi$ the contact term is also suppressed, if 
$\phi=0$, instead, it is unsuppressed. The last possibility offers a very explicit example of a case 
when the $0\nu 2\beta$ transition is {\em entirely} due to heavy neutrino exchange. In the 
language of Schechter-Valle 'theorem', 
we can say that this is one case when the black box of  \cite{sv}, 
that apparently should connect the 
double beta decay transition to non-zero neutrino masses, actually 
does not yield  any contribution to the ee-element of the Majorana neutrino mass matrix.\footnote{
In mathematics, such a case should be called a counterexample, but it should be understood
that the Schechter-Valle 'theorem' is just an illustration of typical expectations and has no 
quantitative aims. In last analysis, and despite the evident fact that an observation of the
$0\nu 2\beta$ decay would imply that electronic lepton number is broken, a true 
understanding of the connection between neutrino masses and neutrinoless double beta decay rate
is possible only within specific extensions of the standard model. 
See \cite{ choi, tello, lindner} for more relevant examples and discussion.}

\paragraph{Suppression of heavy neutrino contribution to $0\nu 2\beta$}
We encountered in the previous discussion a case ($\phi=\pi$ and $m_1\approx 4.5$ meV) where 
the heavy neutrino contribution to $0\nu 2\beta$
is suppressed. The other case when this happens, as we see from Eq.~\ref{contwo}, is when 
$\xi=0$. In order to discuss the viability of this case, we need to restore the
 $\mathcal{O}(1)$  coefficients 
in Eq.~\ref{e1}. First, we note that the expression for $M_D$ can be regarded as a definition of
$m\neq 0$ and $\epsilon\neq 0$, and thus can be left unchanged. Second, we write $M_R$ as 
\be
M_R=M 
\left(
\begin{array}{cc}
\epsilon a  & 1 \\
1 & b
\end{array}
\right)
\label{mde}
\ee
where we introduced
two free parameters and 
provided incidentally a precise definition of the parameter $M\neq 0$. It is easy to show that
\be
M_D M_R^{-1} M_D=\frac{\epsilon m^2}{M} 
\left(
\begin{array}{cc}
0  & 1 \\
1 & -a
\end{array}
\right) +\mathcal{O}(\epsilon^2)
\mbox{ and }
M_D M_R^{-3} M_D=\frac{m^2}{M^3} 
\left(
\begin{array}{cc}
0  & 0 \\
0 & -b
\end{array}
\right) +\mathcal{O}(\epsilon)
\label{eq}
\ee
which is a more precise statement than Eq.~\ref{pm}. 
This calculation shows that not only the mass scales, but also the order-one coefficient of the 
contact term is completely independent from those 
entering the neutrino mass  matrix. In other words,
we have the freedom to take the limit $\xi\to 0$ in Eq.~\ref{contwo}, simply letting $b\to 0$ in Eq.~\ref{mde}. Evidently, this limit amounts to enforcing an approximate global symmetry in the heavy sterile  neutrino mass matrix, which assumes a `quasi Dirac' structure \cite{Ibarra, smirnov}.

\subsubsection{Dominating heavy-neutrino contribution with three flavors\label{c3}}

A large part of the discussion and conclusions for 
the two flavor example can be repeated  for the three flavor case.  
It is easy to show that in all cases A, B, C of Sect.~\ref{cases} the leading part 
in $\epsilon$ of the contact term  has the form
\be
M_D^T M_R^{-3} M_D=\xi\ \frac{m^2}{M^3}
\left(
\begin{array}{ccc}
0 & 0 & 0 \\
0 & 0 & 0 \\
0 & 0 & 1
\end{array}
\right) \label{trf}
\ee
in the basis where Dirac masses are diagonal. Here, $\xi$ is a combination of the order-one coefficients
of the heavy Majorana  and Dirac mass matrices. It is easy to show by direct calculation that this combination can be made independent from the coefficients order-one that regulate the light neutrino mass matrix.

However, we need to determine the contact term {\em in the flavor basis}.
We begin from the light neutrino mass matrix $M_\nu=-M_D^T M_R^{-1} M_D$
in the basis where the Dirac mass matrix is diagonal; then,
we change to mass basis  by  the diagonalization   $O^T M_\nu O=\mbox{diag}( m_i)$; 
finally, we reach the flavor basis simply including the leptonic mixing matrix $U$. The same changes of basis apply to the contact term as well; thus, the quantity that 
regulates the heavy neutrino exchange contribution to $0\nu 2\beta$ is given by,
\be\label{general}
( M_D^T M_R^{-3} M_D)^{\mathrm{(Fl.)}}_{ee}\equiv  
(U^* O^T M_D^T M_R^{-3} M_D O U^\dagger)_{ee}
\ee
By comparing with Eq.~\ref{kapa}, we see that the order one coefficient that correct 
the naive estimation for the contact term is given by 
\be
\kappa=\xi\times \varphi^2, \mbox{ with }\varphi=\sum_{i=1}^{3} U_{ei}^* O_{3i}\label{vvv}
\ee
The quantity $\varphi$ has modulus between 0 and 1, since it
can be thought of as a scalar product between two unit vectors.
In practice, all we need is to calculate is the matrix $O$, 
since the moduli of the leptonic mixing matrix $U_{ei}$ are known experimentally precisely enough: 
$|U_{e2}/U_{e1}|=\tan\theta_{12}$ and  
$|U_{e3}|=\sin\theta_{13}$, 
with $\theta_{12}\approx 34^\circ$ and 
$\theta_{13}\approx 8^\circ$. In the rest of this section,
we provide its evaluation for all relevant cases.

In the cases A and B,   discussed in Sect.~\ref{cases},  working in the leading order $\epsilon$,  and upon 
suitable rotations that do not modify the contact term, the neutrino mass matrix has the same structure
\be
-M_D^T M_R^{-1} M_D
\stackrel{\tiny\mathcal{O}(1)}{=}
\frac{\epsilon m^2}{M}
\left(
\begin{array}{ccc}
0 & 0 & 0 \\
0 & 0 & 1 \\
0 & 1 & 1
\end{array}
\right) \label{tra}
\ee
Again, it is easy to verify that the higher order contribution in $m/M$
can be safely neglected. 
As already remarked, these cases produce a very suppressed lightest neutrino mass: if we assume 
$m_{\mbox{\tiny min}}=m_1$, we consider normal mass hierarchy, if instead  
$m_{\mbox{\tiny min}}=m_3$, we consider inverted mass hierarchy. The later case is in fact already treated; it 
corresponds in all details to the calculation of the previous section, and the result of 
Eq.~\ref{contwo} are almost (up effects due to $\theta_{13}$) unchanged.  In the basis where Dirac 
couplings are diagonal, the first case instead  requires to 
identify the following light neutrino mass matrix, 
\be
-M_D^T M_R^{-1} M_D
=
\left(
\begin{array}{ccc}
0 & 0 & 0 \\
0 & 0 & \sqrt{m_2 m_3} \\
0 & \sqrt{m_2 m_3} & m_3-m_2
\end{array}
\right) \label{trb}
\ee
 To get the contact term in the flavor basis, one can repeat the same steps.  
In summary,  we get the simple and explicit result for the cases A and B ( for case B, considering the neutrino mass matrix 
up to $\mathcal{O}(\epsilon)$)
\be
(M_D^T M_R^{-3} M_D)_{ee}^{\mathrm{(Fl.)}}= 
\xi \frac{m^2}{M^3}\times 
\left\{
\begin{array}{cl}
\frac{ (U^*_{e2} \sqrt{m_2} +U^*_{e3} \sqrt{m_3})^2}{m_2+m_3}
& \mbox{ with normal hierarchy}\\[2ex]
\frac{(U^*_{e2} \sqrt{m_2}+ U^*_{e1} \sqrt{m_1})^2}{m_1+m_2}
& \mbox{ with inverted hierarchy}
\end{array}
\right.
\label{conthree}
\ee
 where the superscript ``$\rm{Fl}$''  represents the flavor basis.  For $\Delta m^2_{12}=7.7 \times 10^{-5} $ $\rm{eV}^2$, $\Delta m^2_{23}=2.4 \times 10^{-3} $ $\rm{eV}^2$, $\theta_{12}=34^{\circ}$, $\theta_{23}=42^{\circ}$ and $\theta_{13}=8^{\circ}$, the modular of the
 order one coefficient in the bracket (i.e., $\varphi^2$ in Eq.~\ref{vvv})
goes from about 0.12  to 0.007  in the case of normal hierarchy, and
from about 0.94 to about 0.03 in the case of inverted hierarchy, depending on the Majorana phases.
Correspondingly, we have the following light neutrino contribution, $|(M_\nu)_{ee}|=| m_3 U_{e3}^2-m_2 U_{e2}^2|$ 
(resp., $|(M_\nu)_{ee}|=| m_2 U_{e2}^2-m_1 U_{e1}^2|$) for normal (resp., inverted) mass hierarchy; note that for 
normal hierarchy we can have an almost complete cancellation, i.e., an insignificant contribution
to $0\nu 2\beta$ from light neutrino exchange, for certain choices of the Majorana phases.

\begin{table}[t]
\begin{center}
\begin{tabular}{|c||c|c|c|c|c||c|}
\hline
 sub-case    &  0 meV & 3 meV &10 meV & 30 meV & 100 meV & hierarchy \\
\hline\hline
$m_1<0$ & .00-.57 & .00-.90 & .06-.97 & .15-.98 & .17-.98 & \\
$m_2<0$ & .08-.57 & .00-.90 & .00-.97 & .00-.98 & .00-.98 & normal\\
$m_3<0$ & .08-.34 & .00-.43 & .00-.54 & .00-.69 & .00-.78 & \\ \hline
$m_3<0$ & .00-.99 & .00-1.0 & .00-.96 & .00-.88 & .00-.82 & \\
$m_1<0$ & .18-.99 & .18-1.0 & .18-1.0 & .18-.99 & .18-.98 & inverted\\
$m_2<0$ & .00-.57 & .00-.97 & .00-.98 & .00-.98 & .00-.98 & \\ \hline
    \hline 
     \end{tabular}
\end{center}
\caption{Maximum and minimum value of the numerical coefficient 
$|\varphi |$ for matrices belonging to 
the case C and in all $2\times 3$ sub-cases
(notation as in  Appendix \ref{ma}, Eq.~\ref{quine}), calculated for 
5 values of the lightest neutrino mass, 
ranging from zero to 100 meV.}
\label{tabbo}
\end{table}%

The calculations of case C of Sect.~\ref{cases} are slightly more complicated, since the 
neutrino mass matrix at  leading 
order in $\epsilon$ and in the basis where the Dirac couplings are diagonal is now
\be
M_{\nu}\stackrel{\tiny\mathcal{O}(1)}{=}\frac{m^2\epsilon^2}{M}\pmatrix{ 0 & 0  & 1 \cr 0 &  1  & 1 \cr 1 & 1 & 1}+{\mathcal {O}} (\epsilon^3)
\label{eq:mnu3a}\label{cic}
\ee
 However, it is still possible to obtain analytical expression  for $\varphi$ as defined in Eq.~\ref{vvv}
 applying the results given in 
 Appendix \ref{ma} for the matrix $O$. The explicit 
expressions are not particularly illuminating and thus  we will omit them here, providing 
some sample calculation in Table~\ref{tabbo}. From this table 
and from the calculations we see that
\begin{enumerate} 
\item
 For any hierarchy and for any value of the lightest neutrino mass $m_{\mbox{\tiny min}}$,
the numerical coefficient $\varphi$ of the contact term  can reach  values close to 1.
\item There is not necessarily a suppression of this coefficient both for normal hierarchy as well as for inverted 
hierarchy.

\item
The special possibility realized for certain masses in normal mass hierarchy 
when the contribution from light neutrino exchange is very small or just zero, 
is compatible with a large coefficient $\varphi$, i.e., it does not contradict 
the hypothesis that $0\nu 2\beta$ is completely due to heavy neutrino exchange.
\end{enumerate}

\subsubsection{ Upper bound on heavy neutrino mass $M$  and $\epsilon$ \label{upbndem}}
Having confirmed the mathematical consistency between Type I seesaw and a large contribution from heavy 
neutrinos to $0\nu 2\beta$ transition, for rather specific Dirac and Majorana mass matrices, we  now want 
to discuss briefly whether these scenarios run necessarily into the objection of fine-tuning, as well as 
we now derive the possible upper bounds on the fine-tuning parameter $\epsilon$ and the heavy neutrino mass scale $M$. 
We focus on non-supersymmetric models, which as discussed in Sect.~\ref{qft}, are more likely to 
encounter such an objection.

  Considering the heavy neutrino contribution  to $0\nu 2\beta$
as the dominant one, we  can assume that the sterile neutrino contribution  
 saturates the present bound (or value) on the lifetime $T_{1/2}=1.9 \times 10^{25}$ yr \cite{epj}. In terms of $m$, $M$ and $\kappa$ this can be interpreted as 
$\kappa m^2/M^3= 7.6\times 10^{-9}$ GeV$^{-1}$, where we have considered the nuclear matrix elements $\mathcal{M}_{\nu}=5.24$ and $\mathcal{M}_N=363$ \cite{f2010}.  Combining with the neutrino mass constraint, $\frac{ \epsilon m^2}{M}< 0.1$ eV, 
which automatically ensures the subdominant contribution from the light neutrino exchange 
as long  as $|p^2| > (120)^2 \rm{MeV}^2$,
the upper bound on $\epsilon$ can be obtained as, 
\be
\epsilon \ltap \kappa \left( \frac{\mbox{100 MeV}}{M}\right)^2.
\ee
This condition remains unchanged if both contributions to $0\nu 2\beta$ transition amplitude 
are scaled down by the same factor; this means that we will  not be forced to abandon the 
 hypothesis that the heavy neutrino contribution is the dominating one, even 
 if the true lifetime will turn out to be lower than the present bound (or value) on the lifetime.

To recollect,   the lower bound on $\epsilon$  given in Eq.~\ref{lowbnd-ep},  ensures the stability of 
tree level neutrino mass matrix. The above condition on $\epsilon$, combined with the lower bound 
on $\epsilon$, given in Eq.~\ref{cza} or Eq.~\ref{lowbnd-ep1},  can also be used to derive upper bound on the 
heavy neutrino mass scale $M$. If we use naively as the minimum value of $\epsilon$ the one given in Eq.~\ref{cza}, we find 
that  $M< \sqrt{\kappa} \times 1$ GeV.  However, for $M<M_{ew}$,  it is clear that the right bound is obtained 
using Eq.~\ref{lowbnd-ep1}, rather than   Eq.~\ref{cza}. Hence,  the previously mentioned tight bound on the 
heavy neutrino mass $M$  is relaxed to 
\be
M\ltap \kappa^{1/4}\times 10\mbox{ GeV}, \label{10gev}
\ee
which is a small subset of region indicated by Eq.~\ref{16tev}.
This is a pretty interesting result: this is the  region which automatically satisfies constraints coming 
from small neutrino mass, as well as from radiative stability of the  tree level neutrino mass matrix; the 
only  stringent bound applicable in this region of parameter space comes just from  $0\nu 2\beta$ 
process.\footnote{Redefining $\epsilon \to \epsilon^2$ into the light neutrino mass matrix (relevant for 
Case C), 
i.e.,  $M_{\nu} \propto {m^2 \epsilon^2}/{M}$,  the lower and upper bound on $\epsilon$  should  be interpreted 
as $ (M/1\rm{TeV})<\epsilon< \sqrt{\kappa} (100 $${\rm{MeV}})$$/ M$. Of course, the upper bound on $M$ remains 
unchanged by this re-interpretation.}
 
Note that,  low mass scales of the sterile neutrinos have been considered for a variety of reasons in the 
literature, for instance: \cite{vis-natural} for heavy neutrinos and naturalness;
 \cite{rub} for a mechanism of baryogenesis;
\cite{shap} for a model of heavy neutrinos ($\nu$MSM) ) to account for 
 baryon asymmetry of the universe  and dark matter;  
 \cite{goran-lep} for low scale leptogenesis in supersymmetry;
 {\em etc.}
See \cite{atre} for a review with a compilation of the
experimental constraints from direct search experiments and earlier bound coming from  from $0\nu 2\beta$,  
but recall  that in view of the revised matrix elements presented in \cite{f2010},   the actual bound 
on $0\nu 2\beta$, as shown in our Fig.~1,  is significantly stronger.

\subsubsection{Numerical Example \label{num-ex}}

 To illustrate  the discussions of the previous sections with  numerical
analysis, in this section we offer  a numerical example,  considering the normal hierarchical 
light neutrino masses  $|m_1|<|m_2|<|m_3|$. This example  clearly shows that it is possible to
achieve  large contact term; i.e., even when  light neutrino contribution is small, the factor 
$\varphi$ can be very large to produce  saturating contribution from the sterile neutrino states.

To proceed further,  we include the coefficients of $M_D$ and $M^{-1}_R$, written in Eq.~\ref{casec}. 
\be\label{petreus1}
M_D=m\pmatrix{ f\epsilon^2 & 0 & 0 \cr 0 & g\epsilon & 0 \cr 0 & 0 & 1};\,\,
M_R^{-1}=M^{-1}\pmatrix{
a & b & k \cr
b & c & d \epsilon \cr
k & d \epsilon & e \epsilon^2}
\ee
The light neutrino masses in leading order is the following,
\be 
M_\nu=\pmatrix{ 0 & 0 & \alpha \cr 0 &  \delta & \beta \cr \alpha & \beta  & \gamma},
\ee
where, $\alpha$, $\beta$, $\gamma$ and $\delta$ satisfy the following relations in terms of $m_i$ and the free parameter
$m_0$,
\be
\alpha = kf \frac{\epsilon^2 m^2}{M}&=&\sqrt{- \frac{m_1 m_2 m_3}{m_0}} \\ \nn
\beta= dg \frac{\epsilon^2 m^2}{M}&=&\sqrt{\frac{(m_1-m_0)(m_2-m_0)(m_3-m_0)}{m_0}} \\ \nn
\gamma= e\frac{\epsilon^2 m^2}{M} &= & m_1+m_2+m_3-m_0\\ \nn
\delta= c g^2\frac{\epsilon^2 m^2}{M}&=&m_0
\ee
The leading order expression of the contact term $M^T_DM^{-3}_RM_D$  is the following,
\be
M^T_DM^{-3}_RM_D =\frac{m^2}{M^3} ak^2 \pmatrix {0 & 0 & 0 \cr 0 & 0 & 0 \cr 0 & 0 & 1}
\ee
For this specific  example, we  present  the different numerical values of the input parameters of $M_D$ and $M^{-1}_R$ in 
Table~\ref{num-sample}. 

\begin{table}[h]
\begin{center}
\begin{tabular}{|c|c|c|c|c|c|c|c|c||c |c|}
\hline
$M$ (GeV)& $m$ (MeV) & $\epsilon$ & $a$ & $k$ & $b$ & $c$ & $d$ & $e$ & $f$ & $g$   \cr
\hline
5.00  &  0.935 & 0.02 & 1.00 & 1.35 & 0.90 &  1.4576 & 0.7942  & 0.2898 & 0.0948 & 0.485 \cr
\hline
\end{tabular}
\caption{\label{num-sample}
The input parameters of $M_D$ and $M^{-1}_R$.
}
\end{center}
\end{table}

The light neutrino mass matrix in the Dirac diagonal basis is the following,
\be
M_{\nu}=\pmatrix{ 0.251 \times 10^{-5} & 0.579 \times 10^{-3} & 0.895 \times 10^{-1} \cr 0.579  \times 10^{-3}  & 0.24  & 0.269  \cr 
0.895 \times 10^{-1} & 0.269  & 0.203 } \times 0.1 \, \rm{eV}
\ee
where,  for completeness, we have shown  the higher order terms as well.  For the   choice of  
input parameters given in Table~\ref{num-sample}, the light as well as heavy neutrino masses and 
the solar and atmospheric mass square differences
has been presented in Table~\ref{num-sample2}.
\begin{table}[h]
\begin{center}
\begin{tabular}{|c|c|c|c|c||c|c|c|}
\hline
$m_1$ (meV)& $m_2$ (meV) & $m_3$ (meV)& $\Delta m^2_{21}$ $\rm{eV^2}$  & $\Delta m^2_{32}$ $\rm{eV^2}$ & $M_1$ (GeV) & $M_2$ (GeV)& $M_3$ (GeV) \cr
\hline
$3.95 $ &  $-9.60 $ & $49.9$ & $ 7.66 \times 10^{-5}$ & $2.40 \times 10^{-3}$ & $1.99$ & $-4.77$ & $5.02$ \cr
\hline
\end{tabular}
\caption{\label{num-sample2}
The light, heavy neutrino masses and the solar as well as atmospheric mass square differences. 
}
\end{center}
\end{table}
To go to the flavor basis, we further choose the mixing angles as  $\theta_{12}=34^{\circ}$, $\theta_{23}=42^{\circ}$ and $\theta_{13}=8^{\circ}$. 
The numerical values of flavor basis contact term, as well as the light neutrino contribution and the enhancement factor $\varphi$  has been shown in Table  \ref{num-sample4}. 

From Table~\ref{num-sample4}, it is clearly evident, that the enhancement factor $\varphi^2=0.55$ 
is quite large. This, together with the other input parameters $m$, $M$ and $a$, $k$ produces a 
saturating contribution \cite{epj, f2010} of the sterile neutrino states in $0\nu 2 \beta$ process, whereas  
the standard model neutrino contribution remains  strongly  suppressed \footnote{ For $\theta_{13}=0$ and the 
same set of other input parameters, the enhancement factor $\varphi^2$ diminishes to 0.42, while $(M_{\nu})_{ee}$ in 
flavor basis  becomes   $(M_{\nu})_{ee}=0.29$ meV.}.

\begin{table}[h]
\begin{center}
\begin{tabular}{|c|c|c|c|}
\hline
$\varphi^2$ & $({M_{\nu}})_{ee}$ (meV) & $({M^T_DM^{-1}_RM_D})_{ee}$ $(\rm{GeV}^{-1})$ & $(M_{\nu})_{ee}/|p^2|$$(\rm{GeV}^{-1})$ \cr
\hline
$0.55$ & $0.69 $  & $7.01 \times 10^{-9}$ & $2.08 \times 10^{-11}$\cr
\hline
\end{tabular}
\caption{\label{num-sample4}
The contact term, enhancement factor $\varphi^2$  and the light neutrino contribution in the flavor basis. In order to calculate
the light neutrino contribution in $0\nu 2\beta$ process, 
$|p^2|=(182)^2\, \rm{ MeV}^2$ has been considered.  }
\end{center}
\end{table}

\subsubsection{Conditions of validity and meaning of Eq.~\ref{10gev}\label{hphp}}

In the last part of this section, we  examine in depth the hypotheses that led to 
Eq.~\ref{10gev} which says that, in the absence of excessive fine-tunings, large contributions to
$0\nu 2\beta$  of sterile neutrinos are possible, 
only if their masses are below 10 GeV or so.

Consider the case when the 
tree-level neutrino mass $M_\nu^{\mbox{\tiny tree}}$ is zero due to a cancellation. 
This can be regarded as an effect of the opposite (and possibly large)
contributions, due to the exchange of  virtual 
sterile neutrinos $N_1$ and $N_2$, 
\begin{equation}
M_\nu^{\mbox{\tiny tree}}= M_\nu^{(1)}+M_\nu^{(2)}=0 \label{ft}
\end{equation}
where, following  \cite{smirnov}, we consider a two-dimensional 
system to simplify the analysis.
This can be made explicit  considering 
the mass matrix in Eq.~\ref{mde}, 
that in the limit $\epsilon\to 0$ has 
\begin{equation}
M^{(1)}_\nu=-M^{(2)}_\nu=\frac{m^2}{M} 
\left(
\begin{array}{cc}
0 & 0 \\
0 & 1
\end{array}
\right) \frac{1}{\sqrt{4+b^2}}
\end{equation}
Indeed, the mass matrix in Eq.~\ref{mde} is constructed just to
obey the condition of Eq.~\ref{ft}.

As already recalled in Sect.~\ref{qft}, one should consider  
the radiative corrections to light neutrino masses, that are potentially large. 
This is true for the one-loop correction estimated in \cite{smirnov}, 
assuming that both sterile neutrino masses $M_2>M_1$ are  
above the electroweak scale. Using again the matrix in Eq.~\ref{mde}, these read 
\begin{equation}
\delta M_\nu\approx \frac{\alpha_2-\alpha_1}{(4\pi)^2} \times 
M^{(1)}_\nu \times 2\ \mbox{sinh}^{-1}(b/2)
\label{loppp}
\end{equation} 
which is the same as Eq.~\ref{rge}, after including all the order-one factors.
The different running of $M_\nu^{(1)}$ and $M_\nu^{(2)}$ is described by the 
order-one factor
$\alpha_2-\alpha_1=\lambda+\frac{3}{2} g^2+\frac{3}{2}{g'}^2$, and the 
last factor in Eq.~\ref{loppp}
is just a compact expression of $\log(M_2/M_1)$ as a function of $b$.
As anticipated in Sect.~\ref{m-M}, such a correction is just 
a couple of orders of magnitude smaller than 
the naive expectation $m^2/M$ and thus it is non-negligible. 

At this point, we can focus on the case of interest.
When we assume that the sterile neutrinos saturate the $0\nu 2\beta$ bound, 
the combination of parameters $\varphi^2 b\ m^2/M^3$ is fixed, see Eq.~\ref{vvv}. 
This permits us to rewrite Eq.~\ref{loppp} as
\begin{equation}
\delta M_\nu\approx 100\mbox{ eV} \left( \frac{M}{100\ \mbox{GeV}}\right)^2
\left( \frac{\varphi^2 b\ m^2/M^3}{7.6\times 10^{-9}\ \mbox{GeV}^{-1}}\right) 
\left( \frac{\lambda(b)/\varphi^2}{0.2}\right)
\left(
\begin{array}{cc}
0 & 0 \\
0 & 1
\end{array}
\right) 
\label{nummmm}
\end{equation}
where $\varphi$ is discussed in Sect.~\ref{c2} and 
\begin{equation}
\lambda(b)=\frac{\mbox{sinh}^{-1}(b/2)}{b \sqrt{1+(b/2)^2}}
\end{equation}
is a decreasing function, with $\lambda(0)=1/2$.

Note that when $b$ grows, the mass of the lighter 
sterile neutrino $M_1=M( \sqrt{1+(b/2)^2}-b/2)$ 
decreases; but this has to be above the 
electroweak scale to ensure the validity of the 
radiative corrections estimated in \cite{smirnov}. 
 The numerical result of Eq.~\ref{nummmm}
means that the one-loop values of the 
light neutrinos will be much larger than the measured value, 
when we assume that the tree level mass matrix is zero and the 
scale of the sterile neutrinos is above $M=100$ GeV.\footnote{This 
is relevant for the works based on    
the assumption that sterile neutrinos with large masses 
provide the main contribution to $0\nu 2\beta$,
e.g., \cite{Ibarra}. A quantitative evaluation of the light-neutrino masses 
was not attempted there, though the authors seem to be aware of the potential issue, 
since after Eq.~(39) they state ``higher order  (one or two loop) contributions 
lead to $m_\nu \neq 0$".}
Let us repeat that 
the crucial hypothesis to reach this conclusion is that the sterile (heavy) neutrinos saturate 
the $0\nu 2\beta$ bound, which implies that the lepton number is strongly violated. In fact, 
as it is clear from Eq.~\ref{nummmm}, the one loop correction $\delta M_\nu$ is just proportional to  
the heavy neutrinos contribution to the $0\nu 2\beta$ amplitude. 


Formally, it is possible to replace the fine-tuning condition operated at tree-level, Eq.~\ref{ft}, with a 
fine-tuning involving also the radiative contributions.  For instance,  we can include
 the logarithmic corrections of \cite{smirnov}, imposing the condition
\begin{equation}
M_\nu^{\mbox{\tiny 1 loop}}= M_\nu^{(1)} \left( 1+ \frac{\alpha_1}{(4\pi)^2} \log\frac{M_2}{M_1}\right)+
M_\nu^{(2)}  \left( 1+ \frac{\alpha_2}{(4\pi)^2} \log\frac{M_2}{M_1} \right) =0 \label{cuculon}
\end{equation}
One can rewrite it approximatively as
\begin{equation}
M_\nu^{(1)}+M_\nu^{(2)}  \approx 
- (M_\nu^{(2)}-M_\nu^{(1)}) (\alpha_2-\alpha_1)\frac{\log({M_2}/{M_1})}{32 \pi^2} 
\end{equation}
which emphasizes
that we have to concoct  
a small and ad hoc tree level term to cancel the one-loop term.

Eq.~\ref{cuculon} permits us to retain 
large sources of lepton number violation in the theory without contradicting the 
observed neutrino masses; moreover, it 
can be extended to include two-loop terms or, possibly, even finite corrections.
While technically acceptable, we believe that Eq.~\ref{cuculon} should be considered 
as an excessive fine-tuning,
since it is not justified in terms of symmetries and it appears to be 
even more artificial than the tree level condition Eq.~\ref{ft}. 

In the present section, 
guided by the phenomenological motivations described in the introduction
and following a vast literature on the subject, we studied the consequences of accepting a moderate amount of fine-tuning, corresponding to Eq.~\ref{ft}.
We could even resort to the fine-tuning condition of Eq.~\ref{cuculon}
eventually, if the data should force us to do so; e.g., if we would have evidence of the existence of sterile neutrinos of 100 GeV or 1 TeV. But, at present, we miss any convincing motivation to analyze the implications of  Eq.~\ref{cuculon} any further, and tentatively, we bar it as unlikely.
If instead one is convinced on a theoretical basis that this type of conditions 
(either Eq.~\ref{ft} or Eq.~\ref{cuculon}) should not be 
accepted,  the conclusions are those already illustrated in Fig.~\ref{xxx}.

\section{Going beyond Type I seesaw \label{ex}}

In the previous sections  we have restricted the discussion of $0\nu 2\beta$ for the simple prototype 
Type I seesaw scenario. As it is well known, smallness of neutrino masses can be well explained by other
 seesaw scenarios as well,  e.g.,  Type II \cite{seesaw-Goran, moha, type-II}, Type III \cite{type-III, goran-typeIII,  typeIIIgmsb, us},
 Inverse seesaw \cite{invo, inv, invdet,  inverseso10, inverseothers}, Extended seesaw \cite{Kang-Kim, Parida}. 
 
 In the case of the Type II seesaw, it is clear that the new contribution due to doubly charged scalars in the triplet
 is always smaller than the neutrino mass one. This is evident from the fact that the new contribution is again
 proportional to $(M_\nu)_{ee}$ element of the neutrino mass matrix, but now suppressed by the large mass
 of the doubly charged scalar. The situation changes in the case of the LR symmetric theory, where the right-handed
 triplet may enter the game since its contribution is proportional to the right-handed analog of $(M_\nu)_{ee}$.
 However, the constraints from lepton flavor violation processes seem to render this contribution sub-leading
 compared to the right-handed gauge boson one \cite{tello}.
 
 The case of Type III proceeds in the same manner as the Type I, since one simply interchanges the fermion
 singlets, i.e. the right-handed neutrinos by the neutral components of the fermionic triplets.  The crucial
 difference lies in the collider aspect of the model, since the triplets can be produced through the gauge couplings.
 We will not discuss this issue any further here; for 
 a recent review of see e.g. \cite{Senjanovic:2011zz}. Also,
 the question lepton flavor violation becomes more intricate \cite{type-Ilfv, type-IIIlfv, miha-lfv}.

 In the cases of the Inverse and Extended seesaws, the
 smallness of the neutrino masses is inherently linked with a small lepton number violating element  of 
the neutral lepton mass matrix. The implementation of Extended seesaw  in low scale leptogenesis has been studied  in literature  \cite{Kang-Kim, Parida}.
 Below, we pursue the question of obtaining a large and dominant sterile neutrino contribution  for  
the Extended seesaw scenario \cite{Kang-Kim,Parida}, where the sterile neutrino sector has been extended 
by additional degrees of freedom. 

  Our discussion on the Extended seesaw  proceeds as follows: Below,  we first describe the Extended seesaw 
mass matrix and the relevant mixing matrix. We present a brief comparison between the analytic and numerical 
result for one generation case in Sect.~\ref{oneg-ex}. After that, in 
Sect.~\ref{ex-0}, we 
discuss the dominant role of sterile neutrino states in $0 \nu 2 \beta$ process. Finally in Sect. \ref{cons-ex}, we 
discuss the major  constraints on the Dirac mixing $m$ and the lightest sterile neutrino mass $m_s$, coming 
from $0\nu 2\beta$ transition, as well as the  heavy Majorana neutrino searches at LHC and the lepton flavor 
violating process. The detail of the diagonalization procedure for Extended seesaw and the higher-dimensional 
considerations of the mass and mixing matrix  has been given  in Appendix \ref{diagonalization-extended}.

\subsection{Extended seesaw \label{ex-see}}

 In Extended seesaw (or Extended double seesaw according to \cite{Kang-Kim}) framework \cite{Kang-Kim, Parida}, 
we have  $n$-generation ($n=3$)  of standard model neutrino $\nu_L$,  $m$-generation of  sterile neutrino state 
$S_L$ and $p$-generation of sterile neutrino state $N_L$. The  Lagrangian describing the mass terms of the neutral 
leptons has the following form, 
\be
L=-\frac{1}{2} \pmatrix {\nu_L & S_L & N_L} \pmatrix {0 & 0 & M^T_D \cr 0 & \mu  & M^T_S \cr M_D & M_S & M_R}\pmatrix{\nu_L \cr S_L \cr N_L}+\rm{h.c}.
\ee
We denote the neutral lepton mass matrix as $M_n$\footnote{Note that, from hereon in the discussion 
of Extended seesaw, $\mu$ implies the Majorana mass of the sterile neutrino state $S_L$.} where,
\be
M_n=\pmatrix {0 & 0 & M^T_D \cr 0 & \mu  & M^T_S \cr M_D & M_S & M_R}.
\label{eq:extended}
\ee
To understand the GUT realization of this seesaw scenario, see \cite{Parida}. In this specific example, 
we work in a basis where the Majorana mass matrix $M_R$ is real  and $M_D$ represents  the mixing between 
the standard model flavored neutrino state $\nu_L$ and the heavy sterile neutrino state $N_L$. Furthermore, 
being  a Majorana mass matrix of the heavy neutrino state    $S_L$, the matrix  $\mu$ is   complex symmetric.
In addition, throughout our analysis we adopt the following few assumptions,
\begin{itemize}
\item
The generation of $N_L$ and $S_L$ are identical, {\it i.e., } $m=p$. As a result, the matrix $M_S$ is  a  square matrix. 
\item
The matrices $M_R$ and $M_S$ are invertible. 
\item
The different sub-matrices of the neutral lepton mass matrix follow this hierarchy, $M_R>M_S>M_D \gg \mu$ and $\mu <
M^T_S M^{-1}_RM_S$, i.e.,  $\mu < \mathcal{O}(\frac{M^2_S}{M_R})$.
\end{itemize}

Note that, this Extended seesaw scenario  is very different from the  inverse seesaw scenario \cite{invo, inv, invdet, inverseso10, inverseothers}, 
due to the simultaneous presence of both the  heavy and small lepton number violating scales  $M_R$ and $\mu$ 
respectively. The later has been widely discussed in the literature for its large contribution in  lepton flavor 
violating processes \cite{invo, inv, invdet,  inverseso10, inverseothers}. In inverse seesaw, there is only one small lepton number violating scale $\mu$ 
and  the lepton number is conserved in  $\mu=0$ limit. Hence,   the $0\nu 2\beta$ transition amplitude also 
vanishes in this  limit. On the contrary,  in  Extended seesaw,   the heavy Majorana neutrino contribution 
can be the dominant contribution, even when the small lepton number violating  scale $\mu$ vanishes.  However, 
the standard model neutrino masses  strongly depend  on the small lepton number violating scale $\mu$ and hence 
in the $\mu=0$ limit, the standard model neutrinos become massless. As a result, the contributions of the 
 standard model neutrinos and the heavy Majorana  neutrinos in $0\nu 2 \beta$ process are  completely decoupled 
from each other. This is the essence  of  our work  on  Extended seesaw, which we discuss in detail in the 
subsequent  sections.

\subsubsection{Mass and Mixing \label{mass-mix}}

 We start with evaluating the mixing of the standard model neutrinos with these extra sterile 
states $S_L$ and $N_L$. The diagonalization of this Extended seesaw mass matrix (Eq.~\ref{eq:extended}) is 
carried out by the $(n+2m) \times (n+2m)$-dimensional matrix $\mathcal{U}$ where,
\be
\mathcal{U}^T M_n \mathcal{U}=M^d_n.
\ee
We decompose the mixing matrix $\mathcal{U}$ as $\mathcal{U}=\mathcal{U}_1\mathcal{U}_2$, where 
$\mathcal{U}_1$ and $\mathcal{U}_2$ satisfy the relations $\mathcal{U}^T_1M_n\mathcal{U}_1=M_{bd}$ and $\mathcal{U}^T_2M_{bd}\mathcal{U}_2=M^d_n$. $M_{bd}$ and $M^d_n$ are respectively the block diagonal and diagonal mass matrices  and are denoted as,
\be
M_{bd}=\pmatrix{m_{\nu} &0 & 0 \cr 0 & m_s & 0 \cr 0 & 0 & m_n}
\label{eq:md}; \, M^d_n =\pmatrix{m^d_{\nu} &0 & 0 \cr 0 & m^d_s & 0 \cr 0 & 0 & m^d_n}.
\label{eq:mdd}
\ee
We define the  block-diagonal  basis and the diagonal mass basis with $\pmatrix{\nu_b,S_b,N_b}^T$ and 
$\pmatrix{\nu_m,S_m,N_m}^T$ respectively. The flavor  state $\pmatrix{\nu_L, S_L, N_L}^T$ is related with 
the mass  state  $\pmatrix{\nu_m, S_m, N_m}^T$ as follows, 
\be
\pmatrix{ \nu_L \cr S_L \cr N_L} =\mathcal{U}\pmatrix{\nu_m \cr S_m \cr N_m}.
\ee
It is clearly evident from Eq.~\ref{eq:mdd}, that $m_{\nu},m_s$ and $m_n $ are the  mass matrices 
corresponding to the intermediate states $\nu_b,S_b$ and $N_b$, while $ m^d_{\nu},m^d_s$ and $m^d_n$ are 
the diagonal matrices containing the physical masses  and correspond  to the  states $\nu_m,S_m$ and $N_m$ 
respectively. For one generation of $S_L$ and $N_L$   evidently,  $m_s=m^d_s$ and $m_n=m^d_n$.  
 
Following the  parameterization in \cite{Grimus-Lavoura},   to the leading order the mixing matrix  $\mathcal{U}_1$  is,
\be
\mathcal{U}_1 \sim \pmatrix {1-\frac{1}{2}M^{\dagger}_D(M^{-1}_S)^{\dagger}M^{-1}_SM_D & M^{\dagger}_D (M^{-1}_S)^{\dagger}  & M^{\dagger}_D M_R^{-1} \cr  -M^{-1}_S M_D & 1-\frac{1}{2}M^{-1}_SM_DM^{\dagger}_D(M^{-1}_S)^{\dagger}-\frac{1}{2}M^{\dagger}_SM^{-2}_RM_S & M^{\dagger}_S M^{-1}_R \cr  {M^T_S}^{-1} \mu M^{-1}_SM_D & -M^{-1}_R M_S &  1-\frac{1}{2} M^{-1}_R M_S M^{\dagger}_S  M^{-1}_R },
\label{eq:mix}
\ee
In the above, we have neglected the  $\mathcal{O}(\frac{M_D}{M_R})^2$ terms as compared 
to $\mathcal{O}(\frac{M_S}{M_R})^2$ and $\mathcal{O}(\frac{M_D}{M_S})^2$ terms, as   
$ \frac{M_D}{M_R} < \frac{M_S}{M_R}, \frac{M_D}{M_S}$. Also,  note that, $(\mathcal{U}_1)_{31}$ is
 of the order $\mathcal{O}(\frac{M_D}{M_R} \frac{\mu}{M^2_S/M_R})$, which comes from the expansion 
of  $\mathcal{O}(\frac{\mu}{M^2_S/M_R})$ elements. In addition, in the diagonal elements $(\mathcal{U}_1)_{11,33}$, 
we have  shown  the dominant sub-leading 
corrections,  and in $(\mathcal{U}_1)_{22}$, we have shown the correction which does not involve  $\mu$. The 
corrections  involving $\mu$ in the other  elements of $\mathcal{U}_1$ are   smaller than the leading order 
terms and hence we do not show them explicitly. To the leading order, the  light neutrino mass matrix $m_{\nu}$, 
and the heavy neutrino mass matrices $m_s$, $m_n$ have these following form,
\be
m_{\nu} & \sim &  M^T_D (M^T_S)^{-1} \mu M^{-1}_S M_D \nn, \\ 
m_s & \sim & -M^T_S M^{-1}_R M_S \nn, \\ 
m_n & \sim &  M_R.
\label{eq:mass}
\ee
Note that, due to   $M_S<M_R$,   the mass matrices  $m_s$ and $m_n$ of the sterile neutrino 
states $S_b$ and $N_b$  satisfy the following inequality $m_s<m_n$. From Eq.~\ref{eq:mix} and 
Eq.~\ref{eq:mass}, one can see that the standard model neutrino mass matrix depends on the 
parameter  $\mu$, whereas  to the leading order, the mixing ($(\mathcal{U}_1)_{12}$ and $(\mathcal{U}_1)_{13})$) 
between the  standard model neutrinos $\nu_L$ and the sterile neutrino states $S_m$, $N_m$  are 
independent of that parameter.  Hence, one can choose $\mu$ to be   small to generate eV neutrino masses,  
still having large active-sterile neutrino mixings. The corresponding eigenvalues of the 
matrices $m_{\nu}$, $m_s$ and $m_n$  can be extracted by further diagonalization with the mixing 
matrix $\mathcal{U}_2$, where the mixing matrix $\mathcal{U}_2$ is denoted as,
\be
\mathcal{U}_2=\pmatrix{U & 0 & 0 \cr 0 & W_S & 0 \cr 0 & 0 & W_N}.
\label{eq:mixb}
\ee
The three matrices $U$, $W_{S,N}$ diagonalize  the light and heavy neutrino matrices $m_{\nu}$ and $m_{s,n}$ respectively,
\be
U^Tm_{\nu}U &=& m^d_{\nu}=diag(m_i)\rm{,}\, \\ \nn     W_S^Tm_sW_S &=& m^d_s=diag(M_{S_i})\rm{,} \\ \nn        W_N^Tm_nW_N &=& m^d_n=diag(M_{N_i}).\nn
\ee 
In the above, $m_i$, $M_{S_i}$ and $M_{N_i}$ are the physical masses of the neutrino 
states $\nu_{m_i}$, $S_{m_i}$ and $N_{m_i}$ respectively. 
From Eq.~\ref{eq:mix} and Eq.~\ref{eq:mixb}, one immediately gets the following form 
of the mixing matrix $\mathcal{U}$,
\be
\pmatrix {(1\!-\!\frac{1}{2}M^{\dagger}_D(M^{-1}_S)^{\dagger}M^{-1}_SM_D) U  & M^{\dagger}_D (M^{-1}_S)^{\dagger}W_S  & M^{\dagger}_D M_R^{-1} W_N \cr  
-M^{-1}_S M_D U & 
(1\!-\!\frac{1}{2}M^{-1}_SM_DM^{\dagger}_D(M^{-1}_S)^{\dagger}\! -\!\frac{1}{2}M^{\dagger}_SM^{-2}_RM_S )W_S & M^{\dagger}_S M^{-1}_R W_N 
\cr  
{M^T_S}^{-1}\! \mu M^{-1}_SM_D  U & -M^{-1}_R M_S W_S &  (1 \!  -\!\frac{1}{2} M^{-1}_R M_S M^{\dagger}_S  M^{-1}_R )W_N } \label{mixA}
\ee
We provide the technical details of the block diagonalization  in  Appendix \ref{diagonalization-extended}. 
Below, we present the comparison between the analytical expression of the mixing matrix with the numerical result, 
considering  one generation of standard model neutrino $\nu_L$ and one extra generation of sterile neutrino states $S_L$ and $N_L$.

\begin{figure}[t]
\begin{center}
\epsfig{file=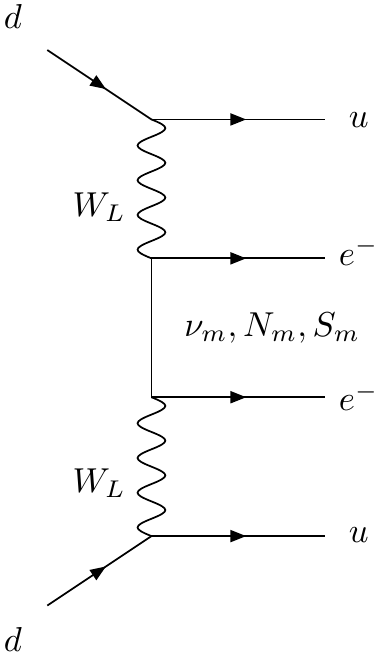,width=0.25\textwidth}
\hspace*{2cm}
\caption{ \label{bb2} Feynman diagram of the $0\nu 2 \beta$ process for Extended seesaw. The  intermediate neutrino states are in mass basis.}  

\end{center}
\end{figure}

\subsubsection{One generation consideration \label{oneg-ex}}
For one generation of standard model neutrino $\nu_L$ and one extra generation of $S_L$ and $N_L$ states, the mass and mixing matrix is a simple $3 \times 3$ matrix. 
We present here a numerical estimation of  the mass and mixing in this simplest case,  
and compare the numerical result with the analytical approximations. 
The mixing matrix is just:
\be
\mathcal{U}=\pmatrix{ 1-\frac{M^2_D}{2 M^2_S} & \frac{M_D}{M_S} & \frac{M_D}{M_R} \cr -\frac{M_D}{M_S} & 1-\frac{1}{2} \frac{M^2_D}{M^2_S}-\frac{1}{2}\frac{M^2_S}{M^2_R} & \frac{M_S}{M_R} \cr
\frac{M_D}{M_R} \frac{\mu}{M^2_S/M_R} & -\frac{M_S}{M_R} & 1-\frac{1}{2} \frac{M^2_S}{M^2_R}}
\label{expvr}
\ee
Few comments are in order,
\begin{itemize}
\item
The leading order estimation of the diagonal terms  is $\mathcal{U}_{{11},{22},{33}} \sim 1$.
\item
$\mathcal{U}_{12,13,23}$ elements are of the following order $\mathcal{U}_{12} \sim \frac{M_D}{M_S}$, $\mathcal{U}_{13} \sim \frac{M_D}{M_R}$ and $\mathcal{U}_{23} \sim \frac{M_S}{M_R}$.
\item
For $\mu =0$,  $\mathcal{O}(\frac{1}{M_R})$ (as well as higher orders  in $\mathcal{O}(M^{-1}_R)$) terms in  
$\mathcal{U}_{31}$ suffers  mutual cancellation, resulting  $\mathcal{U}_{31}=0$.  For non-zero $\mu$, the leading 
order contribution in $\mathcal{U}_{31}$ becomes $\mathcal{U}_{31} \sim \frac{M_D}{M_R} \frac{\mu}{M^2_S/M_R}$. See 
Appendix \ref{high-c} for the detailed discussion about 
this particular feature. 
\end{itemize}
For this simple one generation case, the  numerical analysis has been shown in Table~\ref{sample}. The different 
elements  $M_D$, $M_S$, $M_R$ and $\mu$ are the inputs;  $m_{\nu}$ is the light neutrino mass, whereas the other 
two heavy neutrino masses are denoted by $m_s$ and $m_n$ respectively. Considering the sample values of the input 
 parameters given  in Table~\ref{sample}, the estimation of $\mathcal{U}$ is the following,
\be
\mathcal{U}=\pmatrix{ 1.000 & 0.995  \times 10^{-3} & 0.985 \times 10^{-4} \cr
-0.001 & 0.995 & 0.985 \times 10^{-1} \cr
0.999 \times 10^{-10}  & -0.985 \times 10^{-1}  & 0.995}
\ee
As expected from Eq.~\ref{eq:mass} and Eq.~\ref{expvr}, 
\begin{itemize}
\item
$m_{\nu}=(\frac{M_D}{M_S})^2 \mu \sim 0.1 $eV and the other two heavy masses $m_s$ and $m_n$  are  $99.02$ GeV 
and $10099$ GeV respectively. Note that to the leading order $m_n \sim M_R$. The analytical 
result $m_n \sim M_R+ \frac{M^2_S}{M_R}$ (see Appendix \ref{diagonalization-extended})  resembles 
very closely the numerical estimation. 
\item
The naive estimation of $\mathcal{U}_{12,13,23}$ as  $\mathcal{U}_{12} \sim \frac{M_D}{M_S} \sim 10^{-3}$, $\mathcal{U}_{13} \sim \frac{M_D}{M_R} \sim 10^{-4} $,  $\mathcal{U}_{23} \sim \frac{M_S}{M_R} \sim 10^{-1}$ and $\mathcal{U}_{31} \sim \frac{M_D}{M_R} \frac{\mu}{M^2_S/M_R} \sim 10^{-10} $ matches well the  numerical result. 
\end{itemize}

\begin{table}[h]
\begin{center}
\begin{tabular}{|c|c|c|c||c|c|c|}
\hline
$M_R$ & $M_S$ &$ M_D$ & $\mu $ & $m_{\nu}$ & $m_s$ &$ m_n $   \cr
\hline
$10^4 $  & $ 10^3 $  & $1.0$ & $10^{-4}$ & $10^{-10} $  & $ -99.0195 $  & $10099.0$  \cr
\hline
\end{tabular}
\caption{\label{sample}
The light and heavy neutrino masses in GeV, for one generation Extended seesaw scenario.
}
\end{center}
\end{table}


\subsection{Extended seesaw and $0\nu 2\beta$ transition \label{ex-0}}
In this subsection, we discuss the contributions of heavy Majorana neutrino states in the $0\nu 2\beta$ 
process (see Fig.~\ref{bb2}). Note that, in this case  both the heavy Majorana neutrino states $S_m$ and $N_m$ 
will participate in $0\nu 2\beta$ process. Depending on the  masses and their mixings with the standard model 
neutrinos, the contributions  of the heavy states $S_m$ and $N_m$ will differ.  As for  the Type I seesaw,   we  
work in the following mass regime $m_n^d> m^d_s>200$ MeV. As evident from Eq.~\ref{mixA}, the electron flavor 
neutrino state $\nu_e$  mixes  with the 
different active and sterile neutrino mass states $\nu_m,S_m$ and $N_m$ as follows,
\be
 \nu_e \sim U_{ei} \nu_{m_i} + (M^{\dagger}_D ({M^{-1}_S})^{\dagger} W_S )_{ek}S_{m_k} + (M^{\dagger}_D M_R^{-1} W_N )_{el}N_{m_l},
\ee
where, $i,k,l$ represents the generations of the standard model neutrino state  $\nu_m$ and the heavy sterile  
states  $S_m$ and $N_m$ respectively. For simplicity, in the above,  we have neglected any non-unitary effect associated with $U$.
Note that,  the mixing between the electron neutrino $\nu_e$ and the sterile neutrino mass eigenstates 
$S_m$ and $N_m$  are not constrained  by  the smallness of the standard model light neutrino mass. 

The discussion of sterile neutrino contribution proceeds analogously as for the  Type I seesaw. Denoting the two 
mixings matrices between the active and sterile states  $M^{\dagger}_D ({M^{-1}_S})^{\dagger} W_S$ and $M^{\dagger}_D M_R^{-1} W_N $ 
by the notations $V_{eS}$ and $V_{eN}$ respectively, the half-life time period of $0\nu 2\beta$ transition can be expressed 
as follows,  
$$
\frac{1}{T_{1/2}}=K_{0\nu} \left| \frac{U_{ei}^2 m_i}{\langle p^2 \rangle} - \frac{V_{e{S_i}}^2}{M_{S_i}} - \frac{V_{e{N_i}}^2}{M_{N_i}}\right|^2,
$$
where the definition $K_{0\nu}$ and $ \langle p^2 \rangle $  follow the Type I seesaw considerations, i.e.,  $K_{0\nu}=G_{0\nu} (\mathcal{M}_N m_p)^2$ and  $ \langle p^2 \rangle\equiv -m_e m_p \frac{\mathcal{M}_N}{\mathcal{M}_\nu}
$ (Eq.~\ref{psq} and \cite{tello}). In the above, $\frac{V_{e{S_i}}^2}{M_{S_i}}$ and $ \frac{V_{e{N_i}}^2}{M_{N_i}}$ are the contributions of the sterile  neutrino states $S_{m_i}$ and $N_{m_i}$ in $0\nu 2\beta$ process respectively, and $M_{S_i}$ and $M_{N_i}$ are the corresponding masses . Expressing $V_{eS}$ and $V_{eN}$ back in terms of $M_D$, $M_S$ and $M_R$, the total amplitude of $0\nu 2 \beta$ process is the following, 
\be
\mathcal{A}^*_l = \frac{m_{ee}}{\langle p^2 \rangle }-\left ( M^T_D{M^{-1}_S}^T {m^*_s}^{-1} M^{-1}_SM_D \right )_{ee}-\left ( M^T_DM_R^{-3}M_D \right )_{ee}.
\ee
In the above,  $m_{ee}$ is the standard model neutrino mass contribution and is given by, 
\be
m^*_{ee}=\left (M^T_D (M^{-1}_S)^T \mu M^{-1}_S M_D \right )_{ee}.
\label{eq:sm-cont}
\ee
The quantity $m_s$ represents the mass matrix of the sterile intermediate state $S_b$  and is given in 
Eq.~\ref{eq:mass}. The contribution of the sterile neutrino mass states $S_m$ and $N_m$ to the neutrinoless
double beta decay amplitude, i.e., the contact terms are respectively given by,
\be
\mathcal{A}_S &=&  \left ( M^T_D{M^T_S}^{-1}{m^*_s}^{-1} M^{-1}_SM_D \right )_{ee} \nn\\
\mathcal{A}_N &=&  \left ( M^T_DM_R^{-3}M_D \right )_{ee},
\label{eq:mul-0nu2beta}
\ee
For one generation of standard model neutrino $\nu_e$  and one extra sterile states
 $S_m$ and $N_m$, the amplitudes simplify  to,
\be
\mathcal{A}_S &=& \left ( \frac{M_D}{M_S} \right )^2\frac{1}{m_s}=\frac{M^2_D}{m_n m^2_s}, \nn \\
\mathcal{A}_N &=& \left (\frac{M_D}{M_R} \right )^2 \frac{1}{m_n}=\frac{M^2_D}{m^3_n},
\label{eq:one-0nu2beta}
\ee
where we have used  $m_s \sim \frac{M^2_S}{M_R}$ and $m_n \sim M_R$. 
Note that, for one generation of $\nu_L$ and $S_L$, $N_L$ states $m_s$ and $m_n$ are the physical 
masses of the  sterile states  $S_m$ and $N_m$. Few  comments are in order,
\begin{itemize}
\item
The contributions coming from the extra sterile
 states $S_m$ and $N_m$ are decoupled from the light neutrino contribution. This can be clearly seen 
from Eq.~\ref{eq:mul-0nu2beta}, as well as from Eq.~\ref{eq:one-0nu2beta}.
\item
In the $\mu \to 0$ limit, when 
the contribution of the light neutrino $m_{ee}$  is zero, one can even get a non-zero and 
significant contribution to the  neutrinoless
double beta decay, due to the additional heavy  neutrino states. 
\item
As an example, the choice $M_D=10\sqrt{0.1}$ GeV, $M_S=10^{4}$ GeV and  $M_R=10^{6}$ GeV 
generate $m_s=10^2$ GeV, $m_n=10^{6}$ GeV. The mixing of standard model neutrinos
with the state $S_m$ is $10^{-3}\sqrt{0.1}$. This generates the contribution $\mathcal{A}_S=10^{-9}\,   \rm{GeV}^{-1}$.
\item
The light neutrino mass scale is fixed by the lepton number violating  parameter $\mu$. For the above
 mentioned numerical values, $\rm{eV}$ neutrino mass  is generated by fixing $\mu=10^{-3}$ GeV.

\item
In the present scenario, the mixing between the heavy Majorana neutrino state $N_m$ and the standard model neutrino $\nu_L$ is 
 given by $\frac{M_D}{M_R}=\frac{M_D}{M_S} \frac{M_S}{M_R}< \frac{M_D}{M_S}$, where $\frac{M_D}{M_S}$ represents  
the mixing of the heavy state $S_m$ with the standard model neutrino state $\nu_L$. Also, the mass of the heavy 
states $N_m$ and $S_m$ are related by the following inequality $m_n>m_s$. Hence, excepting  fine-tuning or 
cancellation among the active-sterile ($\nu_L-S_m$) neutrino mixing,  the contribution of the heavy 
states $N_m$ will most likely be much smaller.  We do not address any such cancellations among the 
sterile neutrino states $S_{m_k}$  in this present study. Hence, in this case,  the two 
contributions $\mathcal{A}_S$ and $\mathcal{A}_N$ in Eq.~\ref{eq:mul-0nu2beta} and Eq.~\ref{eq:one-0nu2beta} 
are  related by the following inequality, $$\mathcal{A}_N \ll \mathcal{A}_S.$$  
\end{itemize}

In the next subsection, we present our analysis in detail. Like for the Type I seesaw,  
we   analyze the constraints
on the Dirac mass matrix $M_D$ and the sterile neutrino mass $m_s$,  keeping the mass $M_R$ of 
the sterile neutrino $N_m$  fixed.

\subsection{Constraining $m-m_s$ parameter plane \label{cons-ex}}

\begin{figure}[t]
\begin{center}
\includegraphics[width=0.5\textwidth,angle=0]{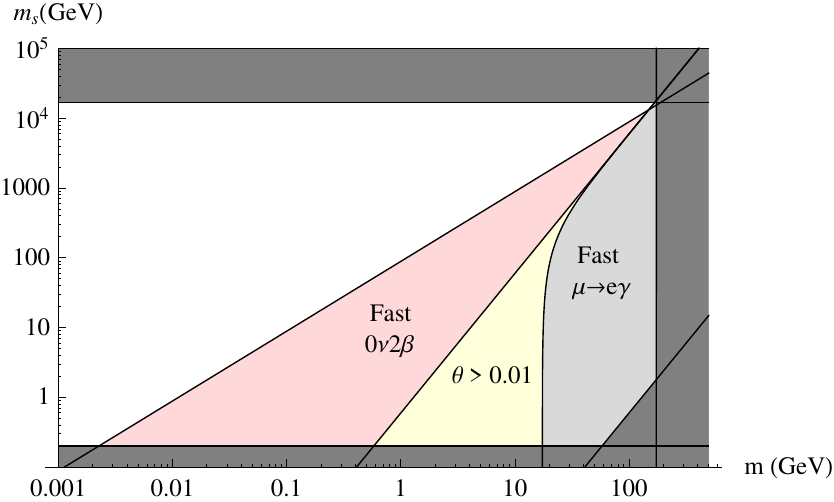}
\end{center}
\caption{\label{fig:extend}
Constrains on $m-m_s$ parameter plane from $0\nu 2\beta$, heavy Majorana neutrino searches at LHC and the lepton flavor violation.
 $m$  and $m_s$ are in GeV. In this figure $M_R$ has been set to $M_R=1.69 \times 10^{4}$ GeV. 
}
\end{figure}

In this section we discuss the different constraints on the lightest sterile neutrino mass $m_s$ and  
Dirac mixing  $m$ coming from $0\nu 2\beta$ process,  and we provide a naive estimation on heavy Majorana 
neutrino searches at LHC as well as  searches for lepton flavor violating processes. Our discussion 
relies on the assumption that $\mu$ is smaller than the light sterile neutrino mass $m_s \sim \frac{M^2_S}{M_R}$.
For simplicity, we adopt the following considerations (compare with Sect.~\ref{nv})
\begin{itemize}
\item
The scale of $M_D$ is referred  as $m$, while the scale of $M_S$ is  denoted  as $M$. Among the two scales, 
the scale of  $M_D$ is  bounded from perturbativity, i.e., $m<174$ GeV; while  being the mixing between two 
sterile neutrino states, the scale of  $M_S$  can take larger  value.  
\item
We denote  the mass scale of the sterile neutrino states $N_m$ by $M_R$,  while  the mass scale of the sterile
 neutrino states $S_m$ is fixed by $m_s$.\footnote{ For more than one generation, the masses $diag(M_{S_i})$ may 
have significant hierarchy, for which further precise definition of $m_s$ as a mass-scale is required. Instead, 
in this example, we consider the masses $diag(M_{S_i})$ are not strongly hierarchical, and hence, can be 
represented  by $m_s$.} This notation perfectly fits  the following scenarios, 
\begin{itemize} 
\item
when  the sterile neutrino masses are not very hierarchical.
\item
when the model is extended by only  two sterile neutrino states $S_L$ and $N_L$. 
\end{itemize}
\end{itemize}
Fig.~\ref{fig:extend} illustrates the naive constraints on  the Dirac mixing $m$ and the sterile 
neutrino mass $m_s$, for  Extended seesaw scenario. The bound on $m_s$ should be interpreted as the 
bound on the absolute value of $m_s$. We present our result for the   sample case, where  the heavy 
Majorana neutrino mass $M_R=1.69 \times 10^4$ GeV. The details of the figure are as follows,
\begin{itemize}
\item
The three grey bands are disallowed by the following considerations,
\begin{itemize}
\item
The upper grey band is disallowed, since it violates  the Extended seesaw condition $M_S<M_R$. In other words,  in 
this region  the mass of the sterile neutrino states $S_m$ and $N_m$ satisfy the following inequality 
relation $m_s> M_R$, which is not permitted  by the Extended seesaw criterion. 
\item
 The disallowed lower grey band corresponds to $m_s< 200$ MeV.
\item
The side grey band is disallowed from perturbativity bound on the mixing between the standard model 
neutrino $\nu_L$ and  the sterile neutrino state $N_L$, i.e., $m<174$ GeV.
\end{itemize}
\item
The oblique line separating the white and pink region corresponds to the saturating $0\nu 2\beta$ contribution from the sterile
neutrino states $S_m$.  The amplitude of the sterile neutrino states $S_m$ across this oblique line satisfies the relation, 
$$ \left(\frac{m}{M}\right)^2\frac{1}{m_s}=\frac{1}{\sqrt{T_{1/2} K_{0\nu}}},$$  where we have considered the 
half-life of germanium 76,  
$T_{1/2}=1.9 \times 10^{25}$ yr \cite{epj}. As already has been discussed in Sect.~\ref{nme}, the factor  $K_{0\nu}=G_{0\nu}(\mathcal{M}_N m_p)^2=9.2 \times 10^{-10} \,\rm{GeV}^2/\rm{yr}$, using the nuclear matrix element of \cite{f2010}. 
Expressing the amplitude in terms of $M_R$, $m_s$ and the Dirac mixing $m$, the sterile neutrino contribution 
across the oblique line satisfies $\frac{m^2}{M_R} \frac{1}{m^2_s}=7.6 \times 10^{-9}$ $\rm {GeV}^{-1}$. The area 
below this oblique line is disallowed, since the sterile neutrino contribution in this region is larger  than 
the above mentioned saturating contribution. The area above this oblique line is however allowed 
from $0\nu 2\beta$ consideration and the sterile neutrino contribution in this region is smaller than the saturating value. 

\item
The oblique line separating the pink and yellow region corresponds to the active-sterile neutrino mixing 
(mixing between $\nu_L$ and  $S_m$ state)  $\theta \sim \frac{m}{M}=0.01$. In terms of $m_s$, $M_R$ and 
the Dirac mixing $m$, this condition can be written as $m_s=\frac{m^2}{M_R} 10^4$.  The area below this 
oblique line corresponds to   the large mixing  $\theta >0.01$ and is the 
favorable region for heavy sterile  neutrino searches at LHC \cite{aguila-LHC-d,aguila-LHC-m,smirnov}. Note that, 
the region is further subdivided into two subregions, where the light gray region is ruled out by the lepton 
flavor violating processes $\mu \to e \gamma$.\footnote{For one generation standard model neutrino $\nu_L$ and 
one generation of extra sterile states $S_L$ and $N_L$, this lepton flavor bound will be absent.} Also  
note that, for this simplified scenario, the $0\nu2\beta$ puts severe constraint on the heavy Majorana neutrino searches at LHC.

\item
The rightmost light gray region, which is disallowed from lepton flavor violating constraint, and 
as well as from $0\nu 2\beta$ consideration, is further subdivided by a oblique line. For  our
 choice $M_R=1.69\times  10^4$ GeV, the smaller  gray region under right most oblique line  violates 
the Extended seesaw condition $M>m$,  i.e., in other words   $m_s > \frac{m^2}{M_R}$, hence also disallowed 
from Extended seesaw criterion. 
\end{itemize}

In our previous discussions  about  the different constraints, we have considered the mass of the sterile 
neutrino state $m_s \sim \frac{M^2}{M_R}$, and the active-sterile mixing angle as $\theta \sim \frac{m}{M}$. 
As, $\mu$ is smaller than $m_s \sim \frac{M^2}{M_R}$, our consideration is perfectly justified. However, for 
completeness,   we discuss   the possible corrections to the previously discussed bounds, which appear because 
of the small lepton number violating parameter $\mu$. Due to the presence of $\mu$, the physical mass $m_s$  of 
the sterile state $S_m$  is changed to $m_s \sim \mu-\frac{M^2}{M_R}$ (see Appendix \ref{diagonalization-extended}), 
where $\mu< \frac{M^2}{M_R}$. Also note that,  the correction to the  active-sterile mixing 
angle ($\nu_L-S_m$ mixing angle   $\frac{m}{M}$) due to a non-zero small  $\mu$ is 
 of $ \mathcal{O}(\frac{m}{M} \frac{\mu}{M^2/M_R})$. As  a result, the bound from $0\nu 2\beta$, heavy
 Majorana neutrino searches at LHC, as well as the theoretical constraints from Extended seesaw  will 
suffer some corrections, which are proportional to the smallness of $\mu$. The different corrections are as follows, 

\begin{itemize}
\item
The correction to the contribution of the sterile state $S_m$ in  $0\nu 2\beta$ process 
is $\delta \mathcal{A}_S \sim \mathcal{O}{(\frac{ m^2 M^2_R \mu}{M^6})}$, 
where the mass of the state $S_m$ is $m_s \sim \mu-\frac{M^2}{M_R}$. In terms of the physical
 mass $m_s$, the correction is $\delta \mathcal{A}_S \sim \frac{m^2}{M_R m_s^2} \frac{\mu}{m_s}$, hence 
suppressed than  the leading order contribution $\frac{m^2}{M_R m^2_s}$ by a factor $\mathcal{O}({\frac{\mu}{m_s}})$, 
which is very small due to $\mu<m_s$. To give an estimation, for $\mu=10^{-3}$ GeV, $m_s=0.1 $ GeV, the 
factor $\frac{\mu}{m_s}=10^{-4}$. Hence, for all practical purpose, is negligible.

\item
Considering the physical mass of the $S_m$ state as $m_s \sim \mu-\frac{M^2}{M_R}$, the Extended seesaw conditions 
$M< M_R$ and $m< M$ is now modified to $-m_s<M_R-\mu$ and $-m_s>\frac{m^2}{M_R}-\mu$ respectively. 

\item
Similarly,  the bound coming from heavy Majorana searches at LHC $\frac{m}{M}> 10^{-2}$,   will be corrected 
by a very small factor  
$\mathcal{O}{(\frac{m^2}{M_R m_s}\frac{\mu}{m_s})}$. 

\end{itemize}

The  discussions of the previous as well as this  section 
clearly shows that  the leading order contribution coming from the  sterile neutrino 
state $S_m$ is independent of the small lepton number 
violating parameter. It is also clearly  evident from the above discussion that the possible correction 
to $0\nu 2\beta$ amplitude due to the  lepton number violating parameter $\mu$ will be extremely small. 
 Hence, one can practically consider the $0\nu 2\beta$ bound as  independent of this parameter. 
However, as pointed out before  in Sect.~\ref{mass-mix} and stressed in the subsequent sections,
 the light neutrino masses  strongly depend on this  parameter. 
Changing  $\mu$ to a relatively larger value, the $0\nu 2\beta$ allowed region  can further be restricted  from 
small neutrino mass constraint. 

\begin{figure}[t]
\begin{center}
\includegraphics[width=0.5\textwidth,angle=0]{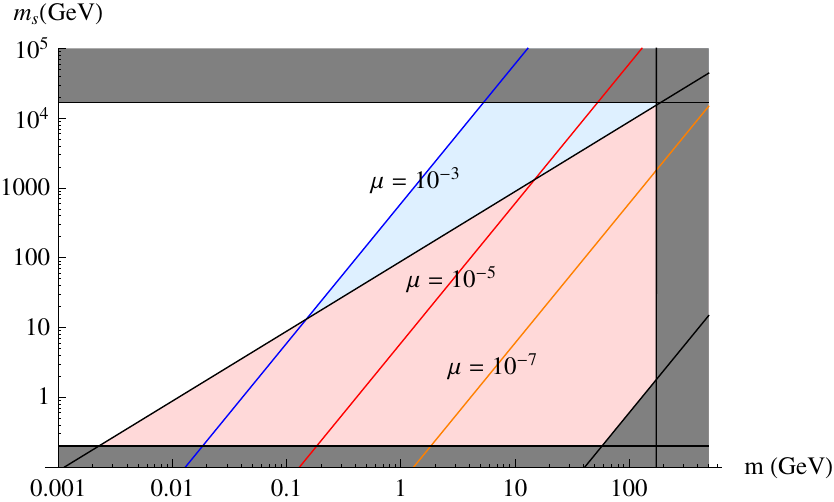}
\end{center}
\caption{\label{fig:comp}
Constrains on $m-m_s$ parameter plane from $0\nu 2\beta$ and neutrino mass. The blue, red and orange oblique line correspond to 
$\mu=10^{-3}, 10^{-5}, 10^{-7}$ GeV. The value of $M_R$ has been set to $M_R=1.69 \times 10^{4}$ GeV.  
}
\end{figure}

We present the possible comparison between the $0\nu 2\beta$ bound  and neutrino mass constraint in Fig.~\ref{fig:comp}. 
For a very small $\mu$ (as shown  in Fig.~\ref{fig:comp}), the grey region   which are excluded from the theoretical 
constraints  $M<M_R$ and $m<M$ remains almost  unchanged as compared to Fig.~\ref{fig:extend}. The pink region is 
disallowed  from $0\nu 2\beta$ consideration \cite{epj}.  Across the oblique black line separating the pink and white region, the contribution of the sterile neutrino state $S_m$ in $0\nu 2\beta$ process  saturates the upper bound of \cite{epj}, i.e.,
$ \frac{m^2}{M_R}\frac{1}{m^2_s}=7.6 \times 10^{-9} \, \rm{GeV}^{-1}$. 

On the other hand, the standard model light neutrino mass is $m_{\nu}=(\frac{m}{M})^2 \mu$; the 
dependence on $\mu$ is clearly evident.  Expressing $m_{\nu}$ in terms of the physical masses $m_s$, $M_R$ 
and the Dirac mixing $m$, the eV neutrino mass constraint can be expressed,
$ \frac{m^2}{m_s M_R} \mu =0.1 \rm{eV}$. The oblique blue line (Fig.~\ref{fig:comp})  separating the blue and 
white region corresponds to $\mu=10^{-3}$ GeV,  and represents the above neutrino mass constraint. The area below 
this line violates $m_{\nu}<0.1$ eV and is strongly disallowed from neutrino mass constraint. It is evident from the 
figure, the blue region which is allowed by the $0\nu 2\beta$ consideration is  ruled out by the  neutrino mass 
constraint. However, considering smaller values of $\mu$, the neutrino mass  constraint on the $m-m_s$ plane 
can be comparatively relaxed. We have given illustrative example for two other  $\mu$ values, the red oblique 
line corresponds to $\mu=10^{-5}$ GeV  and the orange oblique line corresponds to $\mu=10^{-7}$ GeV. For each of 
the oblique lines, the area below the line is disallowed from the neutrino mass constraint. As it is clearly 
evident,  for the smallest of these three values, {\it i.e.,} for   $\mu=10^{-7}$ GeV, the neutrino mass constraint  
does not restrict the $m-m_s$ parameter space any further than the $0\nu 2\beta$ consideration. Hence, we will 
conclude that the possible {\it additional} restriction on $m-m_s$ parameter space   coming from small neutrino 
mass can be evaded with the choice of smaller $\mu$, whereas  the bound coming from lepton number violating 
$0\nu2\beta$ is process possibly the most stringent one.

\label{sec:extended}

\section{Summary and discussion \label{sum}}
Undoubtedly, the study of neutrinoless double beta decay is one of the main available probes of
the lepton number violation. The existing results of the Heidelberg-Moscow collaboration \cite{epj},
 as well as results
from the other experiments, e.g., Cuoricino \cite{cuoricino}, IGEX \cite{igex}, Nemo \cite{nemo}  provide lower  bound on   
the half-life of this process. In addition, there is already an existing experimental hint on $0\nu 2\beta$ 
obtained by Klapdor and collaborators \cite{klapdor}, which  however, according to \cite{fogli-05},  violates 
the bound obtained from cosmology. 
The currently running experiment 
Gerda \cite{Gerda}, as well as the the different future  experiments 
Cuore \cite{Cuore}, 
EXO \cite{exo}, 
SuperNEMO \cite{supernemo},  
Majorana \cite{Majorana0}, 
Lucifer \cite{lucifer}, 
SNO+ \cite{snop},
KamLAND-Zen \cite{klz},
Cobra \cite{cobra}
and
NEXT \cite{next} 
will provide much 
more useful information regarding this process. 

The simplest extension of standard model
includes the heavy sterile neutrino states,  which are  responsible of
generating light neutrino masses via Type I seesaw mechanism. While the
naive expectations  would attribute the violation of lepton number in the 
$0\nu 2\beta$ process totally to the light Majorana neutrino states, with
heavy sterile neutrinos playing a subdominant role, there is however  the
possibility to achieve  the opposite extreme within the minimal Type-I
seesaw  model only.
In this work, this question of how to achieve this possibility has been
analyzed in detail.

The existing  bounds  on the
active-sterile mixing angle have been re-evaluated, and the leading role of $0\nu
2\beta$ reassessed. Due to the improved result of the nuclear matrix
elements given in \cite{f2010}, the bounds on active-sterile mixing coming
from  $0\nu 2\beta$ is very stringent: the previous constraint  on the
mixing angle \cite{f2005} has become one order of magnitude tighter. On
the face of this new result,  the bounds from various meson decay as well as other
experiments \cite{atre} has become relatively less important in all
parameter space, and almost entirely superseded by the $0\nu2\beta$
bound--see Fig.~\ref{plut}.

The question of having a dominant  heavy sterile neutrino
contribution  in $0\nu 2\beta$ process in the minimal Type I seesaw model
(i.e., with three heavy neutrino states) has been explored  in detail.
In the case of one generation standard model neutrino and  one extra
sterile neutrino state, the seesaw structure of the neutrino mass matrix
automatically guarantees the smallness of the extra sterile neutrino
contribution in the $0\nu 2 \beta$ process. The opposite regime, when the $0\nu 2\beta$ transition is
saturated by the heavy sterile neutrino states,
 can be possibly achieved, for more than one generation case.  To obtain this, the light neutrino
mass has to be  smaller than the naive expectation from the ordinary seesaw
formula. This condition has been implemented beginning
from an exact cancellation and then arranging small perturbation in
several possible ways.
 A classification of the interesting Type I seesaw models emerged, with
light neutrino mass matrices of all types but also with sterile neutrino
states dominating the  $0\nu 2 \beta$ transition. All the cases can be
studied by mean of simple analytical formulae. Moreover, in order to
obtain a viable scenario where the light neutrino masses do not suffer
from radiative instabilities, the perturbations as well as the sterile
neutrinos have to obey additional constraints, and in particular, their
masses have to be approximatively lighter than 10 GeV. Several explicit
examples illustrate how the sterile neutrino and light neutrino
contribution in $0\nu 2\beta$ process can possibly  decouple, and the
sterile neutrino contribution become dominant, for the  two flavor and
the three flavor scenarios.  The analytical results have been verified
numerically. The dominant sterile neutrino contribution in $0 \nu 2\beta$ process provide a
way to overcome
 the conflict between cosmology and the 
 experimental hint obtained by Klapdor and collaborators \cite{klapdor}, or more in general to
have a relatively fast $0\nu 2\beta$ transition, even  with very small neutrino masses.

Similar investigations have been carried in more complex seesaw scenarios,
and in particular, in the Extended seesaw \cite{Kang-Kim, Parida}.  In
this model, one has  additional   heavy neutrino states. The lepton number
violation is introduced by two main hierarchical mass scales. The light neutrino masses
depend  on the  small lepton number violating scale, and therefore, the sub-eV neutrino
masses can be explained by the small lepton number violating scale of the
theory.  Since the active-sterile neutrino mixing in leading order is 
independent of this small lepton number violating scale, the
standard model neutrino contribution and the extra sterile state
contribution are completely decoupled in this seesaw scenario. 
As a result, the sterile neutrino contribution can saturate the
present experimental bound on $0\nu 2\beta$ transition \cite{epj},
while the light neutrino contribution can even be subdominant. 
Possible issues, such as, the higher-dimensional corrections to the active-sterile
mixing angle due to the small lepton number violating scale, as well as 
to the $0\nu 2\beta$ transition amplitude have been discussed in some
detail. The details of higher-dimensional correction to the mass and
mixing matrix have been evaluated in Appendix \ref{diagonalization-extended}.

 These results have direct implications for the phenomenology of the
 (minimal and extended) seesaw models. In particular, they are potentially
 relevant for collider physics and rare transitions (such as $\mu\to
e\gamma$) though the exploratory  investigations here presented suggest only a marginal
impact; but more promisingly, these models have interesting implications for
 meson decays, neutrino-decay searches  and cosmology. A systematic study of these issues will be matter of a future work.

\vglue 0.8cm
\noindent
{\Large{\bf Acknowledgments}}\vglue 0.3cm
\noindent
The authors would like to thank Sandhya Choubey, Srubabati Goswami for reading the manuscript, and
Anushree Ghosh for computational help. This work has been partially supported with the INFN-ICTP exchange program and by Centro Fisica Astroparticellare (CFA) supported by POR Abruzzo.


\begin{appendix}

\section{A special type of Majorana mass matrices\label{ma}}

Here 
a special type of Majorana mass matrix is analyzed.
This type of matrix occurs repeatedly in the present study: as sub-block of the $6\times 6$ mass matrix for    
the non-trivial solution of the equation $M_D^T M_R^{-1} M_D=0$, see
Eq.~\ref{ciclone}; in the discussion of $M_R$ and $M_\nu$, see e.g., Eq.~\ref{petreus}; 
in the simplest version of Extended seesaw mass matrix,  the one involving just 3 states
$\nu_L,S_L,N_L$. 
This matrix is a $3\times 3$ lower triangular matrix, say
\be
M=
\left(
\begin{array}{ccc}
0 & 0 & \alpha \\
0 & \delta & \beta \\
\alpha & \beta & \gamma
\end{array}
\right)
\label{param}
\ee 
since this matrix will be considered as a Majorana mass matrix,  the phases of the neutrino 
fields (i.e., the basis in the complex space) can be conveniently chosen in order to make the
 elements $\alpha,\beta,\delta$ real non-negative. 
Assuming the condition of reality, $M_{ij}=M_{ij}^*$ (i.e., also $\gamma$ is real for the same 
choice of phases) the  matrix can be diagonalized simply by an orthogonal change of basis:
$O^T M O=\mbox{diag}(m_i)$. 
Here $m_i$ are the eigenvalues, that are real but not necessarily positive, and 
that can be arranged in an increasing order
\be
|m_1| \le |m_2| \le |m_3|  
\ee

The elements $\alpha,\beta,\gamma$ can be expressed in terms of the eigenvalues
by considering the characteristic polynomial, $p(x)=\mbox{det}(M- x I)=\prod_{i=1}^3 (m_i-x)$ and 
identifying the terms of the same order in $x$
\be
\left\{
\begin{array}{lcl}
\alpha & = & \sqrt{\displaystyle -\frac{m_1 m_2 m_3}{m_0}} \\[2ex]
\beta & = & \sqrt{\displaystyle \frac{(m_1-m_0) (m_2-m_0)(m_3-m_0)}{m_0}} \\[2ex]
\gamma & = & m_1 +m_2+m_3 - m_0
\end{array}
\right.
\ee 
where, of course, $\delta=m_0$ acts as a free parameter
(we changed the notation just to emphasize that it has the dimension of a mass).
The condition that these parameters are real require 
that one of the following mutually exclusive conditions holds true: 
\be
\begin{array}{l}\label{cnd}
m_1 \le 0\le m_2 \le m_0 \le m_3   \mbox{ or}\\[1ex]
m_2 \le 0\le m_1 \le m_0 \le m_3   \mbox{ or}\\[1ex]
m_3 \le 0\le m_1 \le m_0 \le m_2   
\end{array}\label{quine}
\ee
Note incidentally that the conditions of Eqs.~\ref{cnd} imply that in the limit $m_0\to 0$ 
also one eigenvalue is forced to  go to zero, thus this limit does not need to be singular.

A last interesting result is the simple expression for  the 
normalized eigenvectors, namely the vectors satisfying $M e_i=m_i e_i$, that are also 
the columns of orthogonal matrix $O$ that diagonalizes $M$. These are given by
\be
e_i=
\frac{1}{N_i}
\left(
\begin{array}{c}
\displaystyle \frac{\alpha}{m_i} \\[2ex]
\displaystyle \frac{\beta}{m_i-m_0} \\[2ex]
1
\end{array}
\right)
\mbox{ with }
N_i=\sqrt{
\frac{
(m_i-m_{j}) (m_i-m_k)
}{
m_i (m_i-m_0)
}
}
\ee
where $\{i,j,k\}=\{1,2,3\}$.
It is easy to show that any of the condition of Eqs.~\ref{cnd} implies that $N_i$ is real.

%

\section{Derivation of Eq.~(\protect{\ref{eq:mix}})\label{diagonalization-extended}}

In this appendix, we provide the details of the block diagonalization of 
Extended seesaw mass matrix and evaluate the mixing matrix $\mathcal{U}_1$. The neutral fermion mass matrix  
(Eq.~\ref{eq:extended}) is,
\be
M_n=\pmatrix {0 & 0 & M^T_D \cr 0 & \mu  & M^T_S \cr M_D & M_S & M_R}.
\label{eq:cascadean}
\ee
For simplicity we consider the Majorana mass matrix  $M_R$ to be  real. Furthermore,  being 
a Majorana mass matrix, $\mu$ is   complex symmetric.   We assume $M_R>M_S>M_D \gg \mu$ and 
also $\mu< M^T_SM^{-1}_RM_S=\mathcal{O}(\frac{M^2_S}{M_R})$. The block diagonalizing 
matrix $\mathcal{U}_1$ is  $(n+2m) \times (n+2m)$-dimensional and satisfies  $
\mathcal{U}_1^T M_n \mathcal{U}_1=M_{bd}$. The matrix $M_{bd}$ has been written in Eq.~\ref{eq:md}. To 
evaluate $\mathcal{U}_1$, we further decompose $\mathcal{U}_1$ as $\mathcal{U}_1=\mathcal{U}^{\prime}_1\mathcal{U}^{\prime \prime }_1$,
where $\mathcal{U}^{\prime}_{1}$  and $\mathcal{U}^{\prime \prime}_1$ satisfy   ${\mathcal{U}^{\prime}_1}^T M_n  \mathcal{U}^{\prime}_1=\hat{M}_{bd} $ and ${\mathcal{U}^{\prime \prime}_1}^T \hat{M}_{bd}  \mathcal{U}^{\prime \prime}_1=M_{bd}$, $\hat{M}_{bd}$ is the intermediate 
block-diagonal matrix. We follow the parameterization of \cite{Grimus-Lavoura}, i.e., 
\be
\mathcal{U}^{\prime}_1=\pmatrix {\sqrt{1-B B^{\dagger}} & B \cr -B^{\dagger} & \sqrt{1-B^{\dagger}B}}
,\ee 
where $ B=\Sigma_j B_j$  and $B_j$ is $B_j \propto \frac{1}{M^j_R}$.
Up to 2nd order in $M_R^{-1}$, the mixing matrix $\mathcal{U}^{\prime}_1$ has the following form,
\be
\mathcal{U}^{\prime}_1 \sim \pmatrix{ 1-\frac{1}{2} M^{\dagger}_D M^{-2}_R M_D & -\frac{1}{2} M^{\dagger}_D M^{-2}_R M_S & M^{\dagger}_D M^{-1}_R \cr 
-\frac{1}{2} M^{\dagger}_S M^{-2}_R M_D & 1-\frac{1}{2} M^{\dagger}_S M^{-2}_R M_S & M^{\dagger}_S M^{-1}_R+ \mu^* M^{T}_S M^{-2}_R \cr 
-M^{-1}_R M_D & -(M^{-1}_RM_S+M^{-2}_RM_S^* \mu ) & 1-\frac{1}{2} M^{-1}_R (M_D M^{\dagger}_D+M_S M^{\dagger}_S) M^{-1}_R }. 
\ee
The intermediate block diagonalized matrix $\hat{M}_{bd}$ is (up to order ${ M^{-2}_R}$),
\be
\hat{M}_{bd}  \sim \pmatrix{-M^T_DM^{-1}_RM_D & -M^{T}_DM^{-1}_R M_S-\frac{1}{2} M^T_D M^{-2}_RM_S^* \mu & 0 \cr 
-M^T_S M^{-1}_RM_D-\frac{1}{2} \mu  M^{\dagger}_S M^{-2}_R M_D & \mu -M^T_S M^{-1}_R M_S-\frac{1}{2} (M^T_S M^{-2}_R M_S^* \mu +\mu M^{\dagger}_S M^{-2}_R M_S) & 0 \cr 
0 & 0 & M^{\prime}_R}, 
\ee
where up to 2nd order in $M^{-1}_R$, 
\be 
M^{\prime}_R=M_R+[(M_DM^{\dagger}_D+ M_S M^{\dagger}_S)M^{-1}_R+ \frac{1}{2} M_S \mu^* M^T_S M^{-2}_R+Trans.].
\label{expmr}
\ee
Note that, further consideration of $M^{-3}_R$ terms in $\mathcal{U}^{\prime}_1$ will open up the 
relative comparison between  the different terms of 
$\mathcal{O}(\frac{M_D}{M_R})^2$, $\mathcal{O}(\frac{M_S}{M_R})^2$ and $\mathcal{O}(\frac{\mu}{M_R})$. As the 
leading order terms in $\mathcal{U}_1$ is unaffected by the inclusion of sub-leading terms, therefore  we do 
not discuss the detail  dynamics of sub-leading terms here. 

The intermediate block diagonal matrix $\hat{M}_{bd} $ can further be  diagonalized by the 2nd mixing 
matrix $\mathcal{U}^{\prime \prime}_1$. As evident from the above, the leading order terms in $\hat{M}_{bd} $ is,
\be
\hat{M}_{bd}  \sim \pmatrix{-M^T_DM^{-1}_RM_D & -M^{T}_DM^{-1}_R M_S & 0 \cr 
-M^T_S M^{-1}_RM_D & \mu-M^T_S M^{-1}_R M_S  & 0 \cr 
0 & 0 & M_R}.
\ee
For $M_R>M_S>M_D \gg \mu$, we have $M^T_S M^{-1}_R M_S>M^{T}_DM^{-1}_R M_S>M^T_DM^{-1}_RM_D$, hence one can again 
apply the seesaw approximation on $\hat{M}_{bd}$. Assuming further  $\mu< M^T_SM^{-1}_RM_S$,  the block diagonal 
matrix $M_{bd}$ has the following form,
\be
M_{bd} \sim \pmatrix { M^T_D (M^T_S)^{-1} \mu M^{-1}_S M_D  & 0 & 0 \cr 0 & -M^T_S M^{-1}_R M_S & 0 \cr 0 & 0 & M_R}.
\ee
To the leading order, the mixing matrix $\mathcal{U}^{\prime \prime}_1$ is,
\be
\mathcal{U}^{\prime \prime}_1 \sim \pmatrix {1-\frac{1}{2}M^{\dagger}_D(M^{-1}_S)^{\dagger}M^{-1}_SM_D & M^{\dagger}_D (M^{-1}_S)^{\dagger} & 0 \cr -M^{-1}_SM_D & 1-\frac{1}{2}M^{-1}_SM_DM^{\dagger}_D(M^{-1}_S)^{\dagger} & 0 \cr 0 & 0 & 1}.
\label{eq:u1p}
\ee
From the expression of $\mathcal{U}^{\prime}_1$ and $\mathcal{U}^{\prime \prime}_1$ and neglecting the relatively 
smaller $\mathcal{O}(\frac{M_D}{M_R})^2$ terms as compared to $\mathcal{O}(\frac{M_S}{M_R})^2$, $\mathcal{O}(\frac{M_D}{M_S})^2$ one obtains  the  following expression of the mixing matrix $\mathcal{U}_1$, given in 
Eq.~\ref{eq:mix},
\be
\mathcal{U}_1 \sim \pmatrix {1-\frac{1}{2}M^{\dagger}_D(M^{-1}_S)^{\dagger}M^{-1}_SM_D & M^{\dagger}_D (M^{-1}_S)^{\dagger}  & M^{\dagger}_D M_R^{-1} \cr  -M^{-1}_S M_D & 1-\frac{1}{2}M^{-1}_SM_DM^{\dagger}_D(M^{-1}_S)^{\dagger}-\frac{1}{2}M^{\dagger}_SM^{-2}_RM_S & M^{\dagger}_S M^{-1}_R \cr (M^T_S)^{-1}\mu M^{-1}_S M_D  & -M^{-1}_R M_S &  1-\frac{1}{2} M^{-1}_R M_S M^{\dagger}_S  M^{-1}_R }. 
\label{u1}
\ee
In the above, the $(\mathcal{U}_1)_{31}$ term is of the order $\mathcal{O}({\frac{M_D}{M_R} \frac{\mu}{M^2_S/M_R}})$. 
To the leading order, the  light and heavy neutrino mass matrices  $m_{\nu}$, and $m_s$, $m_n$ of Eq.~\ref{eq:md}  are respectively the following,
\be
m_{\nu} & \sim &  M^T_D (M^T_S)^{-1}\mu M^{-1}_S M_D, \nn \\
m_s & \sim & -M^T_S M^{-1}_R M_S,  \\ 
m_n & \sim &  M_R. \nn
\ee
Note that, in the above $(\mathcal{U}_1)_{31}$ strongly depends on $\mu$. This can be very easily seen for 
the one generation case, which we discuss in some detail below.

\subsection{Higher Order Consideration \label{high-c}}
We discuss the possible higher order correction to the block diagonalized matrix $\hat{M}_{bd}$ and as well as to 
the mixing matrices $\mathcal{U}^{\prime}_1$, $\mathcal{U}^{\prime \prime}_1$, considering  one generation $\nu_L$, $S_L$ and $N_L$. The higher order terms are important to understand the possible corrections to the mixing matrix $\mathcal{U}_1$ and 
as well as to understand a  nonzero  $(\mathcal{U}_1)_{31}$. It is straightforward to verify the results for multiple 
generation and hence we do not repeat the task anymore.  We first discuss the higher order corrections  for 
the case $\mu=0$ and then simply extend the discussion for $\mu \neq 0$. 

Case. I) When $\mu=0$,  the intermediate block-diagonalized matrix $\hat{M}_{bd}$ has the following form,
\be
\hat{M}_{bd}=\pmatrix{ M_{ll} & M_{lh} \cr M^T_{lh} & M_{hh}},
\ee
where $ M_{ll}$, $M_{lh}$ and $M_{hh}$ are: 
\be
M_{ll} &=&-M^T_DM^{-1}_RM_D +\frac{1}{2} \left ( M^T_D M^{-1}_RM_D M^{\dagger}_DM^{-2}_RM_D+M^T_DM^{-1}_RM_S M^{\dagger}_SM^{-2}_RM_D+ Trans. \right )  \nn \\
M_{hh}&=&-M^T_SM^{-1}_RM_S +\frac{1}{2} \left ( M^T_S M^{-1}_RM_S M^{\dagger}_SM^{-2}_RM_S+M^T_SM^{-1}_RM_D M^{\dagger}_DM^{-2}_RM_S+ Trans. \right)\nn  \\
M_{lh}&=&-M^T_DM^{-1}_RM_S+ \frac{1}{2} \left (M^T_D M^{-1}_RM_D M^{\dagger}_DM^{-2}_RM_S+ M^T_DM^{-1}_RM_S M^{\dagger}_SM^{-2}_RM_D \right )   \nn \\ 
&& + \frac{1}{2} \left ( M^T_D M^{-2}_R M^*_D M^T_DM^{-1}_RM_S+ M^*_D M^{-2}_RM^*_S M^T_S M^{-1}_R M_S \right ), 
\ee
In the above,  we have shown explicitly up to $\mathcal{O}(M^{-3}_R)$. Considering  one generation of $\nu_L$, $S_L$ and $N_L$, 
the intermediate block-diagonal  matrix $\hat{M}_{bd}$ simplifies to, 
\be
\hat{M}_{bd}=-\pmatrix{ \frac{M_D M_D}{M_R} & \frac{M_D M_S}{M_R} \cr \frac{M_DM_S}{M_R} & \frac{M_S M_S}{M_R}} \left ( 1-\frac{M^2_D}{M^2_R}-\frac{M^2_S}{M^2_R} \right).
\ee 
It is clearly evident, as  the determinant  of this matrix vanishes, the light neutrino mass is zero.  
The mixing matrix $\mathcal{U}^{\prime}_1$ in this case has the following simple form (up to $\mathcal{O}(M^{-3}_R)$ ,
\be
\mathcal{U}^{\prime}_{1} =\pmatrix{ 1-\frac{1}{2} \frac{M^2_D}{M^2_R} & -\frac{1}{2} \frac{M_D M_S}{M^2_R} & \frac{M_D}{M_R}\left (1-\frac{3}{2} \frac{M^2_D}{M^2_R}-\frac{3}{2} \frac{M^2_S}{M^2_R} \right ) \cr
-\frac{1}{2}\frac{M_D M_S}{M^2_R} &  1-\frac{1}{2} \frac{M^2_S}{M^2_R} & \frac{M_S}{M_R} \left (1-\frac{3}{2} \frac{M^2_D}{M^2_R}-\frac{3}{2} \frac{M^2_S}{M^2_R} \right ) \cr
-\frac{M_D}{M_R}\left (1-\frac{3}{2} \frac{M^2_D}{M^2_R}-\frac{3}{2} \frac{M^2_S}{M^2_R} \right ) & -\frac{M_S}{M_R} \left (1-\frac{3}{2} \frac{M^2_D}{M^2_R}-\frac{3}{2} \frac{M^2_S}{M^2_R} \right ) & 1-\frac{1}{2} \left ( \frac{M^2_D}{M^2_R} +\frac{M^2_S}{M^2_R} \right)}
\label{bigeq}
\ee
To calculate the other mixing matrix $\mathcal{U}^{\prime \prime}_1$, we again follow \cite{Grimus-Lavoura}. Up to $\mathcal{O}((\frac{M_D}{M_S})^3)$  the parameter 
$B$ is  $ B=\frac{M_D}{M_S}-\frac{1}{2} \frac{M^3_D}{M^3_S}$, where $B_1=\frac{M_D}{M_S}$ and $B_3=-\frac{1}{2}(\frac{M_D}{M_S})^3$.   The mixing matrix $\mathcal{U}^{\prime \prime}_1$ is (up to $\mathcal{O}(\frac{M_D}{M_S})^3$),
\be
\mathcal{U}^{\prime \prime}_1=\pmatrix{ 1-\frac{M^2_D}{2M^2_S} & \frac{M_D}{M_S} \left( 1- \frac{1}{2}\frac{M^2_D}{M^2_S} \right )  & 0 \cr
- \frac{M_D}{M_S} \left( 1- \frac{1}{2}\frac{M^2_D}{M^2_S} \right ) & 1-\frac{M^2_D}{2M^2_S} & 0 \cr 
0 & 0 & 1}.
\label{bigeq2}
\ee 
Given $\mathcal{U}^{\prime}_1$ and $\mathcal{U}^{\prime \prime}_1$, one can straightforwardly calculate 
$\mathcal{U}_1=\mathcal{U}_1^{\prime} \mathcal{U}_1^{\prime \prime}$. Also note that, as discussed in the 
previous section and as well as in section \ref{mass-mix}, due to mutual cancellation between the 
elements of $\mathcal{U}^{\prime}_1$ and $\mathcal{U}^{\prime \prime}_1$, the element $(\mathcal{U}_1)_{31}=0$. With 
a $\mu \neq 0$, it is possible to obtain a non-zero  $(\mathcal{U}_1)_{31}$.

Case.II) 

For non-zero $\mu$, the  intermediate block diagonal  matrix $\hat{M}_{bd}$  changes to the following, where 
we have written up to $M^{-3}_R$. 

\be
\hat{M}_{bd}= \pmatrix{ 0 & 0 \cr 0 & \mu} - \pmatrix{ \frac{M_DM_D}{M_R} & \frac{M_DM_S}{M_R} \cr \frac{M_DM_S}{M_R} & \frac{M_SM_S}{M_R}} \left(1-\frac{M^2_D}{M^2_R}-\frac{M^2_S}{M^2_R} \right)-\pmatrix{ 0 & \frac{M_D M_S}{M_R} \cr \frac{M_D M_S}{M_R} & \frac{2M_S M_S}{M_R}} \left( \frac{\mu}{2M_R}+ \frac {\mu^2}{2M^2_R} \right ).
\ee
The mixing matrix $\mathcal{U}^{\prime}_1$ (up to $\mathcal{O}(M^{-3}_R)$), described in Eq.~\ref{bigeq} now 
changes to $\mathcal{U}^{\prime}_1=\mathcal{U}^{\prime 0}_1+\delta \mathcal{U}^{\prime 0}_1$, where $\mathcal{U}^{\prime 0}_1$ is the same  as  $\mathcal{U}^{\prime}_1$ of  Eq.~\ref{bigeq}, while  $\delta \mathcal{U}^{\prime 0}_1$ is the following, 
\be
\delta \mathcal{U}^{\prime 0}_1=\pmatrix {0 & -\frac{1}{2} \frac{M_D \mu M_S}{M^3_R} & 0 \cr -\frac{1}{2}\frac{M_D \mu M_S}{M^3_R} & -\frac{M^2_S \mu}{M^3_R} & \frac{\mu M_S}{M^2_R}+\frac{\mu^2M_S}{M^3_R} \cr 0 & -\frac{ \mu M_S}{M^2_R}-\frac{\mu^2M_S}{M^3_R} & -\frac{\mu M^2_S}{M^3_R}}
\ee

The other mixing matrix  mixing matrix ${\mathcal{U}}^{\prime \prime}_1$ can  be evaluated again following the 
parameterization \cite{Grimus-Lavoura}, 
\be
{\mathcal{U}}^{\prime  \prime}_1 =\pmatrix {\sqrt{1-BB^{\dagger} } & B \cr -B^{\dagger} & \sqrt{1-B^{\dagger}B} }
\ee
where $B_j=\mathcal{O}((\frac{M_D}{M_S})^j)$. We conclude the section with the following few remarks, 
\begin{itemize}
\item
For $\mu< \frac{M^2_S}{M_R}$, the light neutrino mass would be $m_{\nu} \sim \frac{M_D}{M_S} \mu \frac{M_D}{M_S}$. 
\item
  For $\mu \neq 0$,  $\mu <  \frac{M^2_S}{M_R}$, the expansion parameters $B_1$ and $B_3$ changes 
by   $\delta B_1 \propto \frac{M_D}{M_S} \frac{\mu}{M^2_S/M_R}$ and $\delta B_3 \propto -\frac{5}{2} \frac{M^3_D}{M^3_S}  \frac{\mu}{M^2_S/M_R}$, where we have only shown the {\it dominant  sub-leading} correction in $\mu$. For $\mu \neq 0$, 
one will obtain the leading order contribution in $(\mathcal{U}_1)_{31} \sim \mathcal{O}(\frac{M_D}{M_R} \frac{\mu} {M^2_S/M_R})$. 
Also, note that the dependency of the active-sterile mixing $(\mathcal{U}_1)_{12}$ on the small lepton number violating parameter $\mu$ is as follows $(\mathcal{U}_1)_{12} \sim \frac{M_D}{M_S} \frac{\mu}{M^2_S/M_R}$. 
\item
Considering leading order  $B \sim \frac{M_D}{M_S}$, the mixing matrix $\mathcal{U}^{\prime \prime}_1$ will have the form given in Eq.~\ref{eq:u1p}.  
\end{itemize}

\end{appendix}


\newpage
{\small
\begin{thebibliography}{99}
\bibitem{epj}
H.~V.~Klapdor-Kleingrothaus, A.~Dietz, L.~Baudis, G.~Heusser, I.~V.~Krivosheina, S.~Kolb, B.~Majorovits, H.~Pas {\it et al.},
  Eur.\ Phys.\ J.\  {\bf A12 } (2001)  147-154 
  [hep-ph/0103062].
  %
\bibitem{cuoricino}
 C.~Arnaboldi {\it et al.} [CUORICINO Collaboration],
  Phys.\ Rev.\  {\bf C78}, 035502 (2008) 
  [arXiv:0802.3439 [hep-ex]].
%
\bibitem{igex}
C.~E.~Aalseth {\it et al.} [ IGEX Collaboration ],
  Phys.\ Rev.\  {\bf D65 } (2002)  092007
  [hep-ex/0202026].
%
\bibitem{nemo}
  X.~Sarazin {\it et al.} [NEMO Collaboration]
  [hep-ex/0006031]; J.~Argyriades {\it et al.} [NEMO Collaboration],
  Phys.\ Rev.\  {\bf C80}, 032501 (2009) 
  [arXiv:0810.0248 [hep-ex]].
%

\bibitem{Gerda}
 I.~Abt, M.~F.~Altmann, A.~Bakalyarov, I.~Barabanov, C.~Bauer, E.~Bellotti, S.~T.~Belyaev, L.~B.~Bezrukov {\it et al.},
  [hep-ex/0404039]; S.~Schonert {\it et al.} [GERDA Collaboration],
  Nucl.\ Phys.\ Proc.\ Suppl.\  {\bf 145}, 242-245 (2005).
%
\bibitem{Cuore}
C.~Arnaboldi {\it et al.} [CUORE Collaboration],
 Nucl.\ Instrum.\ Meth.\  {\bf A518}, 775-798 (2004) 
  [hep-ex/0212053].
%
\bibitem{klapdor}
 H.~V.~Klapdor-Kleingrothaus, I.~V.~Krivosheina, A.~Dietz, O.~Chkvorets,
  Phys.\ Lett.\  {\bf B586}, 198-212 (2004) 
  [hep-ph/0404088]; H.~V.~Klapdor-Kleingrothaus, I.~V.~Krivosheina,
  Mod.\ Phys.\ Lett.\  {\bf A21}, 1547-1566 (2006).
%
\bibitem{exo}
E.~Conti {\it et al.} [EXO Collaboration],
  Phys.\ Rev.\  {\bf B68}, 054201 (2003).
  [hep-ex/0303008].
  
\bibitem{supernemo}
  R.~Arnold {\it et al.} [SuperNEMO Collaboration],
  Eur.\ Phys.\ J.\  {\bf C70}, 927-943 (2010)
  [arXiv:1005.1241 [hep-ex]].
%
\bibitem{Majorana0}
  V.~E.~Guiseppe {\it et al.} [Majorana Collaboration]
  [arXiv:0811.2446 [nucl-ex]].
%

\bibitem{lucifer}
F.~Ferroni,
  J.\ Phys.\ Conf.\ Ser.\  {\bf 293}, 012005 (2011)
  and J.W. Beeman et al.,   arXiv:1106.6286.

\bibitem{snop}
C.~Kraus [SNO+ Collaboration],
http://www.sno.phy.queensu.ca/$\sim$alex/SNOLab.pdf and 
  Prog.\ Part.\ Nucl.\ Phys.\  {\bf 57 } (2006)  150.


\bibitem{klz}
See talk of K. Inoue at Neutrino Telescope 2011, 
{\tt http://neutrino.pd.infn.it/Neutel2011/Program.html}

\bibitem{cobra}
K.~Zuber,
  Phys.\ Lett.\  {\bf B519 } (2001)  1
  [nucl-ex/0105018];
T.~Bloxham {\it et al.} [ COBRA Collaboration ],
  Phys.\ Rev.\  {\bf C76 } (2007)  025501
  [arXiv:0707.2756 [nucl-ex]].

\bibitem{next}
F.~Granena {\it et al.} [The NEXT Collaboration],
  [arXiv:0907.4054 [hep-ex]].
  

\bibitem{solar}
B.~T.~Cleveland {\it et al.},
Astrophys.\ J.\  {\bf 496}, 505 (1998);
%
J.~N.~Abdurashitov {\it et al.}  [SAGE Collaboration],
Zh.\ Eksp.\ Teor.\ Fiz.\  {\bf 122}, 211 (2002)
[J.\ Exp.\ Theor.\ Phys.\  {\bf 95}, 181 (2002)];
%
W.~Hampel {\it et al.}  [GALLEX Collaboration],
Phys.\ Lett.\ B {\bf 447}, 127 (1999); 
%
S.~Fukuda {\it et al.}  [Super-Kamiokande Collaboration],
Phys.\ Lett.\ B {\bf 539}, 179 (2002);
%
B.~Aharmim {\it et al.}  [SNO Collaboration],
Phys.\ Rev.\ C {\bf 72}, 055502 (2005);
%
B.~Collaboration,
  arXiv:0708.2251 [astro-ph].
%
B.~Aharmim {\it et al.} [ SNO Collaboration ],
  Phys.\ Rev.\  {\bf C81}, 055504 (2010) 
  [arXiv:0910.2984 [nucl-ex]].
\bibitem{kl}
K.~Eguchi {\it et al.}, 
  [KamLAND Collaboration],
Phys.Rev.Lett.{\bf 90} (2003) 021802;
%
T.~Araki {\it et al.}  [KamLAND Collaboration],
Phys.\ Rev.\ Lett.\  {\bf 94}, 081801 (2005);  A.~Gando {\it et al.} [ The KamLAND Collaboration ],
  Phys.\ Rev.\  {\bf D83}, 052002 (2011) 
  [arXiv:1009.4771 [hep-ex]].
\bibitem{atm}
 Y.~Fukuda {\it et al.} [ Super-Kamiokande Collaboration ],
  Phys.\ Rev.\ Lett.\  {\bf 81}, 1562-1567 (1998)
  [hep-ex/9807003];
  %
  Y.~Ashie {\it et al.}  [Super-Kamiokande Collaboration],
  Phys.\ Rev.\ D {\bf 71}, 112005 (2005) 
  [arXiv:hep-ex/0501064];
  %
  R.~Wendell {\it et al.} [ Kamiokande Collaboration ],
  Phys.\ Rev.\  {\bf D81}, 092004 (2010)
  [arXiv:1002.3471 [hep-ex]].
\bibitem{macro}
   M.~Ambrosio {\it et al.} [ MACRO Collaboration ],
  Phys.\ Lett.\  {\bf B434}, 451-457 (1998) 
  [hep-ex/9807005]; M.~Ambrosio {\it et al.} [ MACRO Collaboration ],
  Eur.\ Phys.\ J.\  {\bf C36}, 323-339 (2004).
\bibitem{soudan}
 W.~W.~M.~Allison {\it et al.} [ Soudan-2 Collaboration ],
  Phys.\ Lett.\  {\bf B449}, 137-144 (1999)
  [hep-ex/9901024].
\bibitem{k2k}
 E.~Aliu {\it et al.} [ K2K Collaboration ],
  Phys.\ Rev.\ Lett.\  {\bf 94}, 081802 (2005)
  [hep-ex/0411038]; M.~H.~Ahn {\it et al.} [ K2K Collaboration ],
  Phys.\ Rev.\  {\bf D74}, 072003 (2006)
  [hep-ex/0606032].
\bibitem{t2k}
  K.~Abe {\it et al.} [ T2K Collaboration ],
  [arXiv:1106.2822 [hep-ex]].
\bibitem{minos}
 P.~Adamson {\it et al.} [ The MINOS Collaboration ],
  Phys.\ Rev.\ Lett.\  {\bf 106}, 181801 (2011)
  [arXiv:1103.0340 [hep-ex]].
\bibitem{chooz}
M.~Apollonio {\it et al.},
Eur.\ Phys.\ J.\ C {\bf 27}, 331 (2003).
\bibitem{book1}
J.~N.~Bahcall,
  {\em Neutrino Astrophysics,}
  Cambridge Univ. Press (1989) 567p.
%
\bibitem{book2}
R.N.~Mohapatra, P.~B. Pal,
{\em Massive neutrinos in physics and astrophysics,}
World Scientific (1991) 318p.
%
\bibitem{book3}
M.~Fukugita, T.~Yanagida,
{\em Physics of neutrinos and applications to astrophysics,}
Springer (2003) 593p.
%
\bibitem{book4}
C. Giunti, C. W. Kim,
{\em Fundamentals of neutrino physics and astrophysics,}
Oxford University Press (2007) 710p.
%
\bibitem{bookm}
{\em Measurements of Neutrino 
Mass}, Enrico Fermi School, Vol. CLXX, ed. C. Brofferio, 
F. Ferroni and F. Vissani. IOS Press, Amsterdam (2009).
%
\bibitem{bookk}
{\em Seventy 
years of double beta decay: From nuclear physics to beyond-standard-model particle physics,}
H.~V.~Klapdor-Kleingrothaus, Hackensack, USA. World Scientific (2010).
  



\bibitem{rev-ponte}
B.~M.~Pontecorvo,
  Sov.\ Phys.\ Usp.\  {\bf 26 } (1983)  1087-1108.
%
\bibitem{review-osc-petcov}
 S.~M.~Bilenky, S.~T.~Petcov,
  Rev.\ Mod.\ Phys.\  {\bf 59}, 671 (1987).
%
\bibitem{mik-shap}
S.~P.~Mikheyev, A.~Y.~Smirnov,
  Prog.\ Part.\ Nucl.\ Phys.\  {\bf 23 } (1989)  41-136.
%
\bibitem{rev-koshiba}
M.~Koshiba,
  Phys.\ Rept.\  {\bf 220 } (1992)  229-381.
%
\bibitem{gelmini-rev}
 G.~Gelmini, E.~Roulet,
  Rept.\ Prog.\ Phys.\  {\bf 58}, 1207-1266 (1995)
  [hep-ph/9412278].
%
\bibitem{kail}
 K.~ Zuber, 
  Phys.\ Rept.\  {\bf 305 } (1998)  295
  [hep-ph/9811267].
%
\bibitem{nir-garcia-rev}
M.~C.~Gonzalez-Garcia, Y.~Nir,
  Rev.\ Mod.\ Phys.\  {\bf 75}, 345-402 (2003)
  [hep-ph/0202058].
%

%
\bibitem{zuber}
 K.~Zuber,
  Acta Phys.\ Polon.\  {\bf B37 } (2006)  1905,
 nucl-ex/0511009.
%
\bibitem{rev-moha-smir}
R.~N.~Mohapatra, A.~Y.~Smirnov,
  Ann.\ Rev.\ Nucl.\ Part.\ Sci.\  {\bf 56 } (2006)  569-628
  [hep-ph/0603118].
%
\bibitem{visreview}
  A.~Strumia and F.~Vissani,
  arXiv:hep-ph/0606054.
%
\bibitem{Senjanovic:2011zz}
  G.~Senjanovi\'c,
  Riv.\ Nuovo Cim.\  {\bf 034}, 1-68 (2011).
%
\bibitem{Fogli2011osc}
  G.~L.~Fogli, E.~Lisi, A.~Marrone, A.~Palazzo, A.~M.~Rotunno,
  [arXiv:1106.6028 [hep-ph]].
\bibitem{limits}
M.~Maltoni, 
T.~Schwetz, M.~A.~Tortola and J.~W.~F.~Valle,
New J.\ Phys.\  {\bf 6}, 122 (2004);
S.~Goswami, A.~Bandyopadhyay, S.~Choubey,
  Nucl.\ Phys.\ Proc.\ Suppl.\  {\bf 143}, 121-128 (2005)
  [hep-ph/0409224];
%
  A.~Bandyopadhyay, 
 S.~Choubey, S.~Goswami, S.~T.~Petcov and D.~P.~Roy,
  Phys.\ Lett.\ B {\bf 608}, 115 (2005);
%
  G.~L.~Fogli, E.~Lisi, A.~Marrone, A.~Palazzo, A.~M.~Rotunno,
  Phys.\ Rev.\ Lett.\  {\bf 101}, 141801 (2008)
  [arXiv:0806.2649 [hep-ph]];
  M.~C.~Gonzalez-Garcia, M.~Maltoni, J.~Salvado,
  JHEP {\bf 1004}, 056 (2010)
  [arXiv:1001.4524 [hep-ph]].


\bibitem{Majorana}
 E.~Majorana,
  Nuovo Cim.\  {\bf 14}, 171-184 (1937).
\bibitem{0nu2beta-old}
 G.~Racah,
  Nuovo Cim.\  {\bf 14}, 322-328 (1937);
  W.~H.~Furry,
  Phys.\ Rev.\  {\bf 56}, 1184-1193 (1939).
\bibitem{dim5}
 S.~Weinberg,
  Phys.\ Rev.\ Lett.\  {\bf 43}, 1566-1570 (1979);
   F.~Wilczek, A.~Zee,
  Phys.\ Rev.\ Lett.\  {\bf 43}, 1571-1573 (1979).
\bibitem{feinberg}
G.~Feinberg, M.~Goldhaber, Proc.\ Nat.\ Ac.\ Sci.\ USA\ {\bf 45}, 
1301  (1959);
  B.~Pontecorvo,
  Phys.\ Lett.\  {\bf B26}, 630-632 (1968).
  %
\bibitem{ms0nu2beta}
 R.~N.~Mohapatra,
 Phys.\ Rev.\  {\bf D34}, 3457-3461 (1986).
%
\bibitem{mwex}
 K.~S.~Babu, R.~N.~Mohapatra,
  Phys.\ Rev.\ Lett.\  {\bf 75}, 2276-2279 (1995)
  [hep-ph/9506354].
%
\bibitem{pion-ex}
J.~D.~Vergados, Phys. Rev. D 25, 914 917 (1982);
  S.~Bergmann, H.~V.~Klapdor-Kleingrothaus, H.~Pas,
  Phys.\ Rev.\  {\bf D62}, 113002 (2000)
  [hep-ph/0004048]; A.~Faessler, T.~Gutsche, S.~Kovalenko, F.~Simkovic,
   Phys.\ Rev.\  {\bf D77}, 113012 (2008)
   [arXiv:0710.3199 [hep-ph]].
%
\bibitem{hirsch}
  M.~Hirsch, H.~V.~Klapdor-Kleingrothaus, S.~G.~Kovalenko,
  Phys.\ Lett.\  {\bf B352}, 1-7 (1995) 
  [hep-ph/9502315]; M.~Hirsch, H.~V.~Klapdor-Kleingrothaus, S.~G.~Kovalenko,
  Phys.\ Rev.\  {\bf D53}, 1329-1348 (1996)
  [hep-ph/9502385];
   M.~Hirsch, H.~V.~Klapdor-Kleingrothaus, S.~G.~Kovalenko,
  Phys.\ Rev.\  {\bf D54}, 4207-4210 (1996)
  [hep-ph/9603213];
  M.~Hirsch, H.~V.~Klapdor-Kleingrothaus, S.~G.~Kovalenko,
  Phys.\ Rev.\  {\bf D57}, 1947-1961 (1998)
  [hep-ph/9707207];
   M.~Hirsch, J.~W.~F.~Valle,
  Nucl.\ Phys.\  {\bf B557}, 60-78 (1999)
  [hep-ph/9812463].
%
\bibitem{allanach}
 B.~C.~Allanach, C.~H.~Kom, H.~Pas,
  Phys.\ Rev.\ Lett.\  {\bf 103}, 091801 (2009)
  [arXiv:0902.4697 [hep-ph]].
\bibitem{vogel}
 V.~Cirigliano, A.~Kurylov, M.~J.~Ramsey-Musolf, P.~Vogel,
  Phys.\ Rev.\  {\bf D70}, 075007 (2004)
  [hep-ph/0404233];
 V.~Cirigliano, A.~Kurylov, M.~J.~Ramsey-Musolf, P.~Vogel,
  Phys.\ Rev.\ Lett.\  {\bf 93}, 231802 (2004)
  [hep-ph/0406199].
\bibitem{choi}
K.~W.~Choi, K.~S.~Jeong, W.~Y.~Song,
  Phys.\ Rev.\  {\bf D66}, 093007 (2002)
  [hep-ph/0207180].
\bibitem{tello}
V.~Tello, M.~Nemev\v sek, F.~Nesti, G.~Senjanovi\'c, F.~Vissani,
  Phys.\ Rev.\ Lett.\  {\bf 106}, 151801 (2011)
  [arXiv:1011.3522 [hep-ph]].
\bibitem{Ibarra}
  A.~Ibarra, E.~Molinaro, S.~T.~Petcov,
  JHEP {\bf 1009}, 108 (2010)
  [arXiv:1007.2378 [hep-ph]];
 A.~Ibarra, E.~Molinaro, S.~T.~Petcov,
  [arXiv:1101.5778 [hep-ph]].
\bibitem{blennow}
  M.~Blennow, E.~Fernandez-Martinez, J.~Lopez-Pavon, J.~Menendez,
  JHEP {\bf 1007 } (2010)  096
  [arXiv:1005.3240 [hep-ph]].
%
\bibitem{seesawM}  
 P.~Minkowski,
  Phys.\ Lett.\  {\bf B67}, 421 (1977).
\bibitem{seesaw-Goran}
 R.~N.~Mohapatra, G.~Senjanovi\'c,
  Phys.\ Rev.\ Lett.\  {\bf 44}, 912 (1980).
\bibitem{seesaw-Yan}
 T.~T.~Yanagida, in {\it Proceedings of the Workshop on the Unified Theory and the Baryon Number in the Universe} (O.Sawada and A.Sugamoto, eds.), KEK, Tsukuba, Japan, 1979, p.95.
\bibitem{seesaw-Ramond}
 M.~Gell-Mann, P.~Ramond, R.~Slansky, Supergravity (P. van Nieuwenhuizen et al.eds), North Holland, Amsterdam, 1980.
\bibitem{f2010}
  F.~\v{S}imkovic, J.~Vergados, A.~Faessler,
  Phys.\ Rev.\  {\bf D82}, 113015 (2010)
  [arXiv:1006.0571 [hep-ph]].  
\bibitem{f2005}
P.~Benes, A.~Faessler, F.~\v{S}imkovic, S.~Kovalenko,
  Phys.\ Rev.\  {\bf D71 } (2005)  077901
  [hep-ph/0501295].
    %
  \bibitem{atre}
   A.~Atre, T.~Han, S.~Pascoli, B.~Zhang,
  JHEP {\bf 0905 } (2009)  030
  [arXiv:0901.3589 [hep-ph]].
   %
\bibitem{mohabnd}
P.~Bamert, C.~P.~Burgess, R.~N.~Mohapatra,
 Nucl.\ Phys.\  {\bf B438}, 3-16 (1995).
 [hep-ph/9408367].
%
\bibitem{Kang-Kim}
 S.~K.~Kang, C.~S.~Kim,
  Phys.\ Lett.\  {\bf B646}, 248-252 (2007)
  [hep-ph/0607072].
%
%
\bibitem{Parida}
S.~K.~Majee, M.~K.~Parida, A.~Raychaudhuri,
  Phys.\ Lett.\  {\bf B668}, 299-302 (2008)
  [arXiv:0807.3959 [hep-ph]]; M.~K.~Parida, A.~Raychaudhuri,
  Phys.\ Rev.\  {\bf D82}, 093017 (2010)
  [arXiv:1007.5085 [hep-ph]].
%
\bibitem{moha}
  R.~N.~Mohapatra, G.~Senjanovi\'c,
  Phys.\ Rev.\  {\bf D23 } (1981)  165.
  %
\bibitem{aguila-LHC-d}
F.~del Aguila, J.~A.~Aguilar-Saavedra,
  Phys.\ Lett.\  {\bf B672}, 158-165 (2009)
  [arXiv:0809.2096 [hep-ph]].
%
\bibitem{aguila-LHC-m}
F.~del Aguila, J.~A.~Aguilar-Saavedra,
  Nucl.\ Phys.\  {\bf B813}, 22-90 (2009)
  [arXiv:0808.2468 [hep-ph]].
%
%
%
\bibitem{smirnov}
  J.~Kersten, A.~Y.~Smirnov,
  Phys.\ Rev.\  {\bf D76}, 073005 (2007)
  [arXiv:0705.3221 [hep-ph]].
%
\bibitem{mutoegamma}
J.~Adam {\it et al.} [ MEG Collaboration ],
  Nucl.\ Phys.\  {\bf B834}, 1-12 (2010)
  [arXiv:0908.2594 [hep-ex]].
%
\bibitem{fogli-05}
 G.~L.~Fogli, E.~Lisi, A.~Marrone, A.~Melchiorri, A.~Palazzo, A.~M.~Rotunno, P.~Serra, J.~Silk {\it et al.},
  Phys.\ Rev.\  {\bf D78}, 033010 (2008)
  [arXiv:0805.2517 [hep-ph]].
\bibitem{mesus}
  D.~Gorbunov, M.~Shaposhnikov,
  JHEP {\bf 0710 } (2007)  015.
  [arXiv:0705.1729 [hep-ph]].


 \bibitem{MYRG}
 A.~Halprin, P.~Minkowski, H.~Primakoff, S.~P.~Rosen,
  Phys.\ Rev.\  {\bf D13}, 2567 (1976).
 %
  
%
  \bibitem{0nu2beta-pet}
 A.~Halprin, S.~T.~Petcov, S.~P.~Rosen,
  Phys.\ Lett.\  {\bf B125}, 335 (1983);
  C.~N.~Leung, S.~T.~Petcov,
  Phys.\ Lett.\  {\bf B145}, 416 (1984).
%
\bibitem{oli-rev}
O.~Cremonesi,
  Nucl.\ Phys.\ Proc.\ Suppl.\  {\bf 118}, 287-296 (2003), 
  Int.\ J.\ Mod.\ Phys.\  {\bf A21}, 1887-1900 (2006) and 
 arXiv:1002.1437 [hep-ex].

\bibitem{vogel-rev} 
 S.~R.~Elliott, P.~Vogel,
  Ann.\ Rev.\ Nucl.\ Part.\ Sci.\  {\bf 52}, 115-151 (2002)
  [hep-ph/0202264];
  P.~Vogel,
  [hep-ph/0611243].
%
\bibitem{bilenky-rev}
  S.~M.~Bilenky,
  Phys.\ Part.\ Nucl.\  {\bf 41}, 690-715 (2010)
  [arXiv:1001.1946 [hep-ph]].
%
\bibitem{werner-rev}
  W.~Rodejohann,
  [arXiv:1106.1334 [hep-ph]];  W.~Rodejohann,
  [arXiv:1011.4942 [hep-ph]].
%
%
\bibitem{vuso}
F.~Vissani,
  JHEP {\bf 9906 } (1999)  022
  [hep-ph/9906525].
%
\bibitem{vis-fer}
  F.~Feruglio, A.~Strumia, F.~Vissani,
  Nucl.\ Phys.\  {\bf B637}, 345-377 (2002)
  [hep-ph/0201291]
%

%
  \bibitem{Wmap7}
E.~Komatsu {\it et al.} [WMAP Collaboration],
  Astrophys.\ J.\ Suppl.\  {\bf 192}, 18 (2011)
  [arXiv:1001.4538 [astro-ph.CO]].
%


\bibitem{stru}
  A.~Strumia, F.~Vissani,
  Nucl.\ Phys.\  {\bf B726 } (2005)  294-316
  [hep-ph/0503246];
  M.~Cirelli, A.~Strumia,
  JCAP {\bf 0612}, 013 (2006)
  [astro-ph/0607086].

%
\bibitem{cosmobound}
 G.~L.~Fogli {\it et al.}, E.~Lisi, A.~Marrone and A.~Palazzo,
  Prog.\ Part.\ Nucl.\ Phys.\  {\bf 57}, 742 (2006)
  [arXiv:hep-ph/0506083];
  M.~C.~Gonzalez-Garcia, M.~Maltoni, J.~Salvado,
  JHEP {\bf 1008}, 117 (2010)
  [arXiv:1006.3795 [hep-ph]].
%

 \bibitem{kovalenko}
  S.~Kovalenko, Z.~Lu, I.~Schmidt,
  Phys.\ Rev.\  {\bf D80 } (2009)  073014
  [arXiv:0907.2533 [hep-ph]].
%
  \bibitem{s2011}
  A.~Faessler, G.~L.~Fogli, E.~Lisi, A.~M.~Rotunno, F.~\v{S}imkovic,
  [arXiv:1103.2504 [hep-ph]].
%
\bibitem{s2011bis}
   A.~Faessler, A.~Meroni, S.~T.~Petcov, F.~\v{S}imkovic, J.~Vergados,
  [arXiv:1103.2434 [hep-ph]].
%
\bibitem{gomez}
J.~J.~Gomez-Cadenas, J.~Martin-Albo, M.~Sorel, P.~Ferrario, F.~Monrabal, J.~Munoz-Vidal, P.~Novella, A.~Poves,
  [arXiv:1010.5112 [hep-ex]].
%
\bibitem{Grimus-Lavoura}
 W.~Grimus, L.~Lavoura,
  JHEP {\bf 0011}, 042 (2000)
  [hep-ph/0008179].
%
  \bibitem{Keung:1983uu}
  W.~-Y.~Keung, G.~Senjanovi\'c,
  Phys.\ Rev.\ Lett.\  {\bf 50 } (1983)  1427. 
  For a review and further references, see e.g. 
  G.~Senjanovi\'c,
    [arXiv:1012.4104 [hep-ph]];
  G.~Senjanovi\'c,
   [arXiv:0911.0029 [hep-ph]].
%
\bibitem{tommasini}
 D.~Tommasini, G.~Barenboim, J.~Bernabeu, C.~Jarlskog,
  Nucl.\ Phys.\  {\bf B444}, 451-467 (1995)
  [hep-ph/9503228].
%
\bibitem{Pilaftsis}
 A.~Datta, A.~Pilaftsis,
  Phys.\ Lett.\  {\bf B278}, 162-166 (1992);
 A.~Pilaftsis,
  Z.\ Phys.\  {\bf C55}, 275-282 (1992)
  [hep-ph/9901206].
%
\bibitem{ARC} 
 R.~Adhikari, A.~Raychaudhuri,
    [arXiv:1004.5111 [hep-ph]].
%
\bibitem{Rodejohann}
  H.~Hettmansperger, M.~Lindner, W.~Rodejohann,
  [arXiv:1102.3432 [hep-ph]].
%
\bibitem{0-others}
 W.~Buchmuller, C.~Greub,
  Phys.\ Lett.\  {\bf B256}, 465-470 (1991);
J.~Gluza,
  Acta Phys.\ Polon.\  {\bf B33}, 1735-1746 (2002)
  [hep-ph/0201002].
%
\bibitem{rg}
  P.~H.~Chankowski, Z.~Pluciennik,
  Phys.\ Lett.\  {\bf B316 } (1993)  312
  [hep-ph/9306333];
  K.~S.~Babu, C.~N.~Leung, J.~T.~Pantaleone,
  Phys.\ Lett.\  {\bf B319 } (1993)  191
  [hep-ph/9309223];
  S. Antusch, J. Kersten, M. Lindner, and M. Ratz,
  Phys. Lett. B538 (2002) 87
  [hep-ph/0203233]; 
  S. Antusch, J. Kersten, M. Lindner, M. Ratz, and M. A. Schmidt, 
  JHEP 03 (2005) 024
   [hep-ph/0501272]. 
  %
\bibitem{ingelman}
G.~Ingelman, J.~Rathsman,
  Z.\ Phys.\  {\bf C60}, 243-254 (1993).
 %
\bibitem{vs}
  F.~Vissani, A.~Yu.~Smirnov,
  Phys.\ Lett.\  {\bf B341 } (1994)  173-180
  [hep-ph/9405399].
  %
\bibitem{sv}
J. Schechter, J. W. F. Valle, 
Phys.\ Rev.\ {\bf D25} (1982) 2951. 
%
  \bibitem{lindner}
   M.~Duerr, M.~Lindner, A.~Merle,
  arXiv:1105.0901 [hep-ph].
 %
 %
\bibitem{vis-natural}
 F.~Vissani,
  Phys.\ Rev.\  {\bf D57}, 7027-7030 (1998)
  [hep-ph/9709409].
\bibitem{rub}
E.~K.~Akhmedov, V.~A.~Rubakov, A.~Y.~Smirnov,
  Phys.\ Rev.\ Lett.\  {\bf 81 } 1359
(1998)    [hep-ph/9803255].
\bibitem{shap}
A.~Boyarsky, O.~Ruchayskiy, M.~Shaposhnikov,
  Ann.\ Rev.\ Nucl.\ Part.\ Sci.\  {\bf 59}, 191 (2009)
  [arXiv:0901.0011 [hep-ph]].
%
\bibitem{goran-lep}
  L.~Boubekeur,
   [hep-ph/0208003];
 L.~Boubekeur, T.~Hambye, G.~Senjanovi\'c,
  Phys.\ Rev.\ Lett.\  {\bf 93}, 111601 (2004)
  [hep-ph/0404038].
%
%
\bibitem{type-II}
  M.~Magg and C.~Wetterich,
  Phys.\ Lett.\  B {\bf 94}, 61 (1980);
%
%
  C.~Wetterich,
  Nucl.\ Phys.\  B {\bf 187}, 343 (1981).
%
  
%
\bibitem{type-III}
 R.~Foot, H.~Lew, X.~G.~He and G.~C.~Joshi,
  Z.\ Phys.\  C {\bf 44}, 441 (1989);
%
 E.~Ma and D.~P.~Roy,
  Nucl.\ Phys.\  B {\bf 644}, 290 (2002)
  [arXiv:hep-ph/0206150].
%
\bibitem{goran-typeIII}
  B.~Bajc, G.~Senjanovi\'c,
  JHEP {\bf 0708}, 014 (2007)
  [hep-ph/0612029];
  B.~Bajc, M.~Nemev\v sek, G.~Senjanovi\'c,
  Phys.\ Rev.\  {\bf D76 } (2007)  055011
  [hep-ph/0703080];
   A.~Arhrib, B.~Bajc, D.~K.~Ghosh, T.~Han, 
  G.~-Y.~Huang, I.~Puljak, G.~Senjanovi\'c,
  Phys.\ Rev.\  {\bf D82}, 053004 (2010)
  [arXiv:0904.2390 [hep-ph]].

%
\bibitem{typeIIIgmsb}
 R.~N.~Mohapatra, N.~Okada, H.~-B.~Yu,
  Phys.\ Rev.\  {\bf D78}, 075011 (2008)
  [arXiv:0807.4524 [hep-ph]];
  P.~Fileviez Perez, H.~Iminniyaz, G.~Rodrigo, S.~Spinner,
  Phys.\ Rev.\  {\bf D81}, 095013 (2010)
  [arXiv:0911.1360 [hep-ph]].
%
\bibitem{us}
 P.~Bandyopadhyay, S.~Choubey, M.~Mitra,
  JHEP {\bf 0910}, 012 (2009)
  [arXiv:0906.5330 [hep-ph]];
S.~Choubey, M.~Mitra,
  JHEP {\bf 1005}, 021 (2010)
  [arXiv:0911.2030 [hep-ph]]; 
 R.~L.~Awasthi, S.~Choubey, M.~Mitra,
  [arXiv:1009.0509 [hep-ph]].
%
%
\bibitem{invo}
R.~N.~Mohapatra,
Phys.\ Rev.\ Lett.\  {\bf 56}, 561-563 (1986);
 R.~N.~Mohapatra, J.~W.~F.~Valle,
 Phys.\ Rev.\  {\bf D34}, 1642 (1986).
\bibitem{inv}
 D.~Wyler, L.~Wolfenstein,
  Nucl.\ Phys.\  {\bf B218}, 205 (1983); E.~Witten,
  Nucl.\ Phys.\  {\bf B268}, 79 (1986).
  %
 \bibitem{invdet}
J.~L.~Hewett, T.~G.~Rizzo,
  Phys.\ Rept.\  {\bf 183}, 193 (1989).
  %
\bibitem{inverseso10}
   P.~S.~B.~Dev, R.~N.~Mohapatra,
  Phys.\ Rev.\  {\bf D81}, 013001 (2010) 
  [arXiv:0910.3924 [hep-ph]];
  S.~Blanchet, P.~S.~B.~Dev, R.~N.~Mohapatra,
  Phys.\ Rev.\  {\bf D82}, 115025 (2010)
  [arXiv:1010.1471 [hep-ph]].

%
\bibitem{inverseothers}
  A.~Ilakovac, A.~Pilaftsis,
  Nucl.\ Phys.\  {\bf B437}, 491 (1995)
  [hep-ph/9403398];
 F.~Deppisch, J.~W.~F.~Valle,
 Phys.\ Rev.\  {\bf D72}, 036001 (2005) 
  [hep-ph/0406040];  
C.~Arina, F.~Bazzocchi, N.~Fornengo, J.~C.~Romao, J.~W.~F.~Valle,
  Phys.\ Rev.\ Lett.\  {\bf 101}, 161802 (2008)
  [arXiv:0806.3225 [hep-ph]]; 
 M.~Malinsky, T.~Ohlsson, Z.~-z.~Xing, H.~Zhang,
  Phys.\ Lett.\  {\bf B679}, 242-248 (2009)
  [arXiv:0905.2889 [hep-ph]];
M.~Hirsch, T.~Kernreiter, J.~C.~Romao, A.~Villanova del Moral,
  JHEP {\bf 1001}, 103 (2010)
  [arXiv:0910.2435 [hep-ph]];
F.~Deppisch, T.~S.~Kosmas, J.~W.~F.~Valle,
  Nucl.\ Phys.\  {\bf B752}, 80-92 (2006)
 [arXiv:0910.3924 [hep-ph]]. 


%
\bibitem{type-Ilfv}
  A.~Abada, C.~Biggio, F.~Bonnet, M.~B.~Gavela and T.~Hambye,
  JHEP {\bf 0712}, 061 (2007)
  [arXiv:0707.4058 [hep-ph]];
%
\bibitem{type-IIIlfv}
  A.~Abada, C.~Biggio, F.~Bonnet, M.~B.~Gavela and T.~Hambye,
  Phys.\ Rev.\  D {\bf 78}, 033007 (2008)
  [arXiv:0803.0481 [hep-ph]].
%
\bibitem{miha-lfv}
J.~F.~Kamenik, M.~Nemev\v sek,
  JHEP {\bf 0911}, 023 (2009)
  [arXiv:0908.3451 [hep-ph]].
%



  \end{thebibliography}
}
\end{document}